\documentclass[aps,prb,superscriptaddress,twocolumn,longbibliography]{revtex4-1}
\usepackage{bbm}
\usepackage{graphicx}
\usepackage{dcolumn}
\usepackage{bm}
\usepackage{subfigure}
\usepackage{amsmath}
\usepackage{feynmf}
\usepackage{hyperref}
\usepackage{color}

\usepackage{attachfile}
\newcommand{\ua}{\uparrow}
\newcommand{\da}{\downarrow}

\newcommand{\bs}{\boldsymbol}

\newcommand{\bk}{\boldsymbol k}

\newcommand{\bq}{\boldsymbol q}

\newcommand{\bd}{\boldsymbol d}

\newcommand{\bR}{\boldsymbol R}
\newcommand{\zhigang}{\color {black}}
\newcommand{\zhongbo}{\color {black}}
\usepackage{times}

\begin{document}
\title{Topological superconductivity in two-dimensional altermagnetic metals}

\author{Di Zhu}
\affiliation{Guangdong Provincial Key Laboratory of Magnetoelectric Physics and Devices, School of Physics, Sun Yat-sen University, Guangzhou 510275, China}

\author{Zheng-Yang Zhuang}
\affiliation{Guangdong Provincial Key Laboratory of Magnetoelectric Physics and Devices, School of Physics, Sun Yat-sen University, Guangzhou 510275, China}

\author{Zhigang Wu}
\email{wuzg@sustech.edu.cn}
\affiliation{Shenzhen Institute for Quantum Science and Engineering (SIQSE),
Southern University of Science and Technology, Shenzhen, P. R. China.}
\affiliation{International Quantum Academy, Shenzhen 518048, China.}
\affiliation{Guangdong Provincial Key Laboratory of Quantum Science and Engineering,
Southern University of Science and Technology Shenzhen, 518055, China.}

\author{Zhongbo Yan}
\email{yanzhb5@mail.sysu.edu.cn}
\affiliation{Guangdong Provincial Key Laboratory of Magnetoelectric Physics and Devices, School of Physics, Sun Yat-sen University, Guangzhou 510275, China}

\date{\today}

\begin{abstract}
Bringing magnetic metals into superconducting states represents an important approach for realizing unconventional superconductors and
potentially even topological superconductors. Altermagnetism, classified as a third basic collinear magnetic phase, gives rise to intriguing momentum-dependent
spin-splitting of the band structure, and results in an even number of spin-polarized Fermi surfaces due to the symmetry-enforced zero net magnetization. In this work, we investigate the effect of this new magnetic order on the superconductivity of a two-dimensional metal with $d$-wave
altermagnetism and Rashba spin-orbital coupling. Specifically we {\zhongbo consider an extended attractive Hubbard interaction,}
and determine the types of superconducting pairing that can occur in this system and ascertain whether they possess topological properties. Through self-consistent mean-field calculations, we find that the system in general
favors a mixture of spin-singlet $s$-wave and spin-triplet $p$-wave pairings, and
{\zhongbo that the altermagnetism is beneficial to the latter}.
Using symmetry arguments supported by detailed calculations, we show that
a number of topological superconductors, including both first-order
and second-order ones, can emerge when the $p$-wave
pairing dominates. In particular, we find that
the second-order topological superconductor is enforced by a $\mathcal{C}_{4z}\mathcal{T}$ symmetry,
which renders the spin polarization of Majorana corner modes into a unique entangled structure.
Our study demonstrates that altermagnetic metals are fascinating platforms for the exploration of
intrinsic unconventional superconductivity and topological superconductivity.
\end{abstract}

\maketitle

\section{Introduction}

Magnetism and superconductivity are two fundamental phenomena in nature, whose interplay in materials is one of the central topics in condensed matter physics~\cite{Scalapino1986,Moriya1990,Monthoux1991,Mathur1998,Saxena2000,Aoki2001,Mito2003,Akazawa2004,Huy2007,Li2011coexist,Bert2011,Dikin2011}.
Magnetism can influence superconductivity in many  ways, and its effect
on the pairing symmetry is of particular interest~\cite{Bergeret2005,Scalapino2012}. Generally speaking, magnetism is detrimental to spin-singlet superconductivity
 but is conducive to the emergence of unconventional superconductivity. Take ferromagnetism for example.
{Its \zhigang adverse} effect on spin-singlet pairings can be attributed to
the breaking of time-reversal symmetry (TRS) which lifts
the spin degeneracy of the {\zhigang electronic bands; this results in} spin-split Fermi surfaces on which electrons can no longer
find time-reversal partners to form spin-singlet Cooper pairs.
{\zhigang{Fortunately}}, a realistic system {\zhigang admits of} many competing pairing channels~\cite{Sigrist1991}.
While the spin-singlet pairing {\zhigang normally} wins out in time-reversal invariant
systems,  {\zhigang its suppression by magnetism means that other unconventional pairings could stand to benefit}.

Recently, it has been observed in a series of
materials with compensated magnetization that {\zhigang a third basic collinear magnetic order~\cite{Libor2020AM,Hayami2019AM,Hayami2020AM,Yuan2020AM,Yuan2021AM,Mazin2021,Liu2022AM,Feng2022AM,Betancourt2023,Mazin2023AM,Turek2022AM,Guo2023AM,Hariki2023AM,Zhou2023AM}, referred to as altermagnetism (AM)~\cite{Libor2022AMc,Libor2022AMa,Libor2022AMb}, exists beyond
the conventional dichotomy between ferromagnetism and antiferromagnetism}. {\zhigang The nomenclature is intended to convey the most important characteristic of this new magnetic order: that the spin polarization alternates in both coordinate and  momentum spaces. }The effect of AM on the electronic
band structure is {\zhigang rather} different from those of the ferromagnetism or antiferromagnetism.
Unlike usual antiferromagnetism due to symmetry reason~\cite{Yuan2020AM,Yuan2021AM},
AM results in momentum-dependent spin splitting to the band structure,
resembling a spin-orbital coupling effect but without spin-momentum locking~\cite{Libor2022AMb}.
 {\zhigang Although these spin-split bands also lead to spin-polarized Fermi surfaces  like in ferromagnetic metals, the Fermi surfaces are generally anisotropic as a result of the momentum-dependent spin polarization. In addition,
the number of spin-polarized Fermi surfaces in AM metals is constrained to be even due to the symmetry-enforced zero net magnetization.} {\zhigang Because of these unique properties,}
AM metals {\zhigang are emerging} as another intriguing class of systems to study the interaction between magnetism and
superconductivity~\cite{Mazin2022AM}.
{\zhigang Several novel} phenomena, such as orientation-dependent Andreev reflection~\cite{Sun2023AM,Papaj2023}, Josephson effect~\cite{Ouassou2023AM} and
finite-momentum Cooper pairing~\cite{Zhang2023AM}, have already been predicted in heterostructures composed of AM materials and superconductors.
{\zhongbo Notably, some parent compounds of high-temperature superconductors are revealed to be altermagnets~\cite{Libor2022AMa,Libor2022AMb}, raising the prospect that the coexistence of AM and superconductivity may be observed in a single material.}
However, the study of AM in general is still at an  {\zhigang early} stage, and {\zhigang fundamental} questions
such as what types of superconductivity may emerge in AM metals and whether they are topological remain to be answered.

In this work, {\zhigang we address these questions in the context of two-dimensional (2D) metals with $d$-wave AM~\cite{Libor2022AMb} and Rashba spin-orbital coupling (RSOC).  We incorporate the RSOC  because it
arises naturally when the AM metal is grown on a substrate~\cite{bychkov1984}. }Focusing on representative short-range
attractive interactions {\zhigang allowing for both $s$ and $p$-wave pairing channels, we first determine
the pairing phase diagram spanned by the RSOC strength and the relative $s$-to-$p$-wave pairing interaction strength. }
Our calculations show that the AM metal with RSOC favors a mixture of
spin-singlet $s$-wave and spin-triplet $p$-wave pairings, in contrast to the case of pure $s$ or $p$-wave pairings without RSOC; such mixed parity pairings are a result of the simultaneous
breaking of the TRS by AM and the inversion symmetry by RSOC.  For finite RSOC, two mixed parity pairing phases are found, namely the  $s$ $+$ helical $p$-wave phase and the $s$ $+$ chiral $p$-wave phase.  {\zhigang Notably, the former can {\zhigang prevail} over the latter for weak RSOC strengths} even though
the TRS is broken. {\zhigang We further investigate the topological properties of these pairings and identify a crucial set of  symmetries that can be used to delineate various topological phases. }
Using symmetry arguments corroborated by detailed calculations, we show that
the superconducting
phase is topologically trivial when the $s$-wave pairing dominates, regardless of the nature of the $p$-wave component; on the other hand, a multitude of topologically non-trivial phases can be realized when the $p$-wave pairing dominates. Specifically, it realizes a chiral TSC~\cite{read2000,sato2009non,Qi2010chiral,Sau2010TSC,alicea2010} characterized by an even Chern number for dominant chiral-$p$ wave pairings, and a helical TSC or a second-order TSC for dominant helical-$p$ wave pairings. In the latter scenario,  the superconductor is a helical TSC~\cite{Qi2009b,Deng2012,Nakosai2012,zhang2013kramers,wang2014TRI,Midtgaard2017,Huang2018helical,Zhang2021TSC,Feng2022TSC}
if, as in the case of no RSOC, a mirror symmetry or subsystem chiral
symmetry  exists to protect the helical Majorana modes; otherwise, it becomes a second-order
TSC with Majorana corner modes~\cite{Langbehn2017,Geier2018,Khalaf2018,Zhu2018hosc,Yan2018hosc,
Wang2018weak,Wang2018hosc,Liu2018hosc,Wu2019hosc,Yan2019hosca,Volpez2019SOTSC,Zhang2019hoscb,Pan2019SOTSC,
Zhu2019mixed,Hsu2020hosc,Wu2020SOTSC,Majid2020hoscb,wu2020boundaryobstructedb,Qin2022hosc,Zhu2022sublattice,Li2022hosc,Scammell2022hosc}  when these symmetries are broken by finite RSOC. These results spotlight 2D
superconducting AM metals as a remarkable platform in which
both 1D and 0D Majorana modes can be achieved.

\section{Model and results}

\subsection{2D metals with $d$-wave AM and RSOC}

We consider a 2D metal with $d$-wave AM described by the Hamiltonian $\hat H_0 = \sum_{\bk \sigma} h_{\sigma\sigma'}(\bk)c_{\bk\sigma}^\dag c_{\bk\sigma'}$. Expressed in terms of the Pauli matrices $\sigma_{x,y,z}$ and the identity matrix $\sigma_0$, $h_{\sigma\sigma'}(\bk)$ is given by
(lattice constant is set to unity throughout)~\cite{Libor2022AMb}
\begin{eqnarray}
h(\bk)&=&-2t(\cos k_{x}+\cos k_{y})\sigma_{0}+2t_{\rm AM}(\cos k_{x}-\cos k_{y})\sigma_{z}\nonumber\\
&&+2\lambda(\sin k_{y}\sigma_{x}-\sin k_{x}\sigma_{y}),\label{Hamiltonian}
\end{eqnarray}
{\zhigang where the $t_{\rm AM}$ term and the $\lambda$ term describe the exchange field associated with AM and  the RSOC respectively}. We note that the momentum dependence of
the AM exchange field resembles {\zhigang that of} the $d$-wave pairing in high-$T_{c}$
superconductors~\cite{Tsuei2000}.

\begin{figure}[t]
\centering
\includegraphics[width=0.45\textwidth]{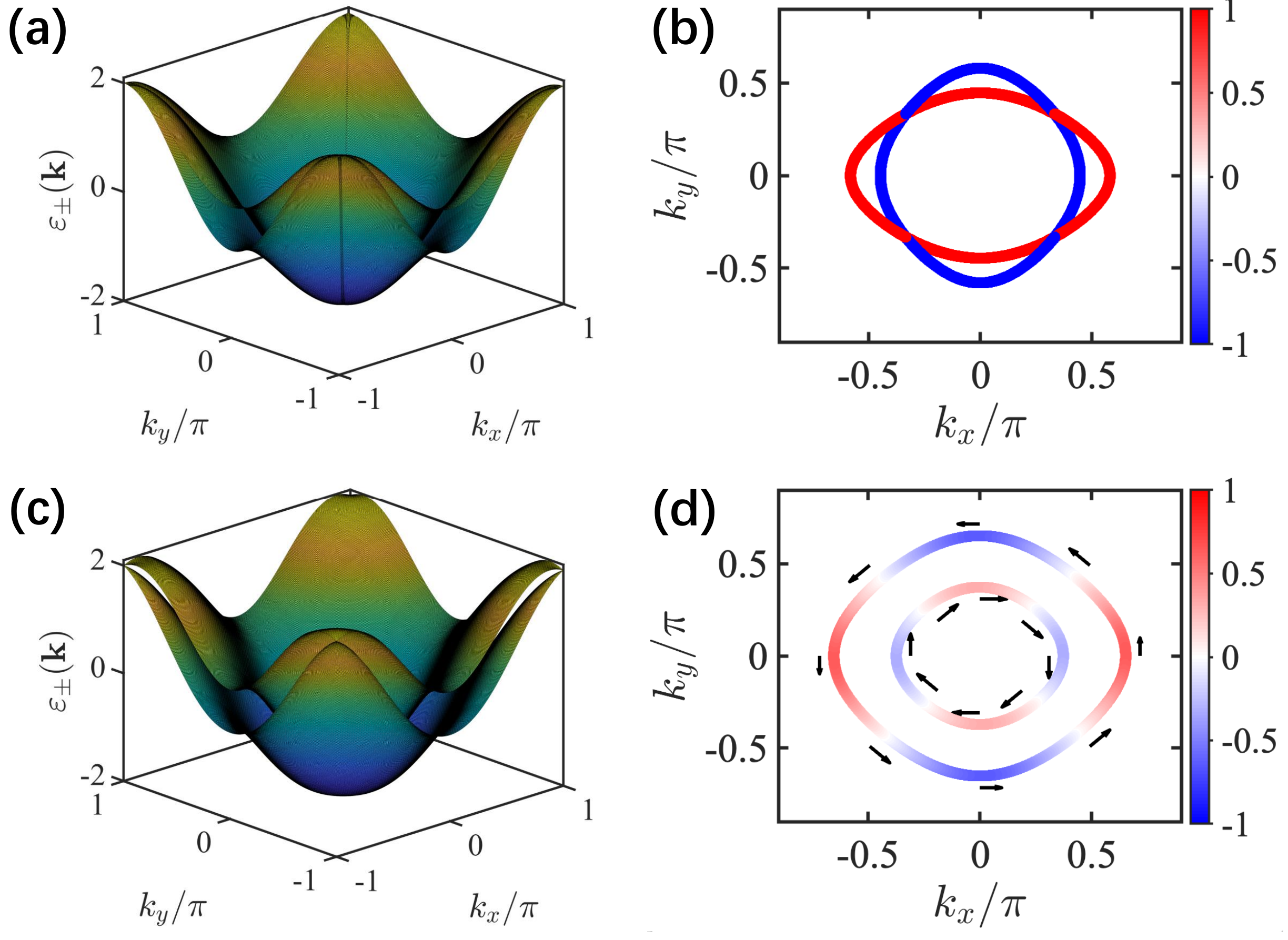}
\caption{(Color online) Band structures (left) and Fermi surfaces (right) of the AM metal with and without RSOC.
(a)-(b), $\lambda=0$; (c)-(d), $\lambda=0.2$.  The color of the Fermi surfaces indicates the magnitude of
$\langle \sigma_{z}\rangle$, and the length and direction of the arrows indicate the magnitude and direction of
the in-plane spin polarization. The spin textures clearly display the $\mathcal{C}_{4z}\mathcal{T}$ symmetry.
Common parameters are $t=0.5$, $t_{\rm AM}=0.1$, and $\mu=-1$.
}\label{fig1}
\end{figure}

{\zhigang The AM exchange field brings significant changes to the original band structure given by the first term of Eq.~(\ref{Hamiltonian}). First,  it breaks the TRS
$\mathcal{T}=-i\sigma_{y} \mathcal{K}$ ($\mathcal{K}$ is the complex conjugation operator) and gives rise to spin-split bands. Second,  its $d$-wave nature breaks the 4-fold rotational symmetry
$\mathcal{C}_{4z}=e^{i\frac{\pi}{4}\sigma_{z}}$ and results in the deformation of the Fermi surfaces. Both features can be clearly seen in Figs.~\ref{fig1}(a) and ~\ref{fig1}(b). Now, the additional RSOC also has important effects on the band structure. Because it breaks both the inversion
symmetry $\mathcal{P}=\sigma_{0}$ and the mirror symmetry $\mathcal{M}_{z}=i\sigma_{z}$ of the AM metal, the remaining degeneracies along the $|k_x|=|k_y|$ axes are removed (see Fig.~\ref{fig1}(c)).
Furthermore, it introduces spin-momentum locking on the spin-polarized Fermi surfaces, as
shown in Fig.\ref{fig1}(d).
All these properties play a role in determining the pairing symmetry and topological properties of the superconducting phases.

Another interesting fact about the Hamiltonian in Eq.~(\ref{Hamiltonian}) is that it preserves the overall $\mathcal{C}_{4z}\mathcal{T}$ symmetry; this can be seen from the fact that all three terms in Eq.~(\ref{Hamiltonian}) respect this symmetry. Two
 important properties of the AM metallic state immediately follow from this observation. First, the net magnetization of the metallic state
must be zero even though the two bands become spin-split. Second, such a state can be viewed as a critical metallic phase. To see this, we first note that the Kramers theorem dictates
the existence of band degeneracies at
the two $\mathcal{C}_{4z}\mathcal{T}$-invariant
momenta, i.e., $\boldsymbol{\Gamma}=(0,0)$ and $\boldsymbol{M}=(\pi,\pi)$, as shown in Fig.\ref{fig1}(c). Thus, an arbitrarily small out-of-plane magnetic field, which breaks the
$\mathcal{C}_{4z}\mathcal{T}$ symmetry, will open a gap at $\boldsymbol{\Gamma}$ and $\boldsymbol{M}$ and
drive the system to be a Chern metal where the two bands will
carry opposite Chern numbers $C=\pm1$~\cite{qi2006QWZ} (see Appendix \ref{Appendixa}). In addition, a reversal of the magnetic field's direction will
reverse the Chern number of the two bands. Such a critical behavior is a manifestation of the fact that
the band structure of AM metals differs drastically from those of ferromagnetic and antiferromagnetic metals.}

\subsection{Pairing phase diagram}

The pairing mechanism in a magnetic metal
is known to be non-unique~\cite{Monthoux2007}, {\zhongbo and the
effects of magnetism on superconductivity are also known to be diverse}.
{\zhongbo Importantly, magnetism and superconductivity are not always exclusive.
Indeed, the coexistence of magnetic and superconducting orders has
been observed in many materials, ranging from heavy-fermion systems~\cite{Saxena2000,Aoki2001,Mito2003,Akazawa2004,Huy2007} to oxide interfaces~\cite{Li2011coexist,Bert2011,Dikin2011}}.
In this work, we do not examine the microscopic origin of the pairing interaction and instead assume the following {\zhongbo extended attractive Hubbard} interaction
\begin{align}
\hat H_{int} = -V_s \sum_i n_{i\uparrow} n_{i\downarrow} -V_p \sum_{\langle ij\rangle,\sigma} n_{i\sigma} n_{j\sigma},\label{interaction}
\end{align}
where $n_{i\sigma}$ is the density operator for  electrons of spin $\sigma$ on site $i$, and $V_s$ and $V_p$ are respectively the strengths of the on-site and nearest-neighbor attraction. Equation (\ref{interaction}) is a minimal form of interaction that is capable of describing the
competition between spin-singlet and spin triplet pairings.

Following the standard BCS theory, we define the gap function as
\begin{align}
\Xi_{\sigma\sigma'}(\bk)  = -\frac{1}{N_L} \sum_{\bk'} V_{{\sigma\sigma'}}(\bk-\bk') \langle c_{\bk' \sigma}c_{-\bk' \sigma'} \rangle,
\label{mgf}
\end{align}
where $N_L$ is the number of lattice sites and $V_{\sigma\sigma'}(\bq) = -(1-\delta_{\sigma,\sigma'})V_s - \delta_{\sigma,\sigma'}2V_p(\cos q_x + \cos q_y )$ is the interaction in Fourier space. The pair correlation can be calculated in terms of the Bogoliubov amplitudes as
\begin{align}
\langle c_{\bk \sigma}c_{-\bk \sigma'}  \rangle =  u_{\sigma,1}(\bs k)v_{\sigma',1}^*(\bs k) + u_{\sigma,2}(\bs k)v_{\sigma',2}^*(\bs k).
\end{align}
These amplitudes are determined by the Bogoliubov-de Gennes {\zhongbo (BdG)} equation
$
\mathcal{H}_{\rm BdG}(\bs k) \chi_{\bs k,l} = E_{\bs k, l} \chi_{\bs k,l}
$, {\zhongbo where $E_{\bs k,l}$ with $l=\{1,2\}$ refer to the two positive eigenenergies,
$\chi_{\bs k,l} \equiv [u_{\uparrow, l}(\bs k) , u_{\downarrow, l}(\bs k), v_{\uparrow, l}(\bs k), v_{\downarrow, l}(\bs k) ]^T$
are the corresponding eigenstates},   and
\begin{equation}
\mathcal{H}_{\rm BdG}(\bs k)=
\begin{bmatrix}
\xi_{\bk \uparrow} & \Lambda(\bk) & \Xi_{\uparrow\uparrow}(\bk) &   \Xi_{\uparrow \downarrow}(\bk) \\
\Lambda^*(\bk) & \xi_{\bk \downarrow} & \Xi_{\downarrow\uparrow}(\bk) &   \Xi_{\downarrow\downarrow}(\bk)  \\
 \Xi^*_{\uparrow\uparrow}(\bk)  & \Xi^*_{\downarrow\uparrow}(\bk) & -\xi_{\bk\uparrow }  & -\Lambda^*(-\bk) \\
\Xi^*_{\uparrow\downarrow}(\bk) &\Xi^*_{\downarrow\downarrow}(\bk)   & -\Lambda(-\bk)& -\xi_{\bk\downarrow}
\end{bmatrix}.
\label{meq:BdGham}
\end{equation}
Here $\Lambda(\bk)=2\lambda (\sin k_y + i\sin k_x)$ and $\xi_{\bk s}=-2(t-st_{\rm AM})\cos k_x - 2(t+st_{\rm AM})\cos k_y -\mu$ where
$s=1$ ($-1$) for spin up (down).

To determine possible pairing channels, one may expand  both $V_{\sigma\sigma'}({\bk-\bk'})$ and $\Xi_{\sigma\sigma'}(\bk)$ in terms of the so-called square lattice harmonics $g_\eta (\bk)$ (see Appendix \ref{Appendixb}), i.e.,
\begin{align}
\label{Vexp}
V_{\sigma\sigma'}({\bk-\bk'}) &= \sum_{\eta} \gamma^\eta_{\sigma\sigma'} g_\eta(\bk) g_\eta^*(\bk'); \\
\Xi_{\sigma\sigma'}(\bk) & = \sum_\eta \Delta_{\sigma\sigma'}^\eta g_\eta (\bk),
\label{Xiexp}
\end{align}
where $\gamma^\eta_{\sigma\sigma'}$ is the strength of the pairing interaction in the $\eta$-channel.
For the {\zhongbo attractive interaction given in Eq.(\ref{interaction})}, the only relevant channels are
the $s$-wave one $g_s (\bk)= 1$, and the $p$-wave ones $g_{p\pm} (\bk)= \sin k_x \pm i\sin k_y$. {\zhongbo
Pairing channels with higher angular momentum, e.g., $d$-wave
or $f$-wave pairing channel, are absent due to the Fermi statistics and the restricted range of the interaction
considered}.
Substituting the expansions in Eqs.~(\ref{Vexp})-(\ref{Xiexp}) into Eqs.~(\ref{mgf})-(\ref{meq:BdGham}), we can then solve for the $\eta$-channel  pairing amplitude  $\Delta_{\sigma\sigma'}^\eta$ self-consistently.

For finite RSOC,  the gap equation admits only mixed parity solutions, of which two specific types are candidates of the ground state. They are (i) mixture of $s$-wave and chiral $p$-wave pairing for which $\Xi_{\uparrow\downarrow}(\bk) = - \Xi_{\downarrow\uparrow}(\bk) =\Delta^s_{\uparrow\downarrow} g_s(\bk)$,  $\Xi_{\uparrow\uparrow}(\bk)  = \Delta^{p+}_{\uparrow\uparrow} g_{p+}(\bk)$ and $\Xi_{\downarrow\downarrow}(\bk)  = \Delta^{p+}_{\downarrow\downarrow} g_{p+}(\bk)$; and (ii) mixture of $s$-wave and helical $p$-wave pairing for which $\Xi_{\uparrow\downarrow}(\bk) = - \Xi_{\downarrow\uparrow}(\bk) =\Delta^s_{\uparrow\downarrow} g_s(\bk)$ and $\Xi_{\uparrow\uparrow}(\bk)  =-\Xi^*_{\downarrow\downarrow}(\bk)  = \Delta^{p+}_{\uparrow\uparrow} g_{p+}(\bk)$. We note that for the chiral $p$-wave pairing,  another degenerate solution exists corresponding to $\Xi_{\uparrow\uparrow}(\bk)  = \Delta^{p-}_{\uparrow\uparrow} g_{p-}(\bk)$ and $\Xi_{\downarrow\downarrow}(\bk)  = \Delta^{p-}_{\downarrow\downarrow} g_{p-}(\bk)$. The fact that the solutions are exclusively mixed parity is a natural consequence of the lack of inversion symmetry in the system~\cite{Gorkov2001}. It is also consistent with our findings that only pure $s$ or $p$-wave solutions are found when the inversion symmetry is restored by letting $\lambda = 0$.

\begin{figure}[t]
\centering
\includegraphics[width=0.48\textwidth]{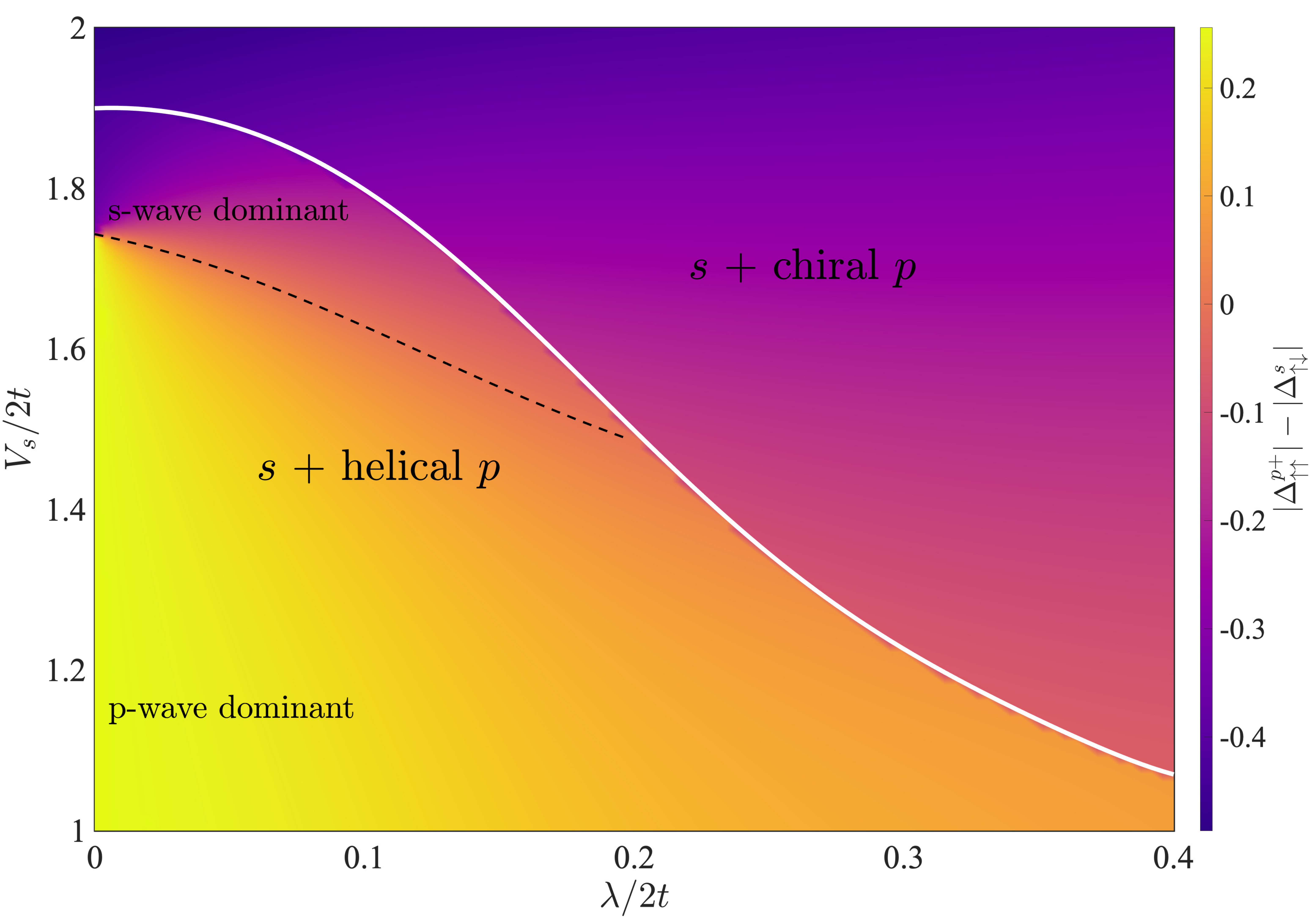}
\caption{(Color online) A representative pairing phase diagram. The white
solid line divides the pairing phase diagram into two regions, with one favoring
the mixed $s+\text{helical}\,p$-wave pairing and the other favoring
the mixed $s+\text{chiral}\,p$-wave pairing. The former preserves
the $\mathcal{C}_{4z}\mathcal{T}$ symmetry, while the latter does not.
Below the black dashed line, along which $|\Delta_{\uparrow\downarrow}^{s}|=|\Delta_{\uparrow\uparrow}^{p+}|$,
is a sizable region with dominant  helical $p$-wave pairing.
Here the parameters are $t=0.5$, $t_{\rm AM}=0.1$, $\mu=-1$ and $V_{p}=1.5$.
}\label{pairing}
\end{figure}

Both types of the pairing solutions are found in the same parameter space and so we need to compare their corresponding condensation energies to determine the pairing ground state. The resulting pairing phase diagram takes
the generic structure shown in Fig.\ref{pairing}. We see that the superconductor favors   a mixed $s+$ chiral $p$-wave pairing for strong RSOC and a mixed $s+$ helical $p$-wave pairing for weak RSOC.  In the former phase, the $s$-wave component is always dominant, whereas in the latter the $p$-wave component can dominate over the $s$-wave one for a significant range of $V_{s}/V_{p}$. The $s$-wave dominant and the $p$-wave dominant pairings indeed regress to the pure $s$-wave and the pure $p$-wave pairings respectively in the $\lambda\rightarrow 0$ limit. However, in the case of pure $p$-wave pairing,   chiral and helical $p$-wave pairings are completely degenerate.

The two phases in Fig.~\ref{pairing} are not only distinguished by the nature of the $p$-wave pairings but also by their different magnetic properties. Since the pairing amplitudes among spin up and spin down electrons are not equal for mixed chiral $p$-wave pairing, a finite net magnetization emerges in this phase. Thus, the fact that this phase is favored for strong RSOC is rather reminiscent of the charge-neutral atomic superfluid with SOC, where a strong SOC  leads to chirality as well as finite magnetization~\cite{Sun2018,Chench2023}.  Lastly, these two phases can also be differentiated by whether they preserves the $\mathcal{C}_{4z}\mathcal{T}$ symmetry. As we shall see immediately, this turns out to be very consequential for the topological properties of the superconducting phases.

\subsection{Diverse topological superconducting phases}

Based on the pairing phase diagram and the BdG
Hamiltonian (\ref{meq:BdGham}), we investigate possible topological superconducting phases. {\zhigang We begin with the $\lambda = 0$ axis on the phase diagram, where only pure parity pairing occurs. Since the three possible pairings can be classified by symmetry properties, the first thing to note is that} if the superconducting state possesses the $\mathcal{C}_{4z}\mathcal{T}$ symmetry it will forbid
the presence of chiral TSC even though the TRS
has been broken by AM. This fact can be intuitively recognized since in such a scenario
two edges related by $\mathcal{C}_{4z}$ rotation will carry chiral Majorana modes with opposite
chiralities due to the concomitant time-reversal operation, implying a zero net Chern number.
This symmetry argument suggests that among the three possible pairings,
only the chiral $p$-wave pairing, which
breaks the $\mathcal{C}_{4z}\mathcal{T}$ symmetry of the normal state, can lead
to the realization of chiral TSCs. For a chiral $p$-wave superconductor, the Chern number $C$ has a simple
relation to the number of Fermi surfaces $N_{\rm FS}$ enclosing one time-reversal
invariant momentum, i.e., $(-1)^{C}=(-1)^{N_{\rm FS}}$~\cite{sato2010odd}.
Since the zero net magnetization of the normal state in the AM metal
demands an even ${N_{\rm FS}}$, an even Chern number
is thus enforced. For example, we find $C=-2$ for the Fermi surface configuration shown in Fig.\ref{fig1}(b).

\begin{figure}[t]
\centering
\includegraphics[width=0.45\textwidth]{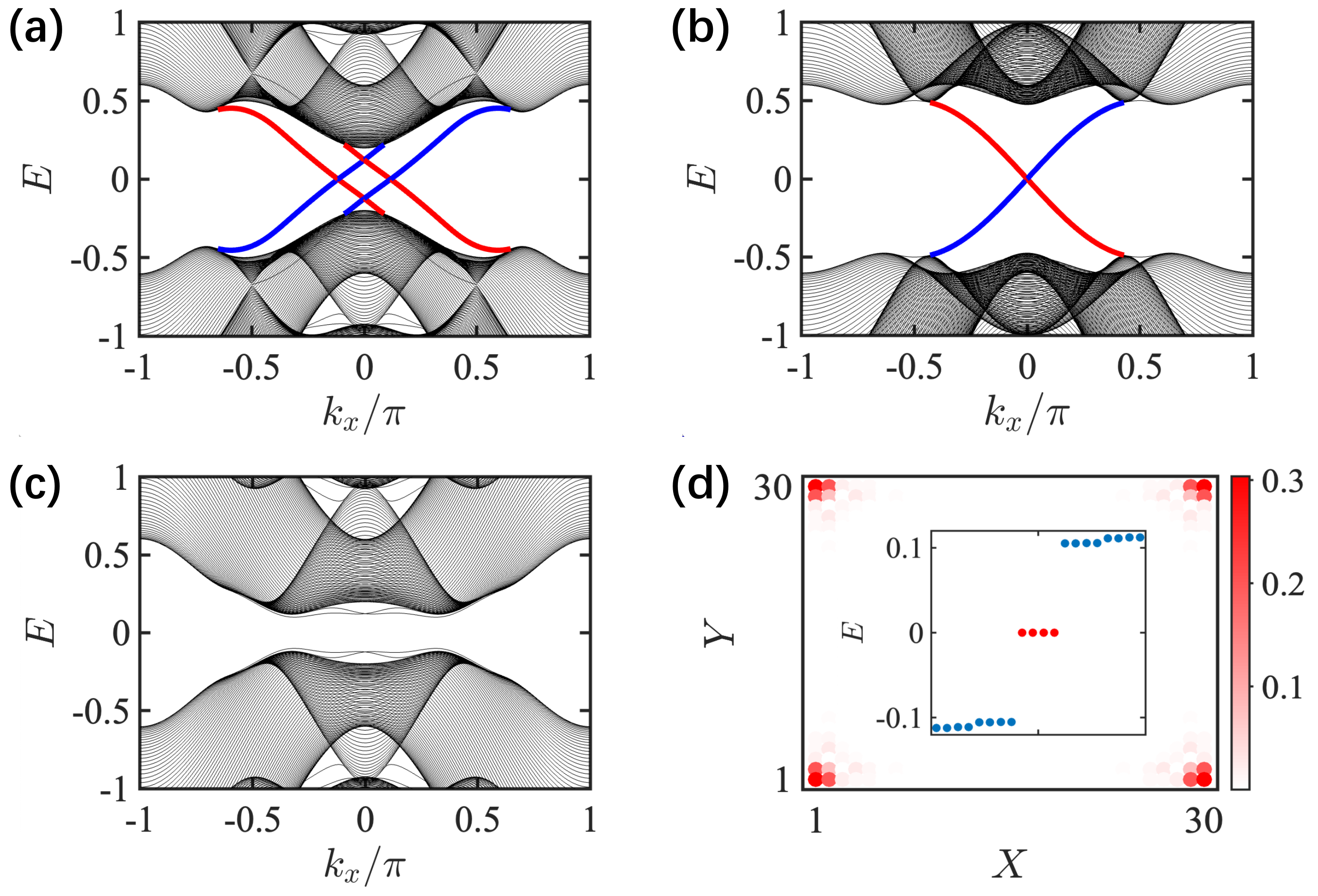}
\caption{(Color online) (a) A chiral TSC with Chern number $C=-2$. The blue
and red solid lines refer to chiral Majorana modes on two opposite edges.
(b) A helical TSC with a pair of helical
Majorana modes. The blue and red solid lines are doubly degenerate. (c)-(d) A second-order TSC with gapped edge spectra and
four Majorana corner modes. The four red dots in the inset of (d) refers to four Majorana zero modes,
with their probability density profiles concentrating around the four corners.
Common parameters are $t=0.5$, $t_{\rm AM}=0.1$, and $\mu=-1$. The rest of parameters are as follows. (a) $\lambda=0.2$, $\Delta_{\uparrow\downarrow}^{s}=0.05$,
and $\Delta_{\uparrow\uparrow}^{p+}=\Delta_{\downarrow\downarrow}^{p+}=0.25$ for the chiral
$p$-wave pairing amplitude;
(b) $\lambda=0$, $\Delta_{\uparrow\downarrow}^{s}=0$;
(c)-(d) $\lambda=0.2$, $\Delta_{\uparrow\downarrow}^{s}=0.05$.
In (b)-(d), the helical $p$-wave pairing amplitude is chosen to be $\Delta_{\uparrow\uparrow}^{p+}=0.25$.
}\label{diverse}
\end{figure}

{\zhigang We focus on the chiral $p$-wave pairing for the moment and move into regions of finite $\lambda$ in the phase diagram, where the pairings are now mixed with an $s$-wave component.} In this case, whether
such a mixed-parity superconducting phase supports
a chiral TSC  hinges on the relative weight between the two pairing components.
When the $s$-wave component dominates, the superconducting phase is topologically
trivial since it is adiabatically
connected to the pure $s$-wave limit with $\mathcal{C}_{4z}\mathcal{T}$ symmetry
as long as the bulk gap remains open.
Similarly, when the chiral $p$-wave pairing dominates, the superconducting
phase is adiabatically connected to the pure chiral $p$-wave limit and retains the topological properties of that limit.
A calculation of the
energy spectrum of a cylindrical sample shows the existence of
two chiral Majorana modes on an open edge,
also confirming the realization of a chiral TSC with $C=-2$ when the chiral
$p$-wave pairing dominates, as shown in
Fig.\ref{diverse}(a). Due to the even $C$ constraint, this mixed chiral $p$-wave phase must transition directly to a trivial phase with $C=0$ when the $s$-wave component gradually increases.

Now let us turn to the helical $p$-wave pairing.
In the $\lambda=0$ limit, we find that the superconductor with
pure helical $p$-wave pairing supports helical Majorana
modes protected by mirror symmetry $\mathcal{M}_{z}$~\cite{teo2008,Feng2022TSC,Zhang2013mirror}
or subsystem chiral symmetries~\cite{Ryu2010,Zhu2023sublattice}.
An example is provided in Fig.\ref{diverse}(b), where we show that
the resulting superconductor carries a pair of helical
Majorana modes on the boundary for  a Fermi surface configuration
shown in Fig.\ref{fig1}(b).
When the RSOC becomes finite, gapless Majorana modes are always absent,
regardless of which component of the mixed $s+\text{helical}\,p$-wave pairing
dominates, as exemplified in Fig.\ref{diverse}(c). Despite being trivial in first-order topology, the superconductor is in fact a second-order TSC when the helical
$p$-wave pairing dominates. Indeed,
considering a square sample with open boundary conditions in both
$x$ and $y$ directions, we find the hallmark of a second-order TSC,
the existence of Majorana corner modes~\cite{Langbehn2017,Geier2018,Khalaf2018,Zhu2018hosc,Yan2018hosc,
Wang2018weak,Wang2018hosc},
as shown in Fig.\ref{diverse}(d).

The arising of a second-order TSC in the region with dominant helical $p$-wave pairing
can also be understood via adiabatic and symmetry arguments.
As mentioned above the helical Majorana modes in the $\lambda=0$ limit
are protected by the mirror symmetry $\mathcal{M}_{z}$
or subsystem chiral symmetries (see Appendix \ref{Appendixc}). However,
all these symmetry protections are removed once $\lambda$ becomes finite,
giving rise to a Dirac mass on the boundary to gap out the helical
Majorana modes. On the other hand,
the $\mathcal{C}_{4z}\mathcal{T}$ symmetry is retained in this
mixed parity superconductor, which forces
the Dirac masses on two nearby $\mathcal{C}_{4z}$-related edges
to be opposite, and hence leads to the emergence of Majorana corner modes
according to the Jackiw-Rebbi theory~\cite{jackiw1976b}; this is
very reminiscent of the scenario for
the $\mathcal{C}_{4z}\mathcal{T}$ symmetry-enforced
second-order topological insulator in 3D~\cite{Schindler2018}.
A rather unique property of this symmetry-enforced second-order TSC is that
the spin polarization of the four
Majorana corner modes are entangled. That is, owing to the constraint from the $\mathcal{C}_{4z}\mathcal{T}$ symmetry, their out-of-plane spin polarizations will
form a quadrupole structure~\cite{Plekhanov2021}, and the in-plane
spin polarization will form a four-hour-clock-like structure.
In experiments, such entangled structures of spin polarization can be detected
by spin-polarized scanning tunneling microscopes as a defining signature
of this second-order TSC~\cite{He2014spin,Sun2016Majorana,Jeon2017}.

\section{Discussions and conclusions}

We have investigated the basic question
of what kind of superconductivity and TSCs may emerge in 2D AM metals.
A set of important symmetries relevant to the 2D AM metal with RSOC are unveiled, which place various constraints on the band structure,
spin textures, and the pairing types for the realization of TSCs. Guided by the symmetry analysis, we have shown that the AM metal favors mixed parity pairings, and a multitude of
TSCs, including both first-order and second-order TSCs, can emerge when the spin-triplet
$p$-wave pairings dominate. We have also shown that the
spin polarizations of Majorana corner modes on a square lattice take
intriguing structures as the second-order TSC is enforced by
the $\mathcal{C}_{4z}\mathcal{T}$ symmetry.
All these findings prove that AM metals have unique band structures
and can give birth to unconventional pairings and TSCs with fascinating properties.

\section*{Acknowledgements}

D. Z., Z.-Y. Z, and Z. Y. are supported by the National Natural Science Foundation of China (Grant No. 12174455) and the Natural Science Foundation of Guangdong Province
(Grant No. 2021B1515020026). Z. W. is supported by National Key R$\&$D Program of China (Grant No. 2022YFA1404103), NSFC (Grant No.~11974161) and Shenzhen Science and Technology Program (Grant No.~KQTD20200820113010023)

\appendix

\section{Chern metals driven by an out-of-plane magnetic field\label{Appendixa}}

By applying an out-of-plane magnetic field and only taking into account the resulting Zeeman-splitting effect, the 
Hamiltonian becomes 
\begin{eqnarray}
h(\bk)&=&-2t(\cos k_{x}+\cos k_{y})\sigma_{0}+2t_{\rm AM}(\cos k_{x}-\cos k_{y})\sigma_{z}\nonumber\\
&&+2\lambda(\sin k_{y}\sigma_{x}-\sin k_{x}\sigma_{y})+B_{z}\sigma_{z},
\end{eqnarray}
where the last term, $B_{z}\sigma_{z}$, denotes the Zeeman field.  It is readily checked that 
the Zeeman field preserves the $\mathcal{C}_{4z}$ rotational symmetry but breaks the time-reversal symmetry, thereby breaking
the combined $\mathcal{C}_{4z}\mathcal{T}$ symmetry. As a result, a nonzero Chern number becomes allowed.

For the two-band Hamiltonian,  the Berry curvature can be simply determined by the following formula~\cite{qi2006QWZ}
\begin{eqnarray}
\Omega_{\pm}(\bk)=\pm\frac{\bd(\bk)\cdot(\partial_{k_{x}}\bd(\bk)\times\partial_{k_{y}}\bd(\bk))}{2d^{3}(\bk)}
\end{eqnarray}
where $\pm$ refer to the upper and lower band respectively,
$\bd(\bk)=(2\lambda \sin k_{y},-2\lambda\sin k_{x},B_{z}+2t_{\rm AM}(\cos k_{x}-\cos k_{y}))$, and $d(\bk)=|\bd(\bk)|$.
A straightforward  calculation gives
\begin{eqnarray}
\Omega_{\pm}(\bk)=\pm\frac{2\lambda^{2}[B_{z}\cos k_{x}\cos k_{y}-2t_{\rm AM}(\cos k_{x}-\cos k_{y})]}
{d^{3}(\bk)}.\nonumber\\
\end{eqnarray}
In the limit of $B_{z}=0$, the Berry curvature has the property $\Omega_{\pm}(k_{x},k_{y})=-\Omega_{\pm}(k_{y},-k_{x})$, also
implying that the Chern number, which is the integral of the Berry curvature over the Brillouin zone, identically vanishes.
For a finite $B_{z}$, a calculation of the Chern number yields
\begin{eqnarray}
C_{\pm}&=&\frac{1}{2\pi}\int_{BZ}\Omega_{\pm}(\bk)d^{2}k\nonumber\\
&=&\pm\left\{\begin{array}{cc}
            sgn(B_{z}t_{\rm AM}), &  0<|B_{z}|<4|t_{\rm AM}|, \\
            0, &  |B_{z}|>4|t_{\rm AM}|.
          \end{array}\right.\label{Chern}
\end{eqnarray}
This result shows that an arbitrarily small out-of-plane magnetic field
will open a gap to the spectrum, and renders a nonzero Chern number to the bands.
In addition,  it is easy to see from Eq.(\ref{Chern}) that
a reversal of the magnetic field's direction will reverse the Chern number of the two bands,
indicating that the Hamiltonian (\ref{Hamiltonian}) describes a critical metallic phase.

\section{The determination of pairing phase diagram\label{Appendixb}}

In this part, we provide more details on the determination of the pairing phase diagram.
We consider density-density interactions between electrons and assume the existence
of discrete translational symmetry. Accordingly, the interaction takes the generic form
\begin{eqnarray}
\hat H_{int} &=& \frac {1}{ 2} \sum_{ij\sigma\sigma'} V_{{\sigma\sigma'}}(\bR_i-\bR_j) n_{i\sigma} n_{j\sigma'} \nonumber\\ &=&\frac{1}{2N_L}\sum_{\bq\sigma\sigma'}V_{{\sigma\sigma'}}(\bq) n_{\bq\sigma}n_{-\bq\sigma'},
\end{eqnarray}
where $\bR_i$ is the lattice site, $N_L$ is the total number of sites,
\begin{eqnarray}
n_{\bq\sigma} = \sum_i n_{i\sigma}e^{-i\bq\cdot\bR_i} = \sum_{\bk} c^\dag_{\bk\sigma}c_{\bk-\bq \sigma},
\end{eqnarray}
and
\begin{eqnarray}
V_{{\sigma\sigma'}}(\bq)  &=& \sum_{j} V_{{\sigma\sigma'}}(\bR_i-\bR_j) e^{-i\bq\cdot(\bR_i - \bR_j)}\nonumber\\
&=&V_{{\sigma\sigma'}}(-\bq).
\end{eqnarray}
For the last equation, we have used the property $V_{{\sigma\sigma'}}(\bR_i-\bR_j)=V_{{\sigma\sigma'}}(\bR_j-\bR_i)$.

Following the Bardeen-Cooper-Schrieffer (BCS) theory, the pairing interaction is simplified as
\begin{eqnarray}
\hat H_{int} \approx \frac{1}{2}\sum_{\bk\sigma\sigma'}\Xi_{\sigma\sigma'}(\bk) c^\dag_{\bk\sigma}c^\dag_{-\bk\sigma'} + h.c.,
\end{eqnarray}
where the gap function is defined as
\begin{eqnarray}
\Xi_{\sigma\sigma'}(\bk)  = -\frac{1}{N_L} \sum_{\bk'} V_{{\sigma\sigma'}}(\bk-\bk') \langle c_{\bk' \sigma}c_{-\bk' \sigma'} \rangle.
\label{gf}
\end{eqnarray}
Fermi statistics and the fact that $V_{\sigma\sigma'}(\bq)= V_{\sigma'\sigma}(-\bq) $ lead to the following property of the gap function,
\begin{eqnarray}
\Xi_{\sigma\sigma'}(\bk) = -  \Xi_{\sigma'\sigma}(-\bk).
\label{sym}
\end{eqnarray}
To determine possible pairing channels, one may expand  $V_{\sigma\sigma'}({\bk-\bk'})$ in terms of the so-called square lattice harmonics $g_\eta (\bk)$
\begin{align}
V_{\sigma\sigma'}({\bk-\bk'}) = \sum_{\eta} \gamma^\eta_{\sigma\sigma'} g_\eta(\bk) g_\eta^*(\bk'),
\end{align}
where $\gamma^\eta_{\sigma\sigma'}$ is the strength of the pairing interaction in the $\eta$-channel. Examples of the harmonics include the $s$-wave $g_s (\bk)= 1$, the extended $s$-wave $g_{es}(\bk)=\cos k_x + \cos k_y$, the $p$-waves $g_{p\pm} (\bk)= \sin k_x \pm i\sin k_y$ and the $d$-wave $g_d (\bk)= \cos k_x - \cos k_y$.  The gap function can similarly be  written as
\begin{align}
\Xi_{\sigma\sigma'}(\bk)  = \sum_\eta \Delta_{\sigma\sigma'}^\eta g_\eta (\bk).
\label{gfexp}
\end{align}
Substituting Eq.~(\ref{gfexp}) into Eq.~(\ref{gf}), we obtain the amplitudes of each pairing channel as
\begin{align}
\Delta_{\sigma\sigma'}^\eta = -\frac{1}{N_L} \sum_{\bk'}\gamma^\eta_{\sigma\sigma'} g_\eta^*(\bk')\langle c_{\bk' \sigma}c_{-\bk' \sigma'} \rangle.
\label{gapeq}
\end{align}
Introducing the Nambu basis $\Psi_{\bs k} = (c_{\bs k \ua}, c_{\bs k \da}, c^{\dagger}_{-\bs k\ua}, c^{\dagger}_{-\bs k \da})^T$, the BCS Hamiltonian can be written as
\begin{equation}
\hat H -\mu \hat{N} = \frac{1}{2} \sum_{\bs k}
\Psi^\dagger_{\bs k}
\mathcal{H}_{\rm BdG}(\bs k)
\Psi_{\bs k},
\end{equation}
with
\begin{equation}
\mathcal{H}_{\rm BdG}(\bs k) =
\begin{bmatrix}
\xi_{\bs k \ua} & \Lambda(\bk) & \Xi_{\ua\ua}(\bs k) &   \Xi_{\ua \da}(\bs k) \\
\Lambda^*(\bk) & \xi_{\bs k \da} & \Xi_{\da\ua}(\bs k) &   \Xi_{\da\da}(\bs k)  \\
 \Xi^*_{\ua\ua}(\bs k)  & \Xi^*_{\da\ua}(\bs k) & -\xi_{\bs k\ua }  & -\Lambda^*(-\bk) \\
\Xi^*_{\ua\da}(\bs k) &\Xi^*_{\da\da}(\bs k)   & -\Lambda(-\bk)& -\xi_{\bs k\da}
\end{bmatrix},
\label{eq:BdGham}
\end{equation}
where
\begin{align}
\xi_{\bk \ua} &= -2(t-t_{\rm AM})\cos k_x - 2(t+t_{\rm AM})\cos k_y -\mu,  \nonumber\\
\xi_{\bk \da} &= -2(t+t_{\rm AM})\cos k_x - 2(t-t_{\rm AM})\cos k_y -\mu,  \nonumber\\
\Lambda(\bk) & = 2\lambda (\sin k_y + i\sin k_x).
\end{align}
The above Hamiltonian can be diagonalized by the following Bogoliubov transformation
\begin{eqnarray}
c_{\bs k \sigma}&=& u_{\sigma,1}(\bs k) b_{\bs k,1} + v_{\sigma,1}^* (-\bs k) b^\dag_{-\bs k,1}  \nonumber\\
&&+ u_{\sigma,2}(\bs k) b_{\bs k,2} + v_{\sigma,2}^* (-\bs k) b^\dag_{-\bs k,2},
\label{BT}
\end{eqnarray}
where $u_{\sigma,l},v_{\sigma,l}$ with $l=\{1,2\}$ are the Bogoliubov amplitudes and $b_{\bs k,l}$ refer to the two quasiparticle operators. The Bogoliubov amplitudes are obtained from the eigenvalue equation
\begin{align}
\mathcal{H}_{\rm BdG}(\bs k) \chi_{\bs k,l} = E_{\bs k, l} \chi_{\bs k,l},
\label{Heig}
\end{align} where $E_{\bs k, l}$ refer to the two positive eigenenergies, and $\chi_{\bs k,l} \equiv [u_{\uparrow,l}(\bs k) , u_{\downarrow,l}(\bs k), v_{\uparrow,l}(\bs k), v_{\downarrow,l}(\bs k) ]^T$ are the corresponding eigenstates.
Using Eq.~(\ref{BT}), the gap equation (\ref{gapeq})  can be written as
\begin{eqnarray}
\Delta_{\sigma\sigma'}^\eta &=& - \frac{1}{N_L} \sum_{\bk}\gamma_{\sigma\sigma'}^\eta  g^*_\eta (\bs k)\nonumber\\
&&\times\left [ u_{\sigma,1}(\bs k)v_{\sigma',1}^*(\bs k) + u_{\sigma,2}(\bs k)v_{\sigma',2}^*(\bs k)\right ].
\label{gapeq1}
\end{eqnarray}
Once the gap functions are obtained, the free energy of the superconductor at zero temperature can be calculated as
\begin{eqnarray}
 F_S &=& -\frac{1}{2} \sum_{\bs k} [E_{\bk, +} + E_{\bk, -} -\xi_{\bk\uparrow} - \xi_{\bk\downarrow}]+ \frac{1}{2}\sum_{\sigma\sigma' {\bk}} \Xi^*_{\sigma\sigma'}(\bk) \nonumber\\
 &&\times\left [ u_{\sigma,1}(\bk)v_{\sigma',1}^*(\bk) + u_{\sigma,2}(\bk)v_{\sigma',2}^*(\bk)\right ].
\end{eqnarray}
The condensation energy is
\begin{align}
\delta  E = F_S - F_N,
\end{align}
where $F_N$ is the free energy of the normal state
\begin{align}
F_N = \sum_{\bk,\alpha=\pm}  (\varepsilon_\alpha(\bk) - \mu) \theta(\mu -\varepsilon_\alpha(\bk)  ).
\end{align}
where the energy spectra $\varepsilon_{\pm}(\bk)=-2t(\cos k_{x}+\cos k_{y})\pm2\sqrt{(t_{\rm AM}(\cos k_{x}-\cos k_{y}))^{2}+\lambda^{2}(\sin^{2}k_{x}+\sin^{2}k_{y})}$.

\begin{figure*}[htb]
\subfigure{\includegraphics[width=7cm]{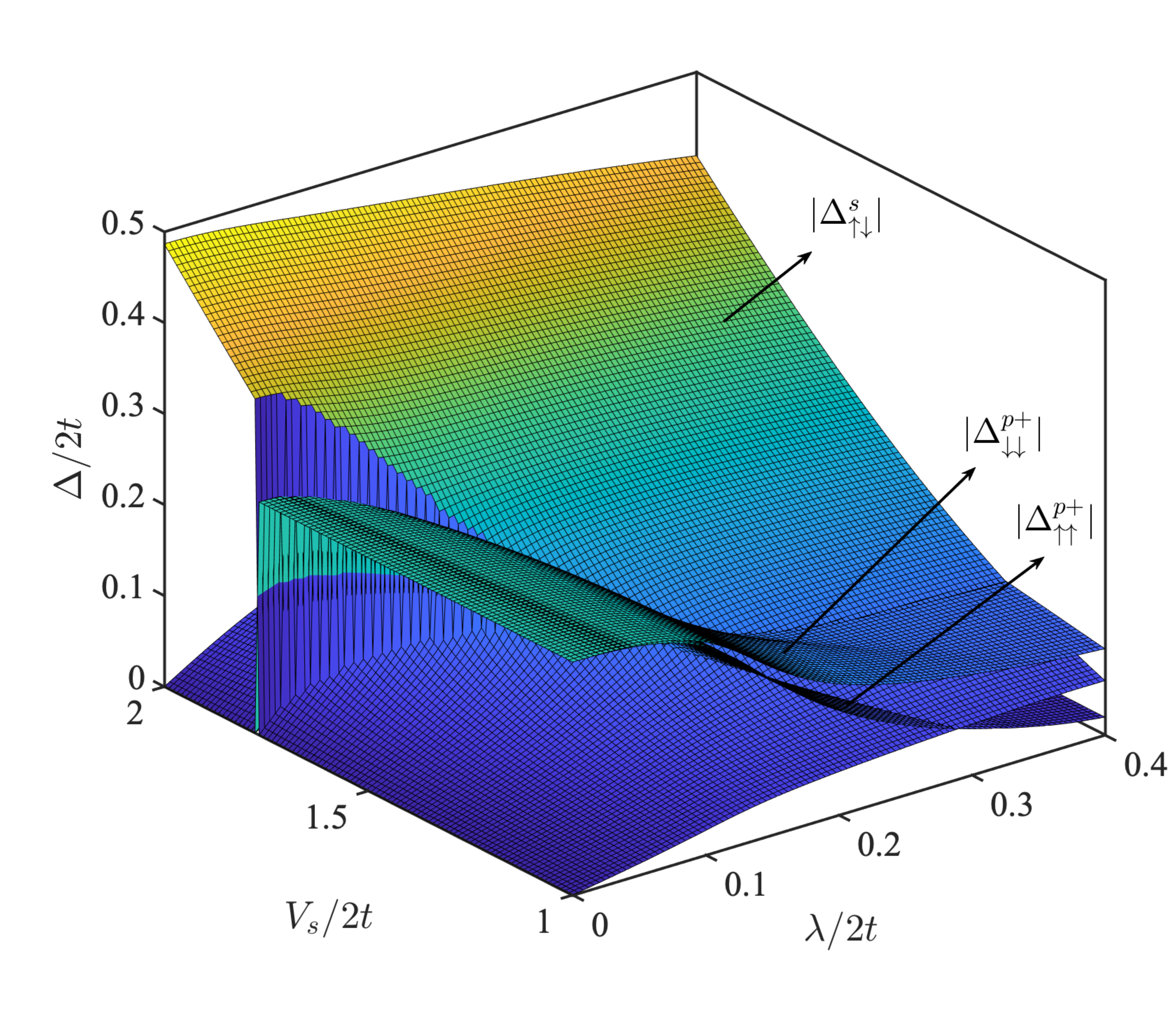}}
\subfigure{\includegraphics[width=7cm]{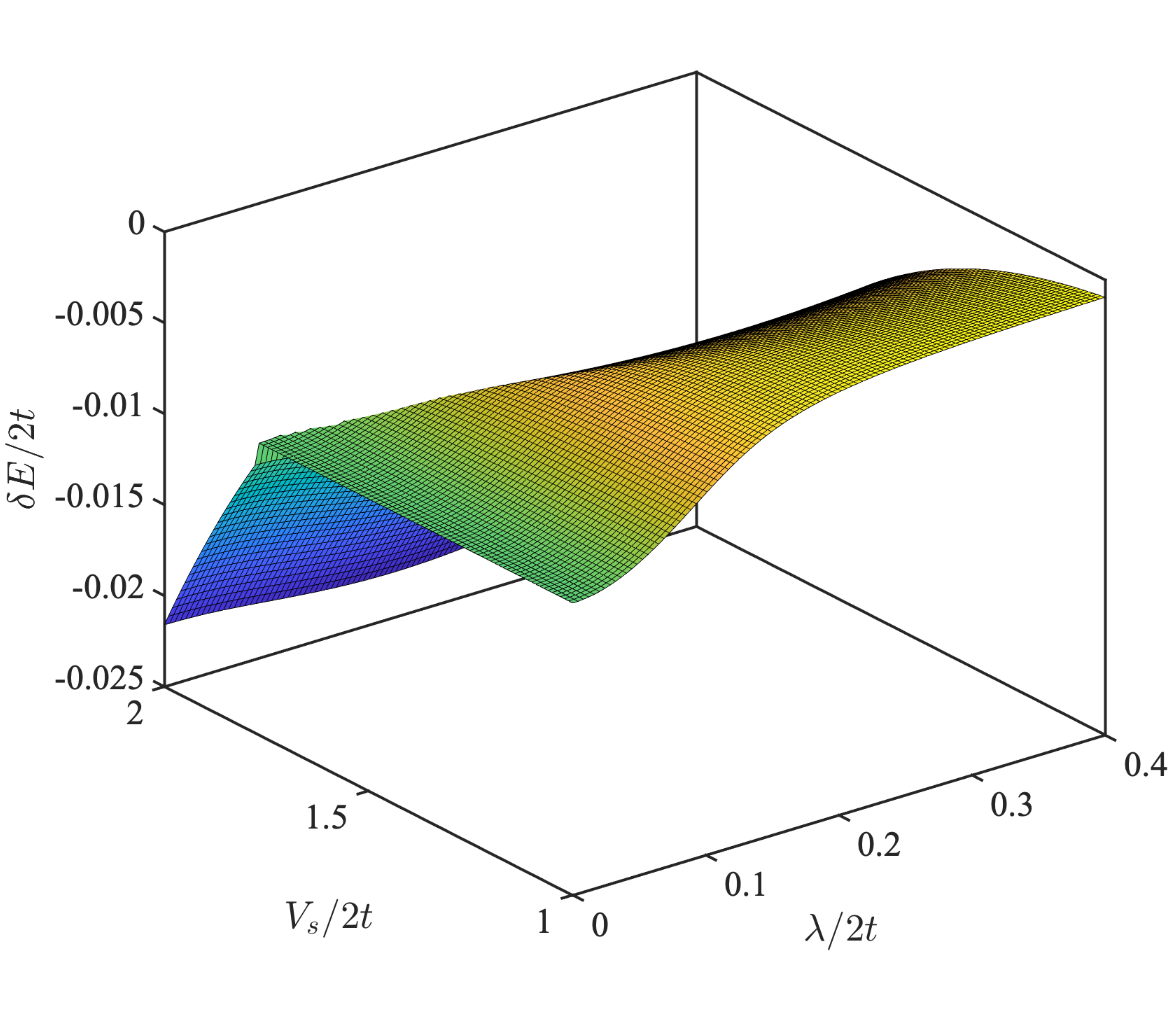}}
\caption{(color online) The $s+$ chiral $p$-wave solution of the gap equations . Here $t=0.5$, $t_{\rm AM} = 0.1$, $V_p = 1.5$, and $\mu = -1$. Left:  the pairing amplitudes. Right: the condensation energy.}
\label{chiralp}
\end{figure*}

Now we turn to our specific short-range interaction (an extended attractive Hubbard interaction)
\begin{align}
\hat H_{int} = -V_s \sum_i n_{i\uparrow} n_{i\downarrow} -V_p \sum_{\langle ij\rangle,\sigma} n_{i\sigma} n_{j\sigma}\label{Hubbard}
\end{align}
for which
\begin{equation}
V_{\sigma\sigma'}(\bq) = \left\{ \begin{tabular}{cc}
            $-V_s, $&  \qquad $\sigma \neq \sigma'$, \\
            $-2V_p\left (\cos q_x + \cos q_y  \right ),$ & \qquad $\sigma = \sigma'$. \\
        \end{tabular} \right.
\end{equation}
In terms of $g_\eta(\bk)$, one finds
\begin{widetext}
\begin{eqnarray}
V_{\uparrow\downarrow}({\bk-\bk'} )&=&V_{\downarrow\uparrow}({\bk-\bk'})= -V_s g_{s}(\bk)g^*_{s}(\bk'),\nonumber \\
V_{\uparrow\uparrow}({\bk-\bk'}) &=&V_{\downarrow\downarrow}({\bk-\bk'})=-V_p \left [ g_{es}(\bk)g^*_{es}(\bk')+g_{d}(\bk)g^*_{d}(\bk')+g_{p+}(\bk)g^*_{p+}(\bk')+g_{p-}(\bk)g^*_{p-}(\bk')\right ].
\end{eqnarray}
\end{widetext}
It is worth mentioning that, in the above decomposition,
$V_{\uparrow\uparrow}$ and $V_{\downarrow\downarrow}$ not only contain
$p$-wave harmonic components, but also contain extended $s$-wave and $d$-wave harmonic
components. However,  because this attractive interaction
occurs between electrons with the same spin, it can not result in extended
$s$-wave pairing or $d$-wave pairing due to the Fermi statistics. As extended $s$-wave and
$d$-wave pairings are spin singlet, their formation requires an attractive interaction
between  two electrons possessing opposite spins and located at two nearest-neighbor sites, i.e.,
$-V_{sd} \sum_{\langle ij\rangle} n_{i\uparrow} n_{j\downarrow}$. To have pairings with even higher angular momentum, such as
an $f$-wave spin-triplet pairing, one needs to further consider  longer-range attractive interactions, such as the
next-nearest-neighbor attractive interaction. In this work, to have a neat understanding of the pairing
phase diagram and the potential topological superconducting phases, we will focus on the simple interaction
given in  Eq.(\ref{Hubbard}). Accordingly, only on-site $s$-wave pairing and $p$-wave pairing will show up.
Despite focusing on this simple interaction, we would like to emphasize that it is sufficient to capture
all key physics, including the competition of even-parity and odd-parity pairings in an inversion-asymmetric
system, and all possible topological superconducting phases at a qualitative level. In the following, we explain
how we determine the superconducting ground state and the pairing phase diagram.

As a result of the pairing interaction and the symmetry property of the gap function in Eq.~(\ref{sym}), we have
\begin{align}
\Xi_{\uparrow\downarrow}(\bk) &= - \Xi_{\downarrow\uparrow}(\bk) =\Delta^s_{\uparrow\downarrow}, \nonumber\\
\Xi_{\uparrow\uparrow}(\bk) & = \Delta^{p+}_{\uparrow\uparrow} g_{p+}(\bk)+ \Delta^{p-}_{\uparrow\uparrow} g_{p-}(\bk), \nonumber\\
\Xi_{\downarrow\downarrow}(\bk) & = \Delta^{p+}_{\downarrow\downarrow} g_{p+}(\bk)+ \Delta^{p-}_{\downarrow\downarrow} g_{p-}(\bk).
\end{align}
The general forms of $\Xi_{\uparrow\uparrow}(\bk)$ and $\Xi_{\downarrow\downarrow}(\bk)$ do not necessarily preserve the lattice symmetry. The two types of solutions that do are: (i) chiral $p$-wave with $(\Xi_{\uparrow\uparrow}(\bk), \Xi_{\downarrow\downarrow}(\bk)) = (\Delta^{p+}_{\uparrow\uparrow} g_{p+}(\bk),\Delta^{p+}_{\downarrow\downarrow} g_{p+}(\bk))$ or $(\Xi_{\uparrow\uparrow}(\bk), \Xi_{\downarrow\downarrow}(\bk)) = (\Delta^{p-}_{\uparrow\uparrow} g_{p-}(\bk),\Delta^{p-}_{\downarrow\downarrow} g_{p-}(\bk))$;
and (ii) helical $p$-wave with $(\Xi_{\uparrow\uparrow}(\bk), \Xi_{\downarrow\downarrow}(\bk)) = (\Delta^{p+}_{\uparrow\uparrow} g_{p+}(\bk),\Delta^{p-}_{\downarrow\downarrow} g_{p-}(\bk))$ or $(\Xi_{\uparrow\uparrow}(\bk), \Xi_{\downarrow\downarrow}(\bk)) = (\Delta^{p-}_{\uparrow\uparrow} g_{p-}(\bk),\Delta^{p+}_{\downarrow\downarrow} g_{p+}(\bk))$.
It is noteworthy that for the chiral $p$-wave pairing, the two situations,
$(\Xi_{\uparrow\uparrow},\Xi_{\downarrow\downarrow})=(\Delta^{p+}_{\uparrow\uparrow} g_{p+},\Delta^{p+}_{\downarrow\downarrow} g_{p+})$
and $(\Xi_{\uparrow\uparrow},\Xi_{\downarrow\downarrow})=(\Delta^{p-}_{\uparrow\uparrow} g_{p-},\Delta^{p-}_{\downarrow\downarrow} g_{p-})$,
lead to the same condensation energy and similar topological property for the superconducting phase (only the sign of the Chern number will
be different since the pairings for the two cases carry opposite angular momentum). For this reason
we only discuss the case of $(\Xi_{\uparrow\uparrow},\Xi_{\downarrow\downarrow})=(\Delta^{p+}_{\uparrow\uparrow} g_{p+},\Delta^{p+}_{\downarrow\downarrow} g_{p+})$. For the helical $p$-wave pairing, similarly, the two situations,
$(\Xi_{\uparrow\uparrow},\Xi_{\downarrow\downarrow})=(\Delta^{p+}_{\uparrow\uparrow} g_{p+},\Delta^{p-}_{\downarrow\downarrow} g_{p-})$
and $(\Xi_{\uparrow\uparrow},\Xi_{\downarrow\downarrow})=(\Delta^{p-}_{\uparrow\uparrow} g_{p-},\Delta^{p+}_{\downarrow\downarrow} g_{p+})$,
lead to the same condensation energy and similar topological property  for the superconducting phase, so
we also only need to focus on one of the cases. Below, we focus on the case $(\Xi_{\uparrow\uparrow},\Xi_{\downarrow\downarrow})=(\Delta^{p+}_{\uparrow\uparrow} g_{p+},\Delta^{p-}_{\downarrow\downarrow} g_{p-})$.
When both chiral $p$-wave pairing and helical $p$-wave pairing are solutions in the same parameter region, we need to compare their corresponding condensation energies in order to determine the ground state.

\begin{figure*}[htb]
\subfigure{\includegraphics[width=7cm]{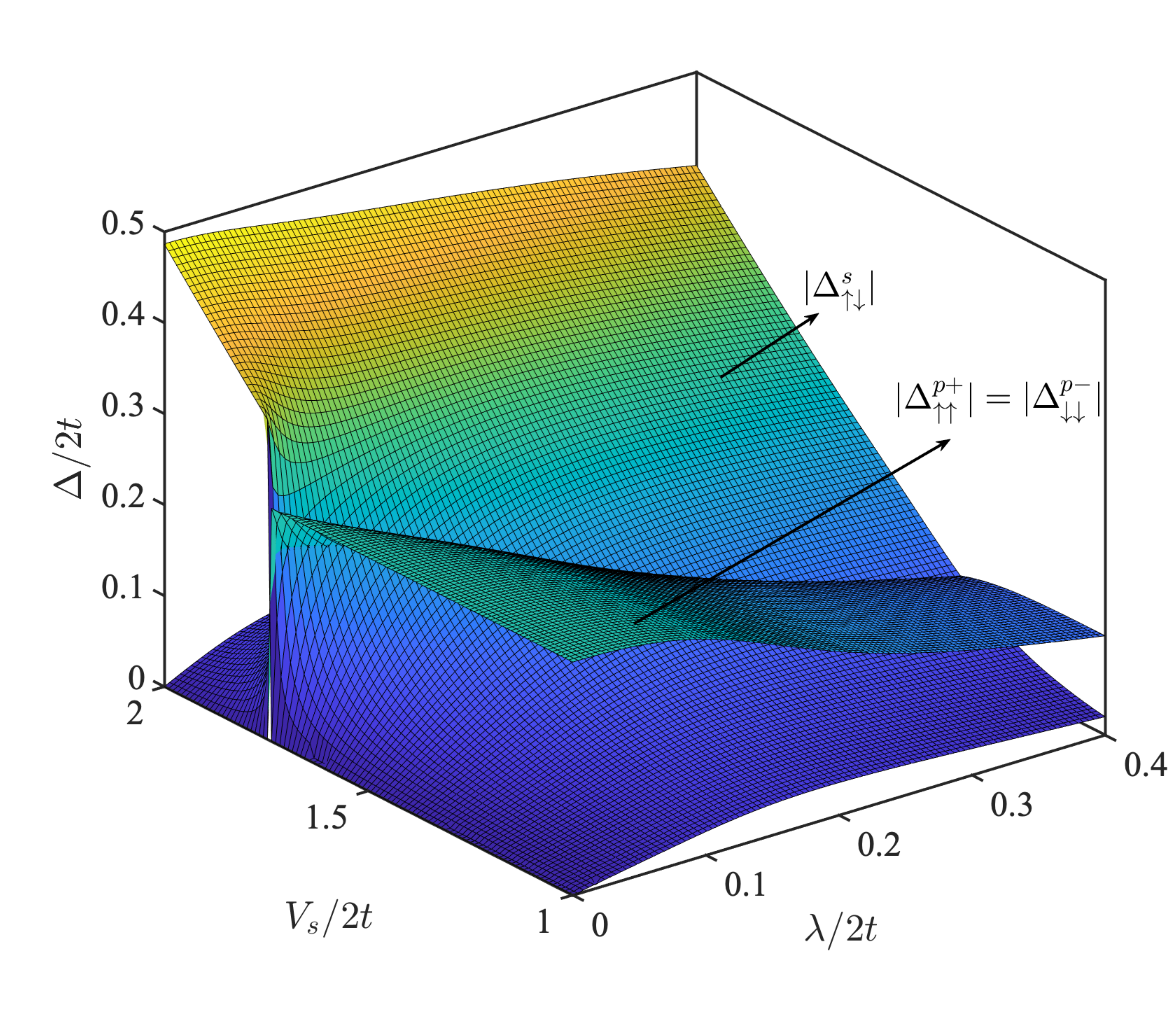}}
\subfigure{\includegraphics[width=7cm]{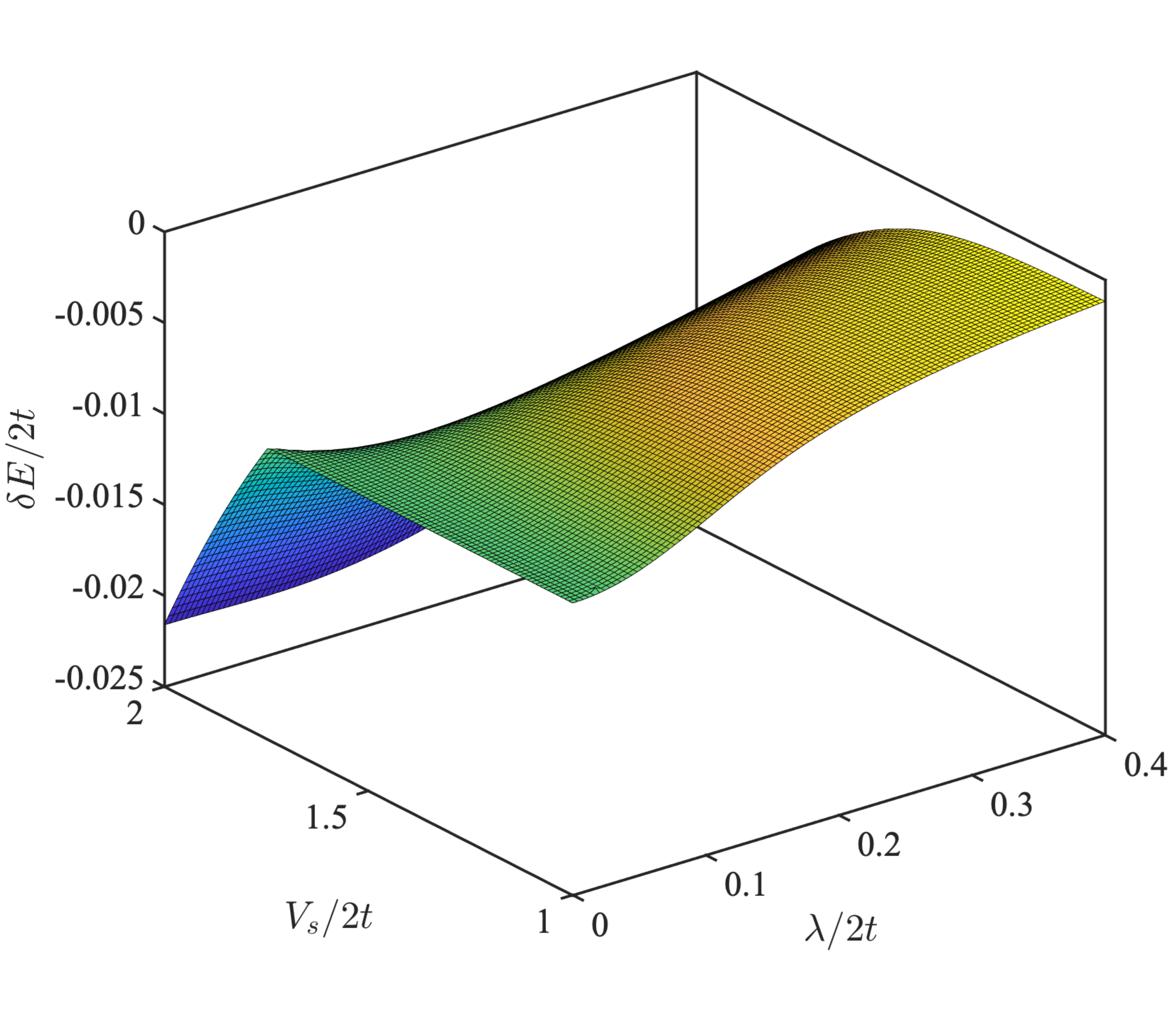}}
\caption{(color online) The $s+$ helical $p$-wave solution of the gap equations . Here $t=0.5$, $t_{\rm AM} = 0.1$, $V_p = 1.5$, and $\mu = -1$.  Left:  the pairing amplitudes. Right: the condensation energy. }
\label{helicalp}
\end{figure*}

\begin{figure*}[htb]
\subfigure{\includegraphics[width=7cm]{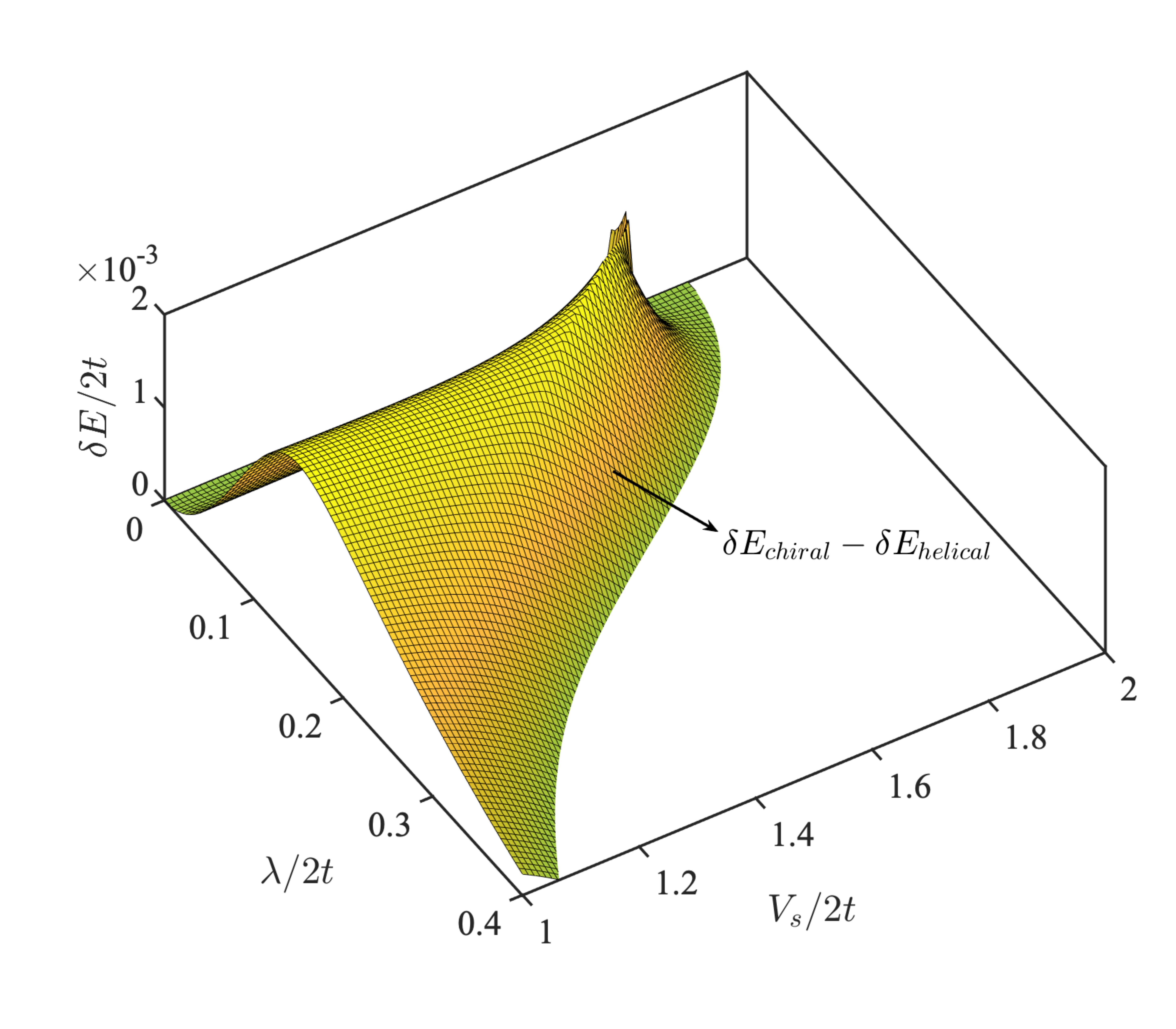}}
\subfigure{\includegraphics[width=7cm]{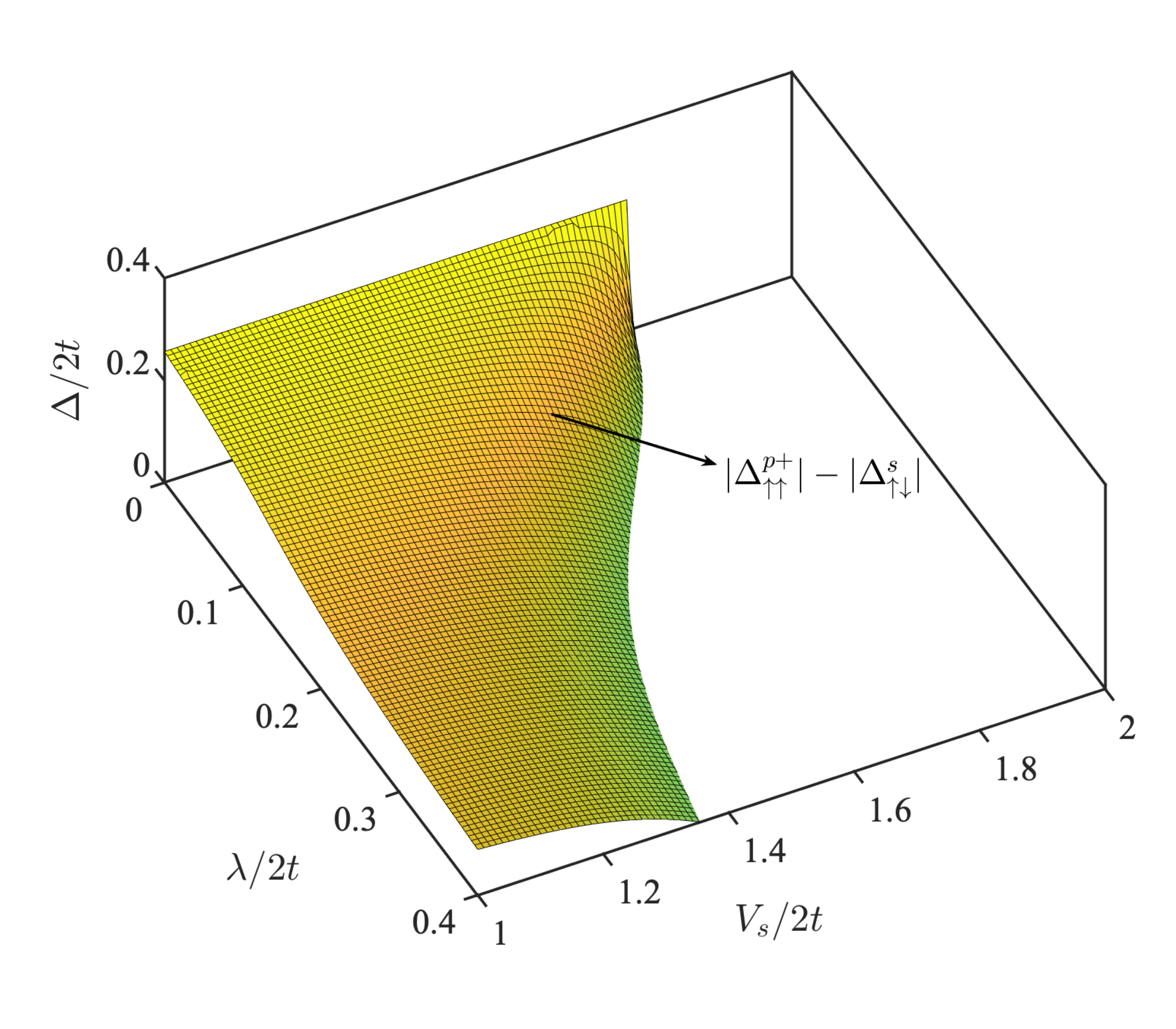}}
\caption{(color online) Left: the difference between the $p$-wave pairing amplitude and the $s$-wave pairing amplitude for the $s+$ helical $p$ solutions (only positive differences are shown). Right: the difference of condensation energies for the $s+$ chiral $p$-wave and the $s+$ helical $p$-wave solutions (only positive differences are shown). }
\label{pd}
\end{figure*}

When the Rashba spin-orbit coupling is absent, we find that the pairing has definitive parity, i.e., either even-parity or odd-parity.
Whether the ground state favors the even-parity spin-singlet $s$-wave pairing or the odd-parity spin-triplet $p$-wave pairing naturally depends on the relative strength of $V_{s}$ and $V_{p}$. It turns out that helical $p$-wave and chiral $p$-wave pairings
are degenerate in ground-state energy in the limit $\lambda=0$.
If both $t_{\rm AM}$ and the spin-orbit coupling constant $\lambda$ are finite, we find that the general solution is a coexistence of $s$-wave and $p$-wave pairings (either chiral or helical); in other words pure $s$-wave or pure $p$-wave pairings  are absent when
the Rashba spin-orbit coupling enters. In Fig.~\ref{chiralp} (left) and (right) we show respectively the pairing amplitudes and the condensation energy for the solution of  mixed $s+$ chiral $p$-wave pairing. We note that the two $p$-wave pairing amplitudes, $\Delta^{p+}_{\uparrow\uparrow}$
 and $\Delta^{p+}_{\downarrow\downarrow}$, are the same in magnitude when the spin-orbit coupling $\lambda$ is absent but they begin to gradually deviate as $\lambda$ increases. In Fig.~\ref{helicalp} (left) and (right) we show respectively the pairing amplitudes and the condensation energy for the solution of  mixed $s+$ helical $p$-wave pairing. In contrast to the chiral solutions, the two $p$-wave pairing amplitudes, $\Delta^{p+}_{\uparrow\uparrow}$
 and $\Delta^{p-}_{\downarrow\downarrow}$, are always the same so that the net angular momentum of this superconducting state is zero.

The phase diagram in the main text is obtained by first comparing the condensation energies of the $s+$ chiral $p$-wave and the $s+$ helical $p$-wave pairing solutions. Shown in Fig.~\ref{pd} (left) is $\delta E_{chiral} - \delta E_{helical}$ in the region where it is positive, namely in the region where the $s+$ helical $p$-wave pairing is the ground state.  This region can be further divided into one where the $p$-wave pairing is dominant and one where $s$-wave pairing is dominant, by a direct comparison of the pairing amplitudes. Shown in Fig.~\ref{pd} (right) is $|\Delta^p_{\uparrow\uparrow}|-|\Delta_{\uparrow\downarrow}^s|$ in the region where it is positive. The phase diagram is then straightforwardly determined from the two plots in Fig.~\ref{pd}.

\section{Topological properties of the superconducting phase with
mixed $s+\text{helical}\,p$-wave pairing\label{Appendixc}}

When the pairing is a mixture of $s$-wave pairing and helical $p$-wave pairing, the BdG Hamiltonian is given by
\begin{eqnarray}
\mathcal{H}_{\rm BdG}(\bk)&=&[-2t(\cos k_{x}+\cos k_{y})-\mu]\tau_{z}\sigma_{0}\nonumber\\
&&+2t_{\rm AM}(\cos k_{x}-\cos k_{y})\tau_{z}\sigma_{z}\nonumber\\
&&+2\lambda(\sin k_{y}\tau_{0}\sigma_{x}-\sin k_{x}\tau_{z}\sigma_{y})\nonumber\\
&&+2\Delta^{p}(\sin k_{x}\tau_{x}\sigma_{z}-\sin k_{y}\tau_{y}\sigma_{0})\nonumber\\
&&+\Delta_{\uparrow\downarrow}^{s}\tau_{y}\sigma_{y},\label{helicalH}
\end{eqnarray}
where $\Delta_{\uparrow\uparrow}^{p+}=-\Delta_{\downarrow\downarrow}^{p-}=\Delta^{p}$ is set for notational
simplicity.
In this section, we give a detailed discussion about the symmetry protection of the topological superconducting
phase in the $\lambda=0$ limit. When $\lambda=0$, our numerical calculations in fact show that
the pairing has fixed parity for the investigated parameter region. Here for a generic discussion
of the symmetry protection of the band topology we ignore
this parity constraint and still assume that the $s$-wave and helical $p$-wave pairings can coexist
even in the $\lambda=0$ limit.

We first consider the $\Delta_{\uparrow\downarrow}^{s}=0$ limit. For this case, the BdG Hamiltonian (\ref{helicalH}) reduces to
\begin{eqnarray}
\mathcal{H}_{\rm p}(\bk)&=&[-2t(\cos k_{x}+\cos k_{y})-\mu]\tau_{z}\sigma_{0}\nonumber\\
&&+2t_{\rm AM}(\cos k_{x}-\cos k_{y})\tau_{z}\sigma_{z}\nonumber\\
&&+2\Delta^{p}(\sin k_{x}\tau_{x}\sigma_{z}-\sin k_{y}\tau_{y}\sigma_{0}).
\end{eqnarray}
This reduced Hamiltonian has mirror
symmetry $\mathcal{M}_{z}$, i.e., $\mathcal{M}_{z}\mathcal{H}_{\rm p}(\bk)\mathcal{M}_{z}^{-1}=\mathcal{H}_{\rm p}(\bk)$
with $\mathcal{M}_{z}=i\tau_{0}\sigma_{z}$. According to the two possible eigenvalues of $\mathcal{M}_{z}$, i.e., $\pm i$,
the Hamiltonian can be decomposed as $\mathcal{H}_{\rm p}(\bk)=\mathcal{H}_{i}(\bk)\oplus\mathcal{H}_{-i}(\bk)$, where
$\mathcal{H}_{\pm i}(\bk)=\bd_{\pm i}(\bk)\cdot\boldsymbol{\tau}$ with 
\begin{eqnarray}
\bd_{i}(\bk)&=&(2\Delta^{p}\sin k_{x},-2\Delta^{p}\sin k_{y},\xi_{\bk\uparrow}),\nonumber\\
\bd_{-i}(\bk)&=&(2\Delta^{p}\sin k_{x},-2\Delta^{p}\sin k_{y},\xi_{\bk\downarrow}).
\end{eqnarray}
Each sector is a chiral $p$-wave superconductor and is accordingly characterized by a Chern number. 
The Chern numbers characterizing the two mirror-graded Hamiltonians are simply given by~\cite{qi2006QWZ}
\begin{eqnarray}
C_{\pm i}=-\frac{1}{4\pi}\int_{BZ}\frac{\bd_{\pm i}\cdot[\partial_{k_{x}}\bd_{\pm i}\times\partial_{k_{y}}\bd_{\pm i}]}{d_{\pm i}^{3}}d^{2}k,
\end{eqnarray}
where $d_{\pm i}=|\bd_{\pm i}|$ denotes the norm of the $\bd_{\pm i}$ vector. Without loss of generality, we take $t>t_{\rm AM} >0$. Then a straightforward calculation yields
\begin{eqnarray}
C_{i}=-C_{-i}=\left\{\begin{array}{cc}
                0, & \mu>4t, \\
                1, & 4t_{\rm AM}<\mu<4t, \\
                0, & -4t_{\rm AM}<\mu<4t_{\rm AM}, \\
                -1, & -4t<\mu<-4t_{\rm AM}, \\
                0, & \mu<-4t.
              \end{array}\right.
\end{eqnarray}
The result suggests that the total Chern number, which is given by the sum of the two mirror-graded
Chern numbers, is always zero. We have previously explained that this is a natural result due to the constraint from the
$\mathcal{C}_{4z}\mathcal{T}$ symmetry. Although the total Chern number is zero, the mirror Chern number,
which is defined as $C_{M}=(C_{i}-C_{-i})/2$~\cite{teo2008}, has an absolute value of $1$ when $4t_{\rm AM}<|\mu|<4t$. When
$|C_{M}|=1$, the superconducting phase is a topological mirror superconductor with a
pair of helical Majorana modes on the open edges~\cite{Zhang2013mirror}.

\begin{figure}[t]
\subfigure{\includegraphics[width=8cm]{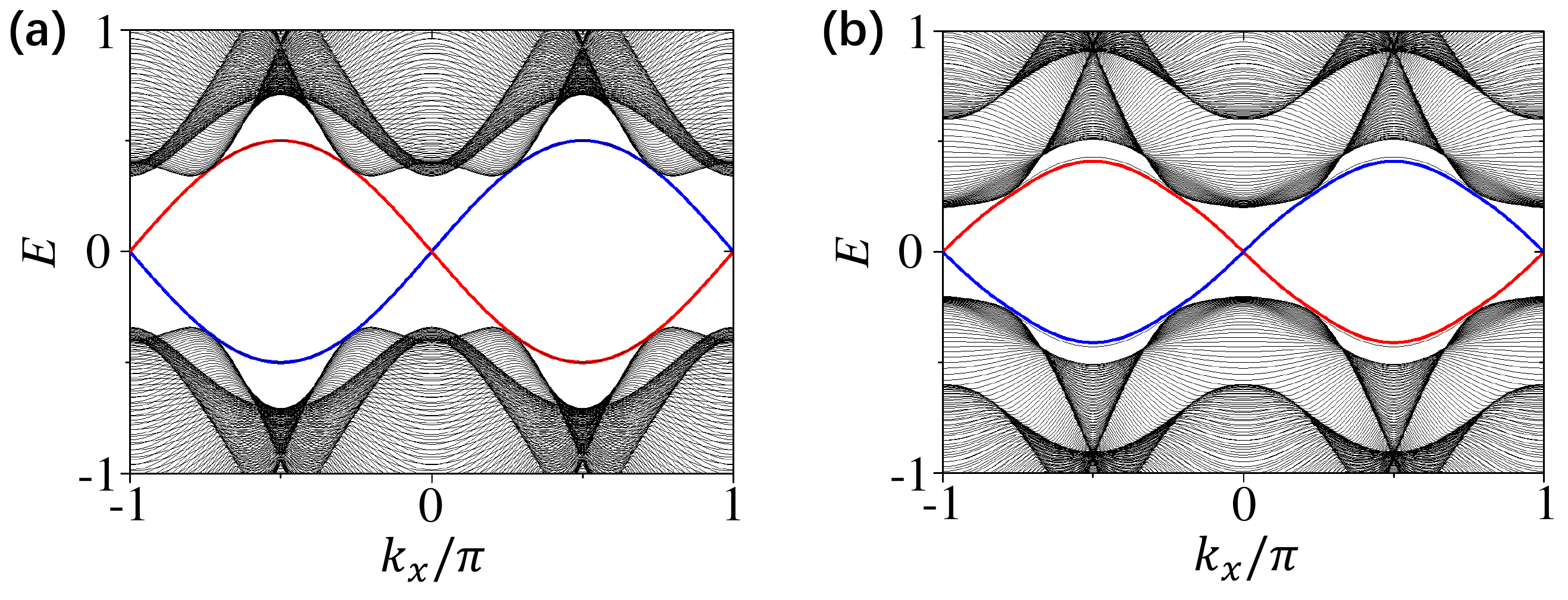}}
\caption{(color online) Common parameters are $t=0.5$, $t_{\rm AM}=0.1$, $\lambda=0.2$, $\mu=-0.0$ and
$\Delta^{p}=0.25$.  (a) $\Delta_{\uparrow\downarrow}^{s}=0.0$, (b) $\Delta_{\uparrow\downarrow}^{s}=0.2$. The red and blue solid lines refer to
mid-gap states on opposite edges. The edge-state energy spectra tangentially touch with the bulk energy spectra when
$\Delta_{\uparrow\downarrow}^{s}=0$, and become floating bands when $\Delta_{\uparrow\downarrow}^{s}$ becomes nonzero. }
\label{weak}
\end{figure}

When $\Delta_{\uparrow\downarrow}^{s}$ is nonzero, the Hamiltonian becomes
\begin{eqnarray}
\mathcal{H}_{\rm sp}(\bk)&=&[-2t(\cos k_{x}+\cos k_{y})-\mu]\tau_{z}\sigma_{0}\nonumber\\
&&+2t_{\rm AM}(\cos k_{x}-\cos k_{y})\tau_{z}\sigma_{z}\nonumber\\
&&+2\Delta^{p}(\sin k_{x}\tau_{x}\sigma_{z}-\sin k_{y}\tau_{y}\sigma_{0})\nonumber\\
&&+\Delta_{\uparrow\downarrow}^{s}\tau_{y}\sigma_{y}.\label{mixed}
\end{eqnarray}
Since the $s$-wave pairing term anticommutes with the mirror symmetry operator,
i.e., $\{\Delta_{\uparrow\downarrow}^{s}\tau_{y}\sigma_{y},i\tau_{0}\sigma_{z}\}=0$, the mirror symmetry is broken. Without
the protection of mirror symmetry, the helical Majorana edge modes are expected to be gapped
due to potential hybridization. However, we find that the helical Majorana edge modes remain
robust, suggesting that there exists some additional symmetry protection. This can be seen as follows. From the view point of dimensional
reduction, we may view the momentum for the direction with periodic boundary conditions as a tuning parameter. Then
the spectrum crossing of the helical Majorana edge modes suggests that the one-dimensional Hamiltonian
is a topological superconductor with two Majorana zero modes at each boundary. To be more specific,
let us take $k_{x}$ as a tuning parameter.  We can then view the two-dimensional Hamiltonian as
a one-dimensional parameter-dependent Hamiltonian $\mathcal{H}_{k_{x}}(k_{y})$. Now only the argument $k_{y}$
in the bracket has the meaning as a momentum. For this parameter-dependent Hamiltonian $\mathcal{H}_{k_{x}}(k_{y})$,
one finds that the Hamiltonian has an emergent chiral symmetry at $k_{x}=0$ and $k_{x}=\pi$, i.e., $\{\mathcal{C},\mathcal{H}_{k_{x}}(k_{y})\}=0$. The explicit form of the chiral symmetry operator is $\mathcal{C}=\tau_{x}\sigma_{0}$.
The existence of chiral symmetry suggests that a winding number can be assigned to characterize
the band topology of $\mathcal{H}_{0}(k_{y})$ and $\mathcal{H}_{\pi}(k_{y})$~\cite{Ryu2010}.
To determine the winding number, the first step is to rewrite the Hamiltonian by changing
the original basis to a new one in which the chiral symmetry operator takes a diagonal
form, i.e., $\tilde{\mathcal{C}}=\tau_{z}\sigma_{0}$. Apparently, this can be realized by
a unitary operation of the form $U=e^{i\pi \tau_{y}\sigma_{0}/4}$, i.e.,
$U\mathcal{C}U^{-1}=\tilde{\mathcal{C}}$. In the new basis, the form of the Hamiltonian becomes
off-diagonal. Let us take $\mathcal{H}_{0}(k_{y})$ as a specific example. It is straightforward to find
\begin{eqnarray}
U\mathcal{H}_{0}(k_{y})U^{-1}=\left(
     \begin{array}{cc}
       0 & Q_{0}(k_{y}) \\
       Q_{0}^{\dag}(k_{y}) & 0 \\
     \end{array}
   \right),\label{1dHamiltonian}
\end{eqnarray}
where $Q_{0}(k_{y})$ is a two-by-two matrix of the form
\begin{eqnarray}
Q_{0}(k_{y})=\left(
           \begin{array}{cc}
             q_{-}(k_{y}) & -\Delta_{\uparrow\downarrow}^{s} \\
             \Delta_{\uparrow\downarrow}^{s} & q_{+}(k_{y}) \\
           \end{array}
         \right),\nonumber\\
\end{eqnarray}
where $q_{\pm}(k_{y})=[2(t\pm t_{\rm AM})+\mu+2(t\mp t_{\rm AM})\cos k_{y}]+2i\Delta^{p}\sin k_{y}$.
The winding number is given by~\cite{Ryu2010}
\begin{eqnarray}
W_{0}=\frac{i}{2\pi}\int_{-\pi}^{\pi}\text{Tr}[Q_{0}^{-1}(k_{y})\partial_{k_{y}}Q_{0}(k_{y})]dk_{y}.
\end{eqnarray}
The winding number $W_{0}$ also does not change its value as long as the energy gap of $\mathcal{H}_{0}(k_{y})$
remains open. Again let us first focus on the limiting case of $\Delta_{\uparrow\downarrow}^{s}=0$, for which
$Q_{0}(k_{y})$ is diagonal, i.e.,
\begin{eqnarray}
Q_{0}(k_{y})=\left(
           \begin{array}{cc}
             q_{-}(k_{y}) & 0 \\
             0 & q_{+}(k_{y}) \\
           \end{array}
         \right).
\end{eqnarray}
Accordingly, it is easy to find that
\begin{eqnarray}
W_{0}=W_{0}^{(-)}+W_{0}^{(+)},
\end{eqnarray}
where
\begin{eqnarray}
W_{0}^{(\pm)}=\frac{i}{2\pi}\int_{-\pi}^{\pi}[q_{\pm}^{-1}(k_{y})\partial_{k_{y}}q_{\pm}(k_{y})]dk_{y}.
\end{eqnarray}
A straightforward calculation yields
\begin{eqnarray}
W_{0}^{(-)}=\left\{\begin{array}{cc}
                          0, & \mu>4t_{\rm AM}, \\
                          -1, & -4t<\mu<4t_{\rm AM}, \\
                          0, & \mu<-4t,
                        \end{array}\right. 
\end{eqnarray}
and 
\begin{eqnarray}                       
W_{0}^{(+)}=\left\{\begin{array}{cc}
                          0, & \mu>-4t_{\rm AM}, \\
                          -1, & -4t<\mu<-4t_{\rm AM}, \\
                          0, & \mu<-4t.
                        \end{array}\right.
\end{eqnarray}
Similar analysis shows that for $\mathcal{H}_{\pi}(k_{y})$ with $\Delta_{\uparrow\downarrow}^{s}=0$,
\begin{eqnarray}
W_{\pi}^{(-)}=\left\{\begin{array}{cc}
                          0, & \mu>4t, \\
                          -1, & -4t_{\rm AM}<\mu<4t, \\
                          0, & \mu<-4t_{\rm AM},
                        \end{array}\right. 
\end{eqnarray}
and 
\begin{eqnarray}
W_{\pi}^{(+)}=\left\{\begin{array}{cc}
                          0, & \mu>4t, \\
                          -1, & 4t_{\rm AM}<\mu<4t, \\
                          0, & \mu<4t_{\rm AM}.
                        \end{array}\right.
\end{eqnarray}
Based on the above analysis, we reach the following result,
\begin{eqnarray}
(W_{0},W_{\pi})=\left\{\begin{array}{cc}
                (0,0), & \mu>4t, \\
                (0,-2), & 4t_{\rm AM}<\mu<4t, \\
                (-1,-1), & -4t_{\rm AM}<\mu<4t_{\rm AM}, \\
                (-2,0), & -4t<\mu<-4t_{\rm AM}, \\
                (0,0), & \mu<-4t.
              \end{array}\right.
\end{eqnarray}
Comparing  with the mirror Chern number, we see that
$C_{M}=1$ and $C_{M}=-1$ correspond to
$(W_{0},W_{\pi})=(-2,0)$ and $(0,-2)$, respectively.
Obviously, the value $-2$ of the winding number also guarantees
the robustness  of the spectrum-crossing feature of the helical Majorana modes.
Interestingly, using the winding number, we find that
the  superconducting phase is also topologically nontrivial in the region
$-4t_{\rm AM}<\mu<4t_{\rm AM}$ even though the mirror Chern number
is zero. In this region, the two winding numbers
$W_{0}$ and $W_{\pi}$ both take value $1$.
It means that if the translational symmetry is preserved
in the $x$ direction, on each $y$-normal edge there are two
branches of chiral Majorana modes with opposite chiralities,
with one branch traversing the gap at $k_{x}=0$, and
the other  traversing the gap at $k_{x}=\pi$, as
shown in Fig.\ref{weak}. This superconducting phase can be categorized as a weak
topological superconducting phase as it is protected by topological invariants defined
in noncontractible subspaces of the Brillouin zone. In this case whether the Majorana modes comes about relies on the orientation of the edges.

When $\Delta_{\uparrow\downarrow}^{s}$ becomes finite, the winding numbers retain their values as long as
the energy gap remains open along those high symmetry lines. For the superconducting
phase with $W_{0}=-2$ or $W_{\pi}=-2$, the robustness of the winding number explains the robust
spectrum-crossing feature of the helical Majorana modes even when the mirror symmetry
is broken by the mixture of $s$-wave and helical $p$-wave pairings.
Due to the existence of $\mathcal{C}_{4z}\mathcal{T}$ symmetry, the physics is similar
when the directions for the open boundary conditions and the periodic boundary conditions are reversed.

Below we provide a discussion on the condition for the change of winding number when
$\Delta_{\uparrow\downarrow}^{s}$ is nonzero. Without loss of generality, we focus on the Hamiltonian
$\mathcal{H}_{0}(k_{y})$ in Eq.(\ref{1dHamiltonian}) for a detailed discussion.
The energy spectra of $\mathcal{H}_{0}(k_{y})$ can be analytically  determined, which read
\begin{eqnarray}
E(k_{y})=\pm\sqrt{h_{1}^{2}+h_{2}^{2}+h_{3}^{2}+h_{4}^{2}\pm2\sqrt{h_{1}^{2}h_{2}^{2}+h_{2}^{2}h_{3}^{2}+h_{3}^{2}h_{4}^{2}}},\nonumber\\
\end{eqnarray}
where $h_{1}=2t(1+\cos k_{y})+\mu$, $h_{2}=2t_{\rm AM}(1-\cos k_{y})$, $h_{3}=\Delta_{\uparrow\downarrow}^{s}$, and $h_{4}=2\Delta^{p}\sin k_{y}$.
The band gap becomes closed when the parameters fulfill either one of the following two conditions,
\begin{eqnarray}
&&(\text{I}): h_{4}=0,\quad h_{2}=\sqrt{h_{1}^{2}+h_{3}^{2}}; \nonumber\\
&&(\text{II}): h_{1}=0,\quad h_{3}=\sqrt{h_{2}^{2}+h_{4}^{2}}.
\end{eqnarray}
For case (I), if $\Delta_{\uparrow\downarrow}^{s}$ is nonzero, the band gap closes only at the time-reversal invariant momentum $k_{y}=\pi$
and only when $4t_{\rm AM}=\sqrt{(\Delta_{\uparrow\downarrow}^{s})^{2}+\mu^{2}}$. For case (II), the band gap closes at the momenta
\begin{eqnarray}
k_{y}=\pm\left(\pi-\arccos\left(\frac{2t+\mu}{2t}\right)\right)
\end{eqnarray}
when
\begin{eqnarray}
\Delta_{\uparrow\downarrow}^{s}=\sqrt{\frac{t_{\rm AM}^{2}\mu^{2}}{t^{2}}+\frac{(\Delta^{p})^{2}(\mu^{2}+4t\mu)}{t^{2}}}.
\end{eqnarray}
According to the number of band-closing points,
the winding number $W_{0}$ will change by $1$ for the case (I) and by 2 for the case (II).

When $\lambda\neq0$, both the mirror symmetry $\mathcal{M}_{z}$ and chiral symmetries along the
high symmetry lines are broken, and the helical Majorana modes become gapped. The gapped
helical Majorana modes on the open edges can effectively be described by a low-energy massive Dirac Hamiltonian
\begin{eqnarray}
h(x_{b})=-iv\partial_{x_{b}}\rho_{z}+m(x_{b})\rho_{y},
\end{eqnarray}
where $v$ denotes the velocity of the Majorana modes, $\rho_{y,z}$ are Pauli matrices acting on the Hilbert space
spanned by the edge states, and $x_{b}$ denotes the coordinate along the edges~\cite{Yan2018hosc}.
Since the Hamiltonian does not have the $\mathcal{C}_{4z}$ rotational symmetry or the  time-reversal
symmetry but
has the $\mathcal{C}_{4z}\mathcal{T}$ symmetry, the Dirac mass $m(x_{b})$
will take opposite signs on two nearby edges related by $\mathcal{C}_{4z}$ rotation~\cite{Schindler2018}. This
leads to the formation of Dirac-mass domain walls at the corners of a square lattice
whose geometry is $\mathcal{C}_{4z}$-rotationally invariant, and hence the appearance of Majorana
corner modes~\cite{jackiw1976b}.

\bibliography{dirac}

\begin{thebibliography}{91}%
\makeatletter
\providecommand \@ifxundefined [1]{%
 \@ifx{#1\undefined}
}%
\providecommand \@ifnum [1]{%
 \ifnum #1\expandafter \@firstoftwo
 \else \expandafter \@secondoftwo
 \fi
}%
\providecommand \@ifx [1]{%
 \ifx #1\expandafter \@firstoftwo
 \else \expandafter \@secondoftwo
 \fi
}%
\providecommand \natexlab [1]{#1}%
\providecommand \enquote  [1]{``#1''}%
\providecommand \bibnamefont  [1]{#1}%
\providecommand \bibfnamefont [1]{#1}%
\providecommand \citenamefont [1]{#1}%
\providecommand \href@noop [0]{\@secondoftwo}%
\providecommand \href [0]{\begingroup \@sanitize@url \@href}%
\providecommand \@href[1]{\@@startlink{#1}\@@href}%
\providecommand \@@href[1]{\endgroup#1\@@endlink}%
\providecommand \@sanitize@url [0]{\catcode `\\12\catcode `\$12\catcode
  `\&12\catcode `\#12\catcode `\^12\catcode `\_12\catcode `\%12\relax}%
\providecommand \@@startlink[1]{}%
\providecommand \@@endlink[0]{}%
\providecommand \url  [0]{\begingroup\@sanitize@url \@url }%
\providecommand \@url [1]{\endgroup\@href {#1}{\urlprefix }}%
\providecommand \urlprefix  [0]{URL }%
\providecommand \Eprint [0]{\href }%
\providecommand \doibase [0]{http://dx.doi.org/}%
\providecommand \selectlanguage [0]{\@gobble}%
\providecommand \bibinfo  [0]{\@secondoftwo}%
\providecommand \bibfield  [0]{\@secondoftwo}%
\providecommand \translation [1]{[#1]}%
\providecommand \BibitemOpen [0]{}%
\providecommand \bibitemStop [0]{}%
\providecommand \bibitemNoStop [0]{.\EOS\space}%
\providecommand \EOS [0]{\spacefactor3000\relax}%
\providecommand \BibitemShut  [1]{\csname bibitem#1\endcsname}%
\let\auto@bib@innerbib\@empty
\bibitem [{\citenamefont {Scalapino}\ \emph {et~al.}(1986)\citenamefont
  {Scalapino}, \citenamefont {Loh},\ and\ \citenamefont
  {Hirsch}}]{Scalapino1986}%
  \BibitemOpen
  \bibfield  {author} {\bibinfo {author} {\bibfnamefont {D.~J.}\ \bibnamefont
  {Scalapino}}, \bibinfo {author} {\bibfnamefont {E.}~\bibnamefont {Loh}}, \
  and\ \bibinfo {author} {\bibfnamefont {J.~E.}\ \bibnamefont {Hirsch}},\
  }\bibfield  {title} {\enquote {\bibinfo {title} {$d$-wave pairing near a
  spin-density-wave instability},}\ }\href {\doibase 10.1103/PhysRevB.34.8190}
  {\bibfield  {journal} {\bibinfo  {journal} {Phys. Rev. B}\ }\textbf {\bibinfo
  {volume} {34}},\ \bibinfo {pages} {8190--8192} (\bibinfo {year}
  {1986})}\BibitemShut {NoStop}%
\bibitem [{\citenamefont {T.~Moriya}\ and\ \citenamefont
  {Ueda}(1990)}]{Moriya1990}%
  \BibitemOpen
  \bibfield  {author} {\bibinfo {author} {\bibfnamefont {Y.~Takahashi}\
  \bibnamefont {T.~Moriya}}\ and\ \bibinfo {author} {\bibfnamefont
  {K.}~\bibnamefont {Ueda}},\ }\bibfield  {title} {\enquote {\bibinfo {title}
  {{Antiferromagnetic Spin Fluctuations and Superconductivity in
  Two-Dimensional Metals--A Possible Model for High-$T_{c}$ Oxides}},}\ }\href
  {\doibase 10.1143/JPSJ.59.2905} {\bibfield  {journal} {\bibinfo  {journal}
  {Journal of the Physical Society of Japan}\ }\textbf {\bibinfo {volume}
  {59}},\ \bibinfo {pages} {2905--2915} (\bibinfo {year} {1990})}\BibitemShut
  {NoStop}%
\bibitem [{\citenamefont {Monthoux}\ \emph {et~al.}(1991)\citenamefont
  {Monthoux}, \citenamefont {Balatsky},\ and\ \citenamefont
  {Pines}}]{Monthoux1991}%
  \BibitemOpen
  \bibfield  {author} {\bibinfo {author} {\bibfnamefont {P.}~\bibnamefont
  {Monthoux}}, \bibinfo {author} {\bibfnamefont {A.~V.}\ \bibnamefont
  {Balatsky}}, \ and\ \bibinfo {author} {\bibfnamefont {D.}~\bibnamefont
  {Pines}},\ }\bibfield  {title} {\enquote {\bibinfo {title} {{Toward a theory
  of high-temperature superconductivity in the antiferromagnetically correlated
  cuprate oxides}},}\ }\href {\doibase 10.1103/PhysRevLett.67.3448} {\bibfield
  {journal} {\bibinfo  {journal} {Phys. Rev. Lett.}\ }\textbf {\bibinfo
  {volume} {67}},\ \bibinfo {pages} {3448--3451} (\bibinfo {year}
  {1991})}\BibitemShut {NoStop}%
\bibitem [{\citenamefont {Mathur}\ \emph {et~al.}(1998)\citenamefont {Mathur},
  \citenamefont {Grosche}, \citenamefont {Julian}, \citenamefont {Walker},
  \citenamefont {Freye}, \citenamefont {Haselwimmer},\ and\ \citenamefont
  {Lonzarich}}]{Mathur1998}%
  \BibitemOpen
  \bibfield  {author} {\bibinfo {author} {\bibfnamefont {N.~D.}\ \bibnamefont
  {Mathur}}, \bibinfo {author} {\bibfnamefont {F.~M.}\ \bibnamefont {Grosche}},
  \bibinfo {author} {\bibfnamefont {S.~R.}\ \bibnamefont {Julian}}, \bibinfo
  {author} {\bibfnamefont {I.~R.}\ \bibnamefont {Walker}}, \bibinfo {author}
  {\bibfnamefont {D.~M.}\ \bibnamefont {Freye}}, \bibinfo {author}
  {\bibfnamefont {R.~K.~W.}\ \bibnamefont {Haselwimmer}}, \ and\ \bibinfo
  {author} {\bibfnamefont {G.~G.}\ \bibnamefont {Lonzarich}},\ }\bibfield
  {title} {\enquote {\bibinfo {title} {{Magnetically mediated superconductivity
  in heavy fermion compounds}},}\ }\href {\doibase 10.1038/27838} {\bibfield
  {journal} {\bibinfo  {journal} {Nature}\ }\textbf {\bibinfo {volume} {394}},\
  \bibinfo {pages} {39--43} (\bibinfo {year} {1998})}\BibitemShut {NoStop}%
\bibitem [{\citenamefont {Saxena}\ \emph {et~al.}(2000)\citenamefont {Saxena},
  \citenamefont {Agarwal}, \citenamefont {Ahilan}, \citenamefont {Grosche},
  \citenamefont {Haselwimmer}, \citenamefont {Steiner}, \citenamefont {Pugh},
  \citenamefont {Walker}, \citenamefont {Julian}, \citenamefont {Monthoux},
  \citenamefont {Lonzarich}, \citenamefont {Huxley}, \citenamefont {Sheikin},
  \citenamefont {Braithwaite},\ and\ \citenamefont {Flouquet}}]{Saxena2000}%
  \BibitemOpen
  \bibfield  {author} {\bibinfo {author} {\bibfnamefont {S.~S.}\ \bibnamefont
  {Saxena}}, \bibinfo {author} {\bibfnamefont {P.}~\bibnamefont {Agarwal}},
  \bibinfo {author} {\bibfnamefont {K.}~\bibnamefont {Ahilan}}, \bibinfo
  {author} {\bibfnamefont {F.~M.}\ \bibnamefont {Grosche}}, \bibinfo {author}
  {\bibfnamefont {R.~K.~W.}\ \bibnamefont {Haselwimmer}}, \bibinfo {author}
  {\bibfnamefont {M.~J.}\ \bibnamefont {Steiner}}, \bibinfo {author}
  {\bibfnamefont {E.}~\bibnamefont {Pugh}}, \bibinfo {author} {\bibfnamefont
  {I.~R.}\ \bibnamefont {Walker}}, \bibinfo {author} {\bibfnamefont {S.~R.}\
  \bibnamefont {Julian}}, \bibinfo {author} {\bibfnamefont {P.}~\bibnamefont
  {Monthoux}}, \bibinfo {author} {\bibfnamefont {G.~G.}\ \bibnamefont
  {Lonzarich}}, \bibinfo {author} {\bibfnamefont {A.}~\bibnamefont {Huxley}},
  \bibinfo {author} {\bibfnamefont {I.}~\bibnamefont {Sheikin}}, \bibinfo
  {author} {\bibfnamefont {D.}~\bibnamefont {Braithwaite}}, \ and\ \bibinfo
  {author} {\bibfnamefont {J.}~\bibnamefont {Flouquet}},\ }\bibfield  {title}
  {\enquote {\bibinfo {title} {{Superconductivity on the border of
  itinerant-electron ferromagnetism in UGe$_{2}$}},}\ }\href {\doibase
  10.1038/35020500} {\bibfield  {journal} {\bibinfo  {journal} {Nature}\
  }\textbf {\bibinfo {volume} {406}},\ \bibinfo {pages} {587--592} (\bibinfo
  {year} {2000})}\BibitemShut {NoStop}%
\bibitem [{\citenamefont {Aoki}\ \emph {et~al.}(2001)\citenamefont {Aoki},
  \citenamefont {Huxley}, \citenamefont {Ressouche}, \citenamefont
  {Braithwaite}, \citenamefont {Flouquet}, \citenamefont {Brison},
  \citenamefont {Lhotel},\ and\ \citenamefont {Paulsen}}]{Aoki2001}%
  \BibitemOpen
  \bibfield  {author} {\bibinfo {author} {\bibfnamefont {Dai}\ \bibnamefont
  {Aoki}}, \bibinfo {author} {\bibfnamefont {Andrew}\ \bibnamefont {Huxley}},
  \bibinfo {author} {\bibfnamefont {Eric}\ \bibnamefont {Ressouche}}, \bibinfo
  {author} {\bibfnamefont {Daniel}\ \bibnamefont {Braithwaite}}, \bibinfo
  {author} {\bibfnamefont {Jacques}\ \bibnamefont {Flouquet}}, \bibinfo
  {author} {\bibfnamefont {Jean-Pascal}\ \bibnamefont {Brison}}, \bibinfo
  {author} {\bibfnamefont {Elsa}\ \bibnamefont {Lhotel}}, \ and\ \bibinfo
  {author} {\bibfnamefont {Carley}\ \bibnamefont {Paulsen}},\ }\bibfield
  {title} {\enquote {\bibinfo {title} {{Coexistence of superconductivity and
  ferromagnetism in URhGe}},}\ }\href {\doibase 10.1038/35098048} {\bibfield
  {journal} {\bibinfo  {journal} {Nature}\ }\textbf {\bibinfo {volume} {413}},\
  \bibinfo {pages} {613--616} (\bibinfo {year} {2001})}\BibitemShut {NoStop}%
\bibitem [{\citenamefont {Mito}\ \emph {et~al.}(2003)\citenamefont {Mito},
  \citenamefont {Kawasaki}, \citenamefont {Kawasaki}, \citenamefont {Zheng},
  \citenamefont {Kitaoka}, \citenamefont {Aoki}, \citenamefont {Haga},\ and\
  \citenamefont {\ifmmode~\bar{O}\else \={O}\fi{}nuki}}]{Mito2003}%
  \BibitemOpen
  \bibfield  {author} {\bibinfo {author} {\bibfnamefont {T.}~\bibnamefont
  {Mito}}, \bibinfo {author} {\bibfnamefont {S.}~\bibnamefont {Kawasaki}},
  \bibinfo {author} {\bibfnamefont {Y.}~\bibnamefont {Kawasaki}}, \bibinfo
  {author} {\bibfnamefont {G.~q.}\ \bibnamefont {Zheng}}, \bibinfo {author}
  {\bibfnamefont {Y.}~\bibnamefont {Kitaoka}}, \bibinfo {author} {\bibfnamefont
  {D.}~\bibnamefont {Aoki}}, \bibinfo {author} {\bibfnamefont {Y~.}\
  \bibnamefont {Haga}}, \ and\ \bibinfo {author} {\bibfnamefont
  {Y.}~\bibnamefont {\ifmmode~\bar{O}\else \={O}\fi{}nuki}},\ }\bibfield
  {title} {\enquote {\bibinfo {title} {{Coexistence of Antiferromagnetism and
  Superconductivity near the Quantum Criticality of the Heavy-Fermion Compound
  ${\mathrm{C}\mathrm{e}\mathrm{R}\mathrm{h}\mathrm{I}\mathrm{n}}_{5}$}},}\
  }\href {\doibase 10.1103/PhysRevLett.90.077004} {\bibfield  {journal}
  {\bibinfo  {journal} {Phys. Rev. Lett.}\ }\textbf {\bibinfo {volume} {90}},\
  \bibinfo {pages} {077004} (\bibinfo {year} {2003})}\BibitemShut {NoStop}%
\bibitem [{\citenamefont {Akazawa}\ \emph {et~al.}(2004)\citenamefont
  {Akazawa}, \citenamefont {Hidaka}, \citenamefont {Kotegawa}, \citenamefont
  {C.~Kobayashi}, \citenamefont {Fujiwara}, \citenamefont {Yamamoto},
  \citenamefont {Haga}, \citenamefont {Settai},\ and\ \citenamefont
  {\={O}nuki}}]{Akazawa2004}%
  \BibitemOpen
  \bibfield  {author} {\bibinfo {author} {\bibfnamefont {Teruhiko}\
  \bibnamefont {Akazawa}}, \bibinfo {author} {\bibfnamefont {Hiroyuki}\
  \bibnamefont {Hidaka}}, \bibinfo {author} {\bibfnamefont {Hisashi}\
  \bibnamefont {Kotegawa}}, \bibinfo {author} {\bibfnamefont {Tatsuo}\
  \bibnamefont {C.~Kobayashi}}, \bibinfo {author} {\bibfnamefont {Tabito}\
  \bibnamefont {Fujiwara}}, \bibinfo {author} {\bibfnamefont {Etsuji}\
  \bibnamefont {Yamamoto}}, \bibinfo {author} {\bibfnamefont {Yoshinori}\
  \bibnamefont {Haga}}, \bibinfo {author} {\bibfnamefont {Rikio}\ \bibnamefont
  {Settai}}, \ and\ \bibinfo {author} {\bibfnamefont {Yoshichika}\ \bibnamefont
  {\={O}nuki}},\ }\bibfield  {title} {\enquote {\bibinfo {title}
  {{Pressure-induced Superconductivity in UIr}},}\ }\href {\doibase
  10.1143/JPSJ.73.3129} {\bibfield  {journal} {\bibinfo  {journal} {Journal of
  the Physical Society of Japan}\ }\textbf {\bibinfo {volume} {73}},\ \bibinfo
  {pages} {3129--3134} (\bibinfo {year} {2004})}\BibitemShut {NoStop}%
\bibitem [{\citenamefont {Huy}\ \emph {et~al.}(2007)\citenamefont {Huy},
  \citenamefont {Gasparini}, \citenamefont {de~Nijs}, \citenamefont {Huang},
  \citenamefont {Klaasse}, \citenamefont {Gortenmulder}, \citenamefont
  {de~Visser}, \citenamefont {Hamann}, \citenamefont {G\"orlach},\ and\
  \citenamefont {L\"ohneysen}}]{Huy2007}%
  \BibitemOpen
  \bibfield  {author} {\bibinfo {author} {\bibfnamefont {N.~T.}\ \bibnamefont
  {Huy}}, \bibinfo {author} {\bibfnamefont {A.}~\bibnamefont {Gasparini}},
  \bibinfo {author} {\bibfnamefont {D.~E.}\ \bibnamefont {de~Nijs}}, \bibinfo
  {author} {\bibfnamefont {Y.}~\bibnamefont {Huang}}, \bibinfo {author}
  {\bibfnamefont {J.~C.~P.}\ \bibnamefont {Klaasse}}, \bibinfo {author}
  {\bibfnamefont {T.}~\bibnamefont {Gortenmulder}}, \bibinfo {author}
  {\bibfnamefont {A.}~\bibnamefont {de~Visser}}, \bibinfo {author}
  {\bibfnamefont {A.}~\bibnamefont {Hamann}}, \bibinfo {author} {\bibfnamefont
  {T.}~\bibnamefont {G\"orlach}}, \ and\ \bibinfo {author} {\bibfnamefont
  {H.~v.}\ \bibnamefont {L\"ohneysen}},\ }\bibfield  {title} {\enquote
  {\bibinfo {title} {{Superconductivity on the Border of Weak Itinerant
  Ferromagnetism in UCoGe}},}\ }\href {\doibase 10.1103/PhysRevLett.99.067006}
  {\bibfield  {journal} {\bibinfo  {journal} {Phys. Rev. Lett.}\ }\textbf
  {\bibinfo {volume} {99}},\ \bibinfo {pages} {067006} (\bibinfo {year}
  {2007})}\BibitemShut {NoStop}%
\bibitem [{\citenamefont {Li}\ \emph {et~al.}(2011)\citenamefont {Li},
  \citenamefont {Richter}, \citenamefont {Mannhart},\ and\ \citenamefont
  {Ashoori}}]{Li2011coexist}%
  \BibitemOpen
  \bibfield  {author} {\bibinfo {author} {\bibfnamefont {Lu}~\bibnamefont
  {Li}}, \bibinfo {author} {\bibfnamefont {C.}~\bibnamefont {Richter}},
  \bibinfo {author} {\bibfnamefont {J.}~\bibnamefont {Mannhart}}, \ and\
  \bibinfo {author} {\bibfnamefont {R.~C.}\ \bibnamefont {Ashoori}},\
  }\bibfield  {title} {\enquote {\bibinfo {title} {{Coexistence of magnetic
  order and two-dimensional superconductivity at LaAlO$_{3}$/SrTiO$_{3}$
  interfaces}},}\ }\href {\doibase 10.1038/nphys2080} {\bibfield  {journal}
  {\bibinfo  {journal} {Nature Physics}\ }\textbf {\bibinfo {volume} {7}},\
  \bibinfo {pages} {762--766} (\bibinfo {year} {2011})}\BibitemShut {NoStop}%
\bibitem [{\citenamefont {Bert}\ \emph {et~al.}(2011)\citenamefont {Bert},
  \citenamefont {Kalisky}, \citenamefont {Bell}, \citenamefont {Kim},
  \citenamefont {Hikita}, \citenamefont {Hwang},\ and\ \citenamefont
  {Moler}}]{Bert2011}%
  \BibitemOpen
  \bibfield  {author} {\bibinfo {author} {\bibfnamefont {Julie~A.}\
  \bibnamefont {Bert}}, \bibinfo {author} {\bibfnamefont {Beena}\ \bibnamefont
  {Kalisky}}, \bibinfo {author} {\bibfnamefont {Christopher}\ \bibnamefont
  {Bell}}, \bibinfo {author} {\bibfnamefont {Minu}\ \bibnamefont {Kim}},
  \bibinfo {author} {\bibfnamefont {Yasuyuki}\ \bibnamefont {Hikita}}, \bibinfo
  {author} {\bibfnamefont {Harold~Y.}\ \bibnamefont {Hwang}}, \ and\ \bibinfo
  {author} {\bibfnamefont {Kathryn~A.}\ \bibnamefont {Moler}},\ }\bibfield
  {title} {\enquote {\bibinfo {title} {{Direct imaging of the coexistence of
  ferromagnetism and superconductivity at LaAlO$_{3}$/SrTiO$_{3}$
  interface}},}\ }\href {\doibase 10.1038/nphys2079} {\bibfield  {journal}
  {\bibinfo  {journal} {Nature Physics}\ }\textbf {\bibinfo {volume} {7}},\
  \bibinfo {pages} {767--771} (\bibinfo {year} {2011})}\BibitemShut {NoStop}%
\bibitem [{\citenamefont {Dikin}\ \emph {et~al.}(2011)\citenamefont {Dikin},
  \citenamefont {Mehta}, \citenamefont {Bark}, \citenamefont {Folkman},
  \citenamefont {Eom},\ and\ \citenamefont {Chandrasekhar}}]{Dikin2011}%
  \BibitemOpen
  \bibfield  {author} {\bibinfo {author} {\bibfnamefont {D.~A.}\ \bibnamefont
  {Dikin}}, \bibinfo {author} {\bibfnamefont {M.}~\bibnamefont {Mehta}},
  \bibinfo {author} {\bibfnamefont {C.~W.}\ \bibnamefont {Bark}}, \bibinfo
  {author} {\bibfnamefont {C.~M.}\ \bibnamefont {Folkman}}, \bibinfo {author}
  {\bibfnamefont {C.~B.}\ \bibnamefont {Eom}}, \ and\ \bibinfo {author}
  {\bibfnamefont {V.}~\bibnamefont {Chandrasekhar}},\ }\bibfield  {title}
  {\enquote {\bibinfo {title} {{Coexistence of Superconductivity and
  Ferromagnetism in Two Dimensions}},}\ }\href {\doibase
  10.1103/PhysRevLett.107.056802} {\bibfield  {journal} {\bibinfo  {journal}
  {Phys. Rev. Lett.}\ }\textbf {\bibinfo {volume} {107}},\ \bibinfo {pages}
  {056802} (\bibinfo {year} {2011})}\BibitemShut {NoStop}%
\bibitem [{\citenamefont {Bergeret}\ \emph {et~al.}(2005)\citenamefont
  {Bergeret}, \citenamefont {Volkov},\ and\ \citenamefont
  {Efetov}}]{Bergeret2005}%
  \BibitemOpen
  \bibfield  {author} {\bibinfo {author} {\bibfnamefont {F.~S.}\ \bibnamefont
  {Bergeret}}, \bibinfo {author} {\bibfnamefont {A.~F.}\ \bibnamefont
  {Volkov}}, \ and\ \bibinfo {author} {\bibfnamefont {K.~B.}\ \bibnamefont
  {Efetov}},\ }\bibfield  {title} {\enquote {\bibinfo {title} {Odd triplet
  superconductivity and related phenomena in superconductor-ferromagnet
  structures},}\ }\href {\doibase 10.1103/RevModPhys.77.1321} {\bibfield
  {journal} {\bibinfo  {journal} {Rev. Mod. Phys.}\ }\textbf {\bibinfo {volume}
  {77}},\ \bibinfo {pages} {1321--1373} (\bibinfo {year} {2005})}\BibitemShut
  {NoStop}%
\bibitem [{\citenamefont {Scalapino}(2012)}]{Scalapino2012}%
  \BibitemOpen
  \bibfield  {author} {\bibinfo {author} {\bibfnamefont {D.~J.}\ \bibnamefont
  {Scalapino}},\ }\bibfield  {title} {\enquote {\bibinfo {title} {A common
  thread: The pairing interaction for unconventional superconductors},}\ }\href
  {\doibase 10.1103/RevModPhys.84.1383} {\bibfield  {journal} {\bibinfo
  {journal} {Rev. Mod. Phys.}\ }\textbf {\bibinfo {volume} {84}},\ \bibinfo
  {pages} {1383--1417} (\bibinfo {year} {2012})}\BibitemShut {NoStop}%
\bibitem [{\citenamefont {Sigrist}\ and\ \citenamefont
  {Ueda}(1991)}]{Sigrist1991}%
  \BibitemOpen
  \bibfield  {author} {\bibinfo {author} {\bibfnamefont {Manfred}\ \bibnamefont
  {Sigrist}}\ and\ \bibinfo {author} {\bibfnamefont {Kazuo}\ \bibnamefont
  {Ueda}},\ }\bibfield  {title} {\enquote {\bibinfo {title} {{Phenomenological
  theory of unconventional superconductivity}},}\ }\href {\doibase
  10.1103/RevModPhys.63.239} {\bibfield  {journal} {\bibinfo  {journal} {Rev.
  Mod. Phys.}\ }\textbf {\bibinfo {volume} {63}},\ \bibinfo {pages} {239--311}
  (\bibinfo {year} {1991})}\BibitemShut {NoStop}%
\bibitem [{\citenamefont {Šmejkal}\ \emph {et~al.}(2020)\citenamefont
  {Šmejkal}, \citenamefont {González-Hernández}, \citenamefont {Jungwirth},\
  and\ \citenamefont {Sinova}}]{Libor2020AM}%
  \BibitemOpen
  \bibfield  {author} {\bibinfo {author} {\bibfnamefont {Libor}\ \bibnamefont
  {Šmejkal}}, \bibinfo {author} {\bibfnamefont {Rafael}\ \bibnamefont
  {González-Hernández}}, \bibinfo {author} {\bibfnamefont {T.}~\bibnamefont
  {Jungwirth}}, \ and\ \bibinfo {author} {\bibfnamefont {J.}~\bibnamefont
  {Sinova}},\ }\bibfield  {title} {\enquote {\bibinfo {title} {{Crystal
  time-reversal symmetry breaking and spontaneous Hall effect in collinear
  antiferromagnets}},}\ }\href {\doibase 10.1126/sciadv.aaz8809} {\bibfield
  {journal} {\bibinfo  {journal} {Science Advances}\ }\textbf {\bibinfo
  {volume} {6}},\ \bibinfo {pages} {eaaz8809} (\bibinfo {year}
  {2020})}\BibitemShut {NoStop}%
\bibitem [{\citenamefont {Hayami}\ \emph {et~al.}(2019)\citenamefont {Hayami},
  \citenamefont {Yanagi},\ and\ \citenamefont {Kusunose}}]{Hayami2019AM}%
  \BibitemOpen
  \bibfield  {author} {\bibinfo {author} {\bibfnamefont {Satoru}\ \bibnamefont
  {Hayami}}, \bibinfo {author} {\bibfnamefont {Yuki}\ \bibnamefont {Yanagi}}, \
  and\ \bibinfo {author} {\bibfnamefont {Hiroaki}\ \bibnamefont {Kusunose}},\
  }\bibfield  {title} {\enquote {\bibinfo {title} {Momentum-dependent spin
  splitting by collinear antiferromagnetic ordering},}\ }\href {\doibase
  10.7566/JPSJ.88.123702} {\bibfield  {journal} {\bibinfo  {journal} {Journal
  of the Physical Society of Japan}\ }\textbf {\bibinfo {volume} {88}},\
  \bibinfo {pages} {123702} (\bibinfo {year} {2019})}\BibitemShut {NoStop}%
\bibitem [{\citenamefont {Hayami}\ \emph {et~al.}(2020)\citenamefont {Hayami},
  \citenamefont {Yanagi},\ and\ \citenamefont {Kusunose}}]{Hayami2020AM}%
  \BibitemOpen
  \bibfield  {author} {\bibinfo {author} {\bibfnamefont {Satoru}\ \bibnamefont
  {Hayami}}, \bibinfo {author} {\bibfnamefont {Yuki}\ \bibnamefont {Yanagi}}, \
  and\ \bibinfo {author} {\bibfnamefont {Hiroaki}\ \bibnamefont {Kusunose}},\
  }\bibfield  {title} {\enquote {\bibinfo {title} {Bottom-up design of
  spin-split and reshaped electronic band structures in antiferromagnets
  without spin-orbit coupling: Procedure on the basis of augmented
  multipoles},}\ }\href {\doibase 10.1103/PhysRevB.102.144441} {\bibfield
  {journal} {\bibinfo  {journal} {Phys. Rev. B}\ }\textbf {\bibinfo {volume}
  {102}},\ \bibinfo {pages} {144441} (\bibinfo {year} {2020})}\BibitemShut
  {NoStop}%
\bibitem [{\citenamefont {Yuan}\ \emph {et~al.}(2020)\citenamefont {Yuan},
  \citenamefont {Wang}, \citenamefont {Luo}, \citenamefont {Rashba},\ and\
  \citenamefont {Zunger}}]{Yuan2020AM}%
  \BibitemOpen
  \bibfield  {author} {\bibinfo {author} {\bibfnamefont {Lin-Ding}\
  \bibnamefont {Yuan}}, \bibinfo {author} {\bibfnamefont {Zhi}\ \bibnamefont
  {Wang}}, \bibinfo {author} {\bibfnamefont {Jun-Wei}\ \bibnamefont {Luo}},
  \bibinfo {author} {\bibfnamefont {Emmanuel~I.}\ \bibnamefont {Rashba}}, \
  and\ \bibinfo {author} {\bibfnamefont {Alex}\ \bibnamefont {Zunger}},\
  }\bibfield  {title} {\enquote {\bibinfo {title} {{Giant momentum-dependent
  spin splitting in centrosymmetric low-$Z$ antiferromagnets}},}\ }\href
  {\doibase 10.1103/PhysRevB.102.014422} {\bibfield  {journal} {\bibinfo
  {journal} {Phys. Rev. B}\ }\textbf {\bibinfo {volume} {102}},\ \bibinfo
  {pages} {014422} (\bibinfo {year} {2020})}\BibitemShut {NoStop}%
\bibitem [{\citenamefont {Yuan}\ \emph {et~al.}(2021)\citenamefont {Yuan},
  \citenamefont {Wang}, \citenamefont {Luo},\ and\ \citenamefont
  {Zunger}}]{Yuan2021AM}%
  \BibitemOpen
  \bibfield  {author} {\bibinfo {author} {\bibfnamefont {Lin-Ding}\
  \bibnamefont {Yuan}}, \bibinfo {author} {\bibfnamefont {Zhi}\ \bibnamefont
  {Wang}}, \bibinfo {author} {\bibfnamefont {Jun-Wei}\ \bibnamefont {Luo}}, \
  and\ \bibinfo {author} {\bibfnamefont {Alex}\ \bibnamefont {Zunger}},\
  }\bibfield  {title} {\enquote {\bibinfo {title} {{Prediction of low-Z
  collinear and noncollinear antiferromagnetic compounds having
  momentum-dependent spin splitting even without spin-orbit coupling}},}\
  }\href {\doibase 10.1103/PhysRevMaterials.5.014409} {\bibfield  {journal}
  {\bibinfo  {journal} {Phys. Rev. Mater.}\ }\textbf {\bibinfo {volume} {5}},\
  \bibinfo {pages} {014409} (\bibinfo {year} {2021})}\BibitemShut {NoStop}%
\bibitem [{\citenamefont {Mazin}\ \emph {et~al.}(2021)\citenamefont {Mazin},
  \citenamefont {Koepernik}, \citenamefont {Johannes}, \citenamefont
  {González-Hernández},\ and\ \citenamefont {\ifmmode~\check{S}\else
  \v{S}\fi{}mejkal}}]{Mazin2021}%
  \BibitemOpen
  \bibfield  {author} {\bibinfo {author} {\bibfnamefont {Igor~I.}\ \bibnamefont
  {Mazin}}, \bibinfo {author} {\bibfnamefont {Klaus}\ \bibnamefont
  {Koepernik}}, \bibinfo {author} {\bibfnamefont {Michelle~D.}\ \bibnamefont
  {Johannes}}, \bibinfo {author} {\bibfnamefont {Rafael}\ \bibnamefont
  {González-Hernández}}, \ and\ \bibinfo {author} {\bibfnamefont {Libor}\
  \bibnamefont {\ifmmode~\check{S}\else \v{S}\fi{}mejkal}},\ }\bibfield
  {title} {\enquote {\bibinfo {title} {Prediction of unconventional magnetism
  in doped {F}e{S}b$_{2}$},}\ }\href {\doibase 10.1073/pnas.2108924118}
  {\bibfield  {journal} {\bibinfo  {journal} {Proceedings of the National
  Academy of Sciences}\ }\textbf {\bibinfo {volume} {118}},\ \bibinfo {pages}
  {e2108924118} (\bibinfo {year} {2021})}\BibitemShut {NoStop}%
\bibitem [{\citenamefont {Liu}\ \emph {et~al.}(2022)\citenamefont {Liu},
  \citenamefont {Li}, \citenamefont {Han}, \citenamefont {Wan},\ and\
  \citenamefont {Liu}}]{Liu2022AM}%
  \BibitemOpen
  \bibfield  {author} {\bibinfo {author} {\bibfnamefont {Pengfei}\ \bibnamefont
  {Liu}}, \bibinfo {author} {\bibfnamefont {Jiayu}\ \bibnamefont {Li}},
  \bibinfo {author} {\bibfnamefont {Jingzhi}\ \bibnamefont {Han}}, \bibinfo
  {author} {\bibfnamefont {Xiangang}\ \bibnamefont {Wan}}, \ and\ \bibinfo
  {author} {\bibfnamefont {Qihang}\ \bibnamefont {Liu}},\ }\bibfield  {title}
  {\enquote {\bibinfo {title} {Spin-group symmetry in magnetic materials with
  negligible spin-orbit coupling},}\ }\href {\doibase
  10.1103/PhysRevX.12.021016} {\bibfield  {journal} {\bibinfo  {journal} {Phys.
  Rev. X}\ }\textbf {\bibinfo {volume} {12}},\ \bibinfo {pages} {021016}
  (\bibinfo {year} {2022})}\BibitemShut {NoStop}%
\bibitem [{\citenamefont {Feng}\ \emph
  {et~al.}(2022{\natexlab{a}})\citenamefont {Feng}, \citenamefont {Zhou},
  \citenamefont {{\v{S}}mejkal}, \citenamefont {Wu}, \citenamefont {Zhu},
  \citenamefont {Guo}, \citenamefont {Gonz{\'a}lez-Hern{\'a}ndez},
  \citenamefont {Wang}, \citenamefont {Yan}, \citenamefont {Qin}, \citenamefont
  {Zhang}, \citenamefont {Wu}, \citenamefont {Chen}, \citenamefont {Meng},
  \citenamefont {Liu}, \citenamefont {Xia}, \citenamefont {Sinova},
  \citenamefont {Jungwirth},\ and\ \citenamefont {Liu}}]{Feng2022AM}%
  \BibitemOpen
  \bibfield  {author} {\bibinfo {author} {\bibfnamefont {Zexin}\ \bibnamefont
  {Feng}}, \bibinfo {author} {\bibfnamefont {Xiaorong}\ \bibnamefont {Zhou}},
  \bibinfo {author} {\bibfnamefont {Libor}\ \bibnamefont {{\v{S}}mejkal}},
  \bibinfo {author} {\bibfnamefont {Lei}\ \bibnamefont {Wu}}, \bibinfo {author}
  {\bibfnamefont {Zengwei}\ \bibnamefont {Zhu}}, \bibinfo {author}
  {\bibfnamefont {Huixin}\ \bibnamefont {Guo}}, \bibinfo {author}
  {\bibfnamefont {Rafael}\ \bibnamefont {Gonz{\'a}lez-Hern{\'a}ndez}}, \bibinfo
  {author} {\bibfnamefont {Xiaoning}\ \bibnamefont {Wang}}, \bibinfo {author}
  {\bibfnamefont {Han}\ \bibnamefont {Yan}}, \bibinfo {author} {\bibfnamefont
  {Peixin}\ \bibnamefont {Qin}}, \bibinfo {author} {\bibfnamefont {Xin}\
  \bibnamefont {Zhang}}, \bibinfo {author} {\bibfnamefont {Haojiang}\
  \bibnamefont {Wu}}, \bibinfo {author} {\bibfnamefont {Hongyu}\ \bibnamefont
  {Chen}}, \bibinfo {author} {\bibfnamefont {Ziang}\ \bibnamefont {Meng}},
  \bibinfo {author} {\bibfnamefont {Li}~\bibnamefont {Liu}}, \bibinfo {author}
  {\bibfnamefont {Zhengcai}\ \bibnamefont {Xia}}, \bibinfo {author}
  {\bibfnamefont {Jairo}\ \bibnamefont {Sinova}}, \bibinfo {author}
  {\bibfnamefont {Tom{\'a}{\v{s}}}\ \bibnamefont {Jungwirth}}, \ and\ \bibinfo
  {author} {\bibfnamefont {Zhiqi}\ \bibnamefont {Liu}},\ }\bibfield  {title}
  {\enquote {\bibinfo {title} {{An anomalous Hall effect in altermagnetic
  ruthenium dioxide}},}\ }\href {\doibase 10.1038/s41928-022-00866-z}
  {\bibfield  {journal} {\bibinfo  {journal} {Nature Electronics}\ }\textbf
  {\bibinfo {volume} {5}},\ \bibinfo {pages} {735--743} (\bibinfo {year}
  {2022}{\natexlab{a}})}\BibitemShut {NoStop}%
\bibitem [{\citenamefont {Gonzalez~Betancourt}\ \emph
  {et~al.}(2023)\citenamefont {Gonzalez~Betancourt}, \citenamefont
  {Zub\'a\ifmmode~\check{c}\else \v{c}\fi{}}, \citenamefont
  {Gonzalez-Hernandez}, \citenamefont {Geishendorf}, \citenamefont {\ifmmode
  \check{S}\else \v{S}\fi{}ob\'a\ifmmode~\check{n}\else \v{n}\fi{}},
  \citenamefont {Springholz}, \citenamefont {Olejn\'{\i}k}, \citenamefont
  {\ifmmode~\check{S}\else \v{S}\fi{}mejkal}, \citenamefont {Sinova},
  \citenamefont {Jungwirth}, \citenamefont {Goennenwein}, \citenamefont
  {Thomas}, \citenamefont {Reichlov\'a}, \citenamefont {\ifmmode~\check{Z}\else
  \v{Z}\fi{}elezn\'y},\ and\ \citenamefont {Kriegner}}]{Betancourt2023}%
  \BibitemOpen
  \bibfield  {author} {\bibinfo {author} {\bibfnamefont {R.~D.}\ \bibnamefont
  {Gonzalez~Betancourt}}, \bibinfo {author} {\bibfnamefont {J.}~\bibnamefont
  {Zub\'a\ifmmode~\check{c}\else \v{c}\fi{}}}, \bibinfo {author} {\bibfnamefont
  {R.}~\bibnamefont {Gonzalez-Hernandez}}, \bibinfo {author} {\bibfnamefont
  {K.}~\bibnamefont {Geishendorf}}, \bibinfo {author} {\bibfnamefont
  {Z.}~\bibnamefont {\ifmmode \check{S}\else
  \v{S}\fi{}ob\'a\ifmmode~\check{n}\else \v{n}\fi{}}}, \bibinfo {author}
  {\bibfnamefont {G.}~\bibnamefont {Springholz}}, \bibinfo {author}
  {\bibfnamefont {K.}~\bibnamefont {Olejn\'{\i}k}}, \bibinfo {author}
  {\bibfnamefont {L.}~\bibnamefont {\ifmmode~\check{S}\else \v{S}\fi{}mejkal}},
  \bibinfo {author} {\bibfnamefont {J.}~\bibnamefont {Sinova}}, \bibinfo
  {author} {\bibfnamefont {T.}~\bibnamefont {Jungwirth}}, \bibinfo {author}
  {\bibfnamefont {S.~T.~B.}\ \bibnamefont {Goennenwein}}, \bibinfo {author}
  {\bibfnamefont {A.}~\bibnamefont {Thomas}}, \bibinfo {author} {\bibfnamefont
  {H.}~\bibnamefont {Reichlov\'a}}, \bibinfo {author} {\bibfnamefont
  {J.}~\bibnamefont {\ifmmode~\check{Z}\else \v{Z}\fi{}elezn\'y}}, \ and\
  \bibinfo {author} {\bibfnamefont {D.}~\bibnamefont {Kriegner}},\ }\bibfield
  {title} {\enquote {\bibinfo {title} {{Spontaneous Anomalous Hall Effect
  Arising from an Unconventional Compensated Magnetic Phase in a
  Semiconductor}},}\ }\href {\doibase 10.1103/PhysRevLett.130.036702}
  {\bibfield  {journal} {\bibinfo  {journal} {Phys. Rev. Lett.}\ }\textbf
  {\bibinfo {volume} {130}},\ \bibinfo {pages} {036702} (\bibinfo {year}
  {2023})}\BibitemShut {NoStop}%
\bibitem [{\citenamefont {Mazin}(2023)}]{Mazin2023AM}%
  \BibitemOpen
  \bibfield  {author} {\bibinfo {author} {\bibfnamefont {I.~I.}\ \bibnamefont
  {Mazin}},\ }\bibfield  {title} {\enquote {\bibinfo {title} {{Altermagnetism
  in MnTe: Origin, predicted manifestations, and routes to detwinning}},}\
  }\href {\doibase 10.1103/PhysRevB.107.L100418} {\bibfield  {journal}
  {\bibinfo  {journal} {Phys. Rev. B}\ }\textbf {\bibinfo {volume} {107}},\
  \bibinfo {pages} {L100418} (\bibinfo {year} {2023})}\BibitemShut {NoStop}%
\bibitem [{\citenamefont {Turek}(2022)}]{Turek2022AM}%
  \BibitemOpen
  \bibfield  {author} {\bibinfo {author} {\bibfnamefont {Ilja}\ \bibnamefont
  {Turek}},\ }\bibfield  {title} {\enquote {\bibinfo {title} {Altermagnetism
  and magnetic groups with pseudoscalar electron spin},}\ }\href {\doibase
  10.1103/PhysRevB.106.094432} {\bibfield  {journal} {\bibinfo  {journal}
  {Phys. Rev. B}\ }\textbf {\bibinfo {volume} {106}},\ \bibinfo {pages}
  {094432} (\bibinfo {year} {2022})}\BibitemShut {NoStop}%
\bibitem [{\citenamefont {Guo}\ \emph {et~al.}(2023)\citenamefont {Guo},
  \citenamefont {Liu}, \citenamefont {Janson}, \citenamefont {Fulga},
  \citenamefont {{van den Brink}},\ and\ \citenamefont {Facio}}]{Guo2023AM}%
  \BibitemOpen
  \bibfield  {author} {\bibinfo {author} {\bibfnamefont {Yaqian}\ \bibnamefont
  {Guo}}, \bibinfo {author} {\bibfnamefont {Hui}\ \bibnamefont {Liu}}, \bibinfo
  {author} {\bibfnamefont {Oleg}\ \bibnamefont {Janson}}, \bibinfo {author}
  {\bibfnamefont {Ion~Cosma}\ \bibnamefont {Fulga}}, \bibinfo {author}
  {\bibfnamefont {Jeroen}\ \bibnamefont {{van den Brink}}}, \ and\ \bibinfo
  {author} {\bibfnamefont {Jorge~I.}\ \bibnamefont {Facio}},\ }\bibfield
  {title} {\enquote {\bibinfo {title} {Spin-split collinear antiferromagnets: A
  large-scale ab-initio study},}\ }\href {\doibase
  https://doi.org/10.1016/j.mtphys.2023.100991} {\bibfield  {journal} {\bibinfo
   {journal} {Materials Today Physics}\ }\textbf {\bibinfo {volume} {32}},\
  \bibinfo {pages} {100991} (\bibinfo {year} {2023})}\BibitemShut {NoStop}%
\bibitem [{\citenamefont {{Hariki}}\ \emph {et~al.}(2023)\citenamefont
  {{Hariki}}, \citenamefont {{Yamaguchi}}, \citenamefont {{Kriegner}},
  \citenamefont {{Edmonds}}, \citenamefont {{Wadley}}, \citenamefont {{Dhesi}},
  \citenamefont {{Springholz}}, \citenamefont {{{\v{S}}mejkal}}, \citenamefont
  {{V{\'y}born{\'y}}}, \citenamefont {{Jungwirth}},\ and\ \citenamefont
  {{Kune{\v{s}}}}}]{Hariki2023AM}%
  \BibitemOpen
  \bibfield  {author} {\bibinfo {author} {\bibfnamefont {A.}~\bibnamefont
  {{Hariki}}}, \bibinfo {author} {\bibfnamefont {T.}~\bibnamefont
  {{Yamaguchi}}}, \bibinfo {author} {\bibfnamefont {D.}~\bibnamefont
  {{Kriegner}}}, \bibinfo {author} {\bibfnamefont {K.~W.}\ \bibnamefont
  {{Edmonds}}}, \bibinfo {author} {\bibfnamefont {P.}~\bibnamefont {{Wadley}}},
  \bibinfo {author} {\bibfnamefont {S.~S.}\ \bibnamefont {{Dhesi}}}, \bibinfo
  {author} {\bibfnamefont {G.}~\bibnamefont {{Springholz}}}, \bibinfo {author}
  {\bibfnamefont {L.}~\bibnamefont {{{\v{S}}mejkal}}}, \bibinfo {author}
  {\bibfnamefont {K.}~\bibnamefont {{V{\'y}born{\'y}}}}, \bibinfo {author}
  {\bibfnamefont {T.}~\bibnamefont {{Jungwirth}}}, \ and\ \bibinfo {author}
  {\bibfnamefont {J.}~\bibnamefont {{Kune{\v{s}}}}},\ }\bibfield  {title}
  {\enquote {\bibinfo {title} {{X-ray Magnetic Circular Dichroism in
  Altermagnetic $\alpha$-MnTe}},}\ }\href {\doibase 10.48550/arXiv.2305.03588}
  {\bibfield  {journal} {\bibinfo  {journal} {arXiv e-prints}\ ,\ \bibinfo
  {eid} {arXiv:2305.03588}} (\bibinfo {year} {2023})},\ \Eprint
  {http://arxiv.org/abs/2305.03588} {arXiv:2305.03588 [cond-mat.mtrl-sci]}
  \BibitemShut {NoStop}%
\bibitem [{\citenamefont {{Zhou}}\ \emph {et~al.}(2023)\citenamefont {{Zhou}},
  \citenamefont {{Feng}}, \citenamefont {{Zhang}}, \citenamefont {{Smejkal}},
  \citenamefont {{Sinova}}, \citenamefont {{Mokrousov}},\ and\ \citenamefont
  {{Yao}}}]{Zhou2023AM}%
  \BibitemOpen
  \bibfield  {author} {\bibinfo {author} {\bibfnamefont {Xiaodong}\
  \bibnamefont {{Zhou}}}, \bibinfo {author} {\bibfnamefont {Wanxiang}\
  \bibnamefont {{Feng}}}, \bibinfo {author} {\bibfnamefont {Run-Wu}\
  \bibnamefont {{Zhang}}}, \bibinfo {author} {\bibfnamefont {Libor}\
  \bibnamefont {{Smejkal}}}, \bibinfo {author} {\bibfnamefont {Jairo}\
  \bibnamefont {{Sinova}}}, \bibinfo {author} {\bibfnamefont {Yuriy}\
  \bibnamefont {{Mokrousov}}}, \ and\ \bibinfo {author} {\bibfnamefont {Yugui}\
  \bibnamefont {{Yao}}},\ }\bibfield  {title} {\enquote {\bibinfo {title}
  {{Crystal Thermal Transport in Altermagnetic RuO2}},}\ }\href {\doibase
  10.48550/arXiv.2305.01410} {\bibfield  {journal} {\bibinfo  {journal} {arXiv
  e-prints}\ ,\ \bibinfo {eid} {arXiv:2305.01410}} (\bibinfo {year} {2023})},\
  \Eprint {http://arxiv.org/abs/2305.01410} {arXiv:2305.01410
  [cond-mat.mtrl-sci]} \BibitemShut {NoStop}%
\bibitem [{\citenamefont {\ifmmode~\check{S}\else \v{S}\fi{}mejkal}\ \emph
  {et~al.}(2022{\natexlab{a}})\citenamefont {\ifmmode~\check{S}\else
  \v{S}\fi{}mejkal}, \citenamefont {Hellenes}, \citenamefont
  {Gonz\'alez-Hern\'andez}, \citenamefont {Sinova},\ and\ \citenamefont
  {Jungwirth}}]{Libor2022AMc}%
  \BibitemOpen
  \bibfield  {author} {\bibinfo {author} {\bibfnamefont {Libor}\ \bibnamefont
  {\ifmmode~\check{S}\else \v{S}\fi{}mejkal}}, \bibinfo {author} {\bibfnamefont
  {Anna~Birk}\ \bibnamefont {Hellenes}}, \bibinfo {author} {\bibfnamefont
  {Rafael}\ \bibnamefont {Gonz\'alez-Hern\'andez}}, \bibinfo {author}
  {\bibfnamefont {Jairo}\ \bibnamefont {Sinova}}, \ and\ \bibinfo {author}
  {\bibfnamefont {Tomas}\ \bibnamefont {Jungwirth}},\ }\bibfield  {title}
  {\enquote {\bibinfo {title} {{Giant and Tunneling Magnetoresistance in
  Unconventional Collinear Antiferromagnets with Nonrelativistic Spin-Momentum
  Coupling}},}\ }\href {\doibase 10.1103/PhysRevX.12.011028} {\bibfield
  {journal} {\bibinfo  {journal} {Phys. Rev. X}\ }\textbf {\bibinfo {volume}
  {12}},\ \bibinfo {pages} {011028} (\bibinfo {year}
  {2022}{\natexlab{a}})}\BibitemShut {NoStop}%
\bibitem [{\citenamefont {\ifmmode~\check{S}\else \v{S}\fi{}mejkal}\ \emph
  {et~al.}(2022{\natexlab{b}})\citenamefont {\ifmmode~\check{S}\else
  \v{S}\fi{}mejkal}, \citenamefont {Sinova},\ and\ \citenamefont
  {Jungwirth}}]{Libor2022AMa}%
  \BibitemOpen
  \bibfield  {author} {\bibinfo {author} {\bibfnamefont {Libor}\ \bibnamefont
  {\ifmmode~\check{S}\else \v{S}\fi{}mejkal}}, \bibinfo {author} {\bibfnamefont
  {Jairo}\ \bibnamefont {Sinova}}, \ and\ \bibinfo {author} {\bibfnamefont
  {Tomas}\ \bibnamefont {Jungwirth}},\ }\bibfield  {title} {\enquote {\bibinfo
  {title} {{Beyond Conventional Ferromagnetism and Antiferromagnetism: A Phase
  with Nonrelativistic Spin and Crystal Rotation Symmetry}},}\ }\href {\doibase
  10.1103/PhysRevX.12.031042} {\bibfield  {journal} {\bibinfo  {journal} {Phys.
  Rev. X}\ }\textbf {\bibinfo {volume} {12}},\ \bibinfo {pages} {031042}
  (\bibinfo {year} {2022}{\natexlab{b}})}\BibitemShut {NoStop}%
\bibitem [{\citenamefont {\ifmmode~\check{S}\else \v{S}\fi{}mejkal}\ \emph
  {et~al.}(2022{\natexlab{c}})\citenamefont {\ifmmode~\check{S}\else
  \v{S}\fi{}mejkal}, \citenamefont {Sinova},\ and\ \citenamefont
  {Jungwirth}}]{Libor2022AMb}%
  \BibitemOpen
  \bibfield  {author} {\bibinfo {author} {\bibfnamefont {Libor}\ \bibnamefont
  {\ifmmode~\check{S}\else \v{S}\fi{}mejkal}}, \bibinfo {author} {\bibfnamefont
  {Jairo}\ \bibnamefont {Sinova}}, \ and\ \bibinfo {author} {\bibfnamefont
  {Tomas}\ \bibnamefont {Jungwirth}},\ }\bibfield  {title} {\enquote {\bibinfo
  {title} {{Emerging Research Landscape of Altermagnetism}},}\ }\href {\doibase
  10.1103/PhysRevX.12.040501} {\bibfield  {journal} {\bibinfo  {journal} {Phys.
  Rev. X}\ }\textbf {\bibinfo {volume} {12}},\ \bibinfo {pages} {040501}
  (\bibinfo {year} {2022}{\natexlab{c}})}\BibitemShut {NoStop}%
\bibitem [{\citenamefont {{Mazin}}(2022)}]{Mazin2022AM}%
  \BibitemOpen
  \bibfield  {author} {\bibinfo {author} {\bibfnamefont {Igor~I.}\ \bibnamefont
  {{Mazin}}},\ }\bibfield  {title} {\enquote {\bibinfo {title} {{Notes on
  altermagnetism and superconductivity}},}\ }\href {\doibase
  10.48550/arXiv.2203.05000} {\bibfield  {journal} {\bibinfo  {journal} {arXiv
  e-prints}\ ,\ \bibinfo {eid} {arXiv:2203.05000}} (\bibinfo {year} {2022})},\
  \Eprint {http://arxiv.org/abs/2203.05000} {arXiv:2203.05000
  [cond-mat.supr-con]} \BibitemShut {NoStop}%
\bibitem [{\citenamefont {Sun}\ \emph {et~al.}(2023)\citenamefont {Sun},
  \citenamefont {Brataas},\ and\ \citenamefont {Linder}}]{Sun2023AM}%
  \BibitemOpen
  \bibfield  {author} {\bibinfo {author} {\bibfnamefont {Chi}\ \bibnamefont
  {Sun}}, \bibinfo {author} {\bibfnamefont {Arne}\ \bibnamefont {Brataas}}, \
  and\ \bibinfo {author} {\bibfnamefont {Jacob}\ \bibnamefont {Linder}},\
  }\bibfield  {title} {\enquote {\bibinfo {title} {Andreev reflection in
  altermagnets},}\ }\href {\doibase 10.1103/PhysRevB.108.054511} {\bibfield
  {journal} {\bibinfo  {journal} {Phys. Rev. B}\ }\textbf {\bibinfo {volume}
  {108}},\ \bibinfo {pages} {054511} (\bibinfo {year} {2023})}\BibitemShut
  {NoStop}%
\bibitem [{\citenamefont {{Papaj}}(2023)}]{Papaj2023}%
  \BibitemOpen
  \bibfield  {author} {\bibinfo {author} {\bibfnamefont {Micha{\l}}\
  \bibnamefont {{Papaj}}},\ }\bibfield  {title} {\enquote {\bibinfo {title}
  {{Andreev reflection at altermagnet/superconductor interface}},}\ }\href
  {\doibase 10.48550/arXiv.2305.03856} {\bibfield  {journal} {\bibinfo
  {journal} {arXiv e-prints}\ ,\ \bibinfo {eid} {arXiv:2305.03856}} (\bibinfo
  {year} {2023})},\ \Eprint {http://arxiv.org/abs/2305.03856} {arXiv:2305.03856
  [cond-mat.supr-con]} \BibitemShut {NoStop}%
\bibitem [{\citenamefont {Ouassou}\ \emph {et~al.}(2023)\citenamefont
  {Ouassou}, \citenamefont {Brataas},\ and\ \citenamefont
  {Linder}}]{Ouassou2023AM}%
  \BibitemOpen
  \bibfield  {author} {\bibinfo {author} {\bibfnamefont {Jabir~Ali}\
  \bibnamefont {Ouassou}}, \bibinfo {author} {\bibfnamefont {Arne}\
  \bibnamefont {Brataas}}, \ and\ \bibinfo {author} {\bibfnamefont {Jacob}\
  \bibnamefont {Linder}},\ }\bibfield  {title} {\enquote {\bibinfo {title} {dc
  josephson effect in altermagnets},}\ }\href {\doibase
  10.1103/PhysRevLett.131.076003} {\bibfield  {journal} {\bibinfo  {journal}
  {Phys. Rev. Lett.}\ }\textbf {\bibinfo {volume} {131}},\ \bibinfo {pages}
  {076003} (\bibinfo {year} {2023})}\BibitemShut {NoStop}%
\bibitem [{\citenamefont {{Zhang}}\ \emph {et~al.}(2023)\citenamefont
  {{Zhang}}, \citenamefont {{Hu}},\ and\ \citenamefont
  {{Neupert}}}]{Zhang2023AM}%
  \BibitemOpen
  \bibfield  {author} {\bibinfo {author} {\bibfnamefont {Song-Bo}\ \bibnamefont
  {{Zhang}}}, \bibinfo {author} {\bibfnamefont {Lun-Hui}\ \bibnamefont {{Hu}}},
  \ and\ \bibinfo {author} {\bibfnamefont {Titus}\ \bibnamefont {{Neupert}}},\
  }\bibfield  {title} {\enquote {\bibinfo {title} {{Finite-momentum Cooper
  pairing in proximitized altermagnets}},}\ }\href {\doibase
  10.48550/arXiv.2302.13185} {\bibfield  {journal} {\bibinfo  {journal} {arXiv
  e-prints}\ ,\ \bibinfo {eid} {arXiv:2302.13185}} (\bibinfo {year} {2023})},\
  \Eprint {http://arxiv.org/abs/2302.13185} {arXiv:2302.13185
  [cond-mat.supr-con]} \BibitemShut {NoStop}%
\bibitem [{\citenamefont {Bychkov}\ and\ \citenamefont
  {Rashba}(1984)}]{bychkov1984}%
  \BibitemOpen
  \bibfield  {author} {\bibinfo {author} {\bibfnamefont {Yua~A}\ \bibnamefont
  {Bychkov}}\ and\ \bibinfo {author} {\bibfnamefont {{\'E}~I}\ \bibnamefont
  {Rashba}},\ }\bibfield  {title} {\enquote {\bibinfo {title} {Properties of a
  2d electron gas with lifted spectral degeneracy},}\ }\href@noop {} {\bibfield
   {journal} {\bibinfo  {journal} {JETP lett}\ }\textbf {\bibinfo {volume}
  {39}},\ \bibinfo {pages} {78} (\bibinfo {year} {1984})}\BibitemShut {NoStop}%
\bibitem [{\citenamefont {Read}\ and\ \citenamefont {Green}(2000)}]{read2000}%
  \BibitemOpen
  \bibfield  {author} {\bibinfo {author} {\bibfnamefont {N.}~\bibnamefont
  {Read}}\ and\ \bibinfo {author} {\bibfnamefont {Dmitry}\ \bibnamefont
  {Green}},\ }\bibfield  {title} {\enquote {\bibinfo {title} {{Paired states of
  fermions in two dimensions with breaking of parity and time-reversal
  symmetries and the fractional quantum Hall effect}},}\ }\href {\doibase
  10.1103/PhysRevB.61.10267} {\bibfield  {journal} {\bibinfo  {journal} {Phys.
  Rev. B}\ }\textbf {\bibinfo {volume} {61}},\ \bibinfo {pages} {10267--10297}
  (\bibinfo {year} {2000})}\BibitemShut {NoStop}%
\bibitem [{\citenamefont {Sato}\ \emph {et~al.}(2009)\citenamefont {Sato},
  \citenamefont {Takahashi},\ and\ \citenamefont {Fujimoto}}]{sato2009non}%
  \BibitemOpen
  \bibfield  {author} {\bibinfo {author} {\bibfnamefont {Masatoshi}\
  \bibnamefont {Sato}}, \bibinfo {author} {\bibfnamefont {Yoshiro}\
  \bibnamefont {Takahashi}}, \ and\ \bibinfo {author} {\bibfnamefont {Satoshi}\
  \bibnamefont {Fujimoto}},\ }\bibfield  {title} {\enquote {\bibinfo {title}
  {{Non-Abelian Topological Order in $s$-Wave Superfluids of Ultracold
  Fermionic Atoms}},}\ }\href {\doibase 10.1103/PhysRevLett.103.020401}
  {\bibfield  {journal} {\bibinfo  {journal} {Phys. Rev. Lett.}\ }\textbf
  {\bibinfo {volume} {103}},\ \bibinfo {pages} {020401} (\bibinfo {year}
  {2009})}\BibitemShut {NoStop}%
\bibitem [{\citenamefont {Qi}\ \emph {et~al.}(2010)\citenamefont {Qi},
  \citenamefont {Hughes},\ and\ \citenamefont {Zhang}}]{Qi2010chiral}%
  \BibitemOpen
  \bibfield  {author} {\bibinfo {author} {\bibfnamefont {Xiao-Liang}\
  \bibnamefont {Qi}}, \bibinfo {author} {\bibfnamefont {Taylor~L.}\
  \bibnamefont {Hughes}}, \ and\ \bibinfo {author} {\bibfnamefont {Shou-Cheng}\
  \bibnamefont {Zhang}},\ }\bibfield  {title} {\enquote {\bibinfo {title}
  {{Chiral topological superconductor from the quantum Hall state}},}\ }\href
  {\doibase 10.1103/PhysRevB.82.184516} {\bibfield  {journal} {\bibinfo
  {journal} {Phys. Rev. B}\ }\textbf {\bibinfo {volume} {82}},\ \bibinfo
  {pages} {184516} (\bibinfo {year} {2010})}\BibitemShut {NoStop}%
\bibitem [{\citenamefont {Sau}\ \emph {et~al.}(2010)\citenamefont {Sau},
  \citenamefont {Lutchyn}, \citenamefont {Tewari},\ and\ \citenamefont
  {Das~Sarma}}]{Sau2010TSC}%
  \BibitemOpen
  \bibfield  {author} {\bibinfo {author} {\bibfnamefont {Jay~D.}\ \bibnamefont
  {Sau}}, \bibinfo {author} {\bibfnamefont {Roman~M.}\ \bibnamefont {Lutchyn}},
  \bibinfo {author} {\bibfnamefont {Sumanta}\ \bibnamefont {Tewari}}, \ and\
  \bibinfo {author} {\bibfnamefont {S.}~\bibnamefont {Das~Sarma}},\ }\bibfield
  {title} {\enquote {\bibinfo {title} {{Generic New Platform for Topological
  Quantum Computation Using Semiconductor Heterostructures}},}\ }\href
  {\doibase 10.1103/PhysRevLett.104.040502} {\bibfield  {journal} {\bibinfo
  {journal} {Phys. Rev. Lett.}\ }\textbf {\bibinfo {volume} {104}},\ \bibinfo
  {pages} {040502} (\bibinfo {year} {2010})}\BibitemShut {NoStop}%
\bibitem [{\citenamefont {Alicea}(2010)}]{alicea2010}%
  \BibitemOpen
  \bibfield  {author} {\bibinfo {author} {\bibfnamefont {Jason}\ \bibnamefont
  {Alicea}},\ }\bibfield  {title} {\enquote {\bibinfo {title} {{Majorana
  fermions in a tunable semiconductor device}},}\ }\href@noop {} {\bibfield
  {journal} {\bibinfo  {journal} {Phys. Rev. B}\ }\textbf {\bibinfo {volume}
  {81}},\ \bibinfo {pages} {125318} (\bibinfo {year} {2010})}\BibitemShut
  {NoStop}%
\bibitem [{\citenamefont {Qi}\ \emph {et~al.}(2009)\citenamefont {Qi},
  \citenamefont {Hughes}, \citenamefont {Raghu},\ and\ \citenamefont
  {Zhang}}]{Qi2009b}%
  \BibitemOpen
  \bibfield  {author} {\bibinfo {author} {\bibfnamefont {Xiao-Liang}\
  \bibnamefont {Qi}}, \bibinfo {author} {\bibfnamefont {Taylor~L.}\
  \bibnamefont {Hughes}}, \bibinfo {author} {\bibfnamefont {S.}~\bibnamefont
  {Raghu}}, \ and\ \bibinfo {author} {\bibfnamefont {Shou-Cheng}\ \bibnamefont
  {Zhang}},\ }\bibfield  {title} {\enquote {\bibinfo {title}
  {{Time-Reversal-Invariant Topological Superconductors and Superfluids in Two
  and Three Dimensions}},}\ }\href {\doibase 10.1103/PhysRevLett.102.187001}
  {\bibfield  {journal} {\bibinfo  {journal} {Phys. Rev. Lett.}\ }\textbf
  {\bibinfo {volume} {102}},\ \bibinfo {pages} {187001} (\bibinfo {year}
  {2009})}\BibitemShut {NoStop}%
\bibitem [{\citenamefont {Deng}\ \emph {et~al.}(2012)\citenamefont {Deng},
  \citenamefont {Viola},\ and\ \citenamefont {Ortiz}}]{Deng2012}%
  \BibitemOpen
  \bibfield  {author} {\bibinfo {author} {\bibfnamefont {Shusa}\ \bibnamefont
  {Deng}}, \bibinfo {author} {\bibfnamefont {Lorenza}\ \bibnamefont {Viola}}, \
  and\ \bibinfo {author} {\bibfnamefont {Gerardo}\ \bibnamefont {Ortiz}},\
  }\bibfield  {title} {\enquote {\bibinfo {title} {{Majorana Modes in
  Time-Reversal Invariant $s$-Wave Topological Superconductors}},}\ }\href
  {\doibase 10.1103/PhysRevLett.108.036803} {\bibfield  {journal} {\bibinfo
  {journal} {Phys. Rev. Lett.}\ }\textbf {\bibinfo {volume} {108}},\ \bibinfo
  {pages} {036803} (\bibinfo {year} {2012})}\BibitemShut {NoStop}%
\bibitem [{\citenamefont {Nakosai}\ \emph {et~al.}(2012)\citenamefont
  {Nakosai}, \citenamefont {Tanaka},\ and\ \citenamefont
  {Nagaosa}}]{Nakosai2012}%
  \BibitemOpen
  \bibfield  {author} {\bibinfo {author} {\bibfnamefont {Sho}\ \bibnamefont
  {Nakosai}}, \bibinfo {author} {\bibfnamefont {Yukio}\ \bibnamefont {Tanaka}},
  \ and\ \bibinfo {author} {\bibfnamefont {Naoto}\ \bibnamefont {Nagaosa}},\
  }\bibfield  {title} {\enquote {\bibinfo {title} {{Topological
  Superconductivity in Bilayer Rashba System}},}\ }\href {\doibase
  10.1103/PhysRevLett.108.147003} {\bibfield  {journal} {\bibinfo  {journal}
  {Phys. Rev. Lett.}\ }\textbf {\bibinfo {volume} {108}},\ \bibinfo {pages}
  {147003} (\bibinfo {year} {2012})}\BibitemShut {NoStop}%
\bibitem [{\citenamefont {Zhang}\ \emph
  {et~al.}(2013{\natexlab{a}})\citenamefont {Zhang}, \citenamefont {Kane},\
  and\ \citenamefont {Mele}}]{zhang2013kramers}%
  \BibitemOpen
  \bibfield  {author} {\bibinfo {author} {\bibfnamefont {Fan}\ \bibnamefont
  {Zhang}}, \bibinfo {author} {\bibfnamefont {C.~L.}\ \bibnamefont {Kane}}, \
  and\ \bibinfo {author} {\bibfnamefont {E.~J.}\ \bibnamefont {Mele}},\
  }\bibfield  {title} {\enquote {\bibinfo {title} {{Time-Reversal-Invariant
  Topological Superconductivity and Majorana Kramers Pairs}},}\ }\href
  {\doibase 10.1103/PhysRevLett.111.056402} {\bibfield  {journal} {\bibinfo
  {journal} {Phys. Rev. Lett.}\ }\textbf {\bibinfo {volume} {111}},\ \bibinfo
  {pages} {056402} (\bibinfo {year} {2013}{\natexlab{a}})}\BibitemShut
  {NoStop}%
\bibitem [{\citenamefont {Wang}\ \emph {et~al.}(2014)\citenamefont {Wang},
  \citenamefont {Xu},\ and\ \citenamefont {Zhang}}]{wang2014TRI}%
  \BibitemOpen
  \bibfield  {author} {\bibinfo {author} {\bibfnamefont {Jing}\ \bibnamefont
  {Wang}}, \bibinfo {author} {\bibfnamefont {Yong}\ \bibnamefont {Xu}}, \ and\
  \bibinfo {author} {\bibfnamefont {Shou-Cheng}\ \bibnamefont {Zhang}},\
  }\bibfield  {title} {\enquote {\bibinfo {title} {{Two-dimensional
  time-reversal-invariant topological superconductivity in a doped quantum
  spin-Hall insulator}},}\ }\href {\doibase 10.1103/PhysRevB.90.054503}
  {\bibfield  {journal} {\bibinfo  {journal} {Phys. Rev. B}\ }\textbf {\bibinfo
  {volume} {90}},\ \bibinfo {pages} {054503} (\bibinfo {year}
  {2014})}\BibitemShut {NoStop}%
\bibitem [{\citenamefont {Midtgaard}\ \emph {et~al.}(2017)\citenamefont
  {Midtgaard}, \citenamefont {Wu},\ and\ \citenamefont
  {Bruun}}]{Midtgaard2017}%
  \BibitemOpen
  \bibfield  {author} {\bibinfo {author} {\bibfnamefont {Jonatan~Melk\ae{}r}\
  \bibnamefont {Midtgaard}}, \bibinfo {author} {\bibfnamefont {Zhigang}\
  \bibnamefont {Wu}}, \ and\ \bibinfo {author} {\bibfnamefont {G.~M.}\
  \bibnamefont {Bruun}},\ }\bibfield  {title} {\enquote {\bibinfo {title}
  {{Time-reversal-invariant topological superfluids in Bose-Fermi mixtures}},}\
  }\href {\doibase 10.1103/PhysRevA.96.033605} {\bibfield  {journal} {\bibinfo
  {journal} {Phys. Rev. A}\ }\textbf {\bibinfo {volume} {96}},\ \bibinfo
  {pages} {033605} (\bibinfo {year} {2017})}\BibitemShut {NoStop}%
\bibitem [{\citenamefont {Huang}\ and\ \citenamefont
  {Chiu}(2018)}]{Huang2018helical}%
  \BibitemOpen
  \bibfield  {author} {\bibinfo {author} {\bibfnamefont {Yingyi}\ \bibnamefont
  {Huang}}\ and\ \bibinfo {author} {\bibfnamefont {Ching-Kai}\ \bibnamefont
  {Chiu}},\ }\bibfield  {title} {\enquote {\bibinfo {title} {Helical majorana
  edge mode in a superconducting antiferromagnetic quantum spin hall
  insulator},}\ }\href {\doibase 10.1103/PhysRevB.98.081412} {\bibfield
  {journal} {\bibinfo  {journal} {Phys. Rev. B}\ }\textbf {\bibinfo {volume}
  {98}},\ \bibinfo {pages} {081412} (\bibinfo {year} {2018})}\BibitemShut
  {NoStop}%
\bibitem [{\citenamefont {Zhang}\ and\ \citenamefont
  {Das~Sarma}(2021)}]{Zhang2021TSC}%
  \BibitemOpen
  \bibfield  {author} {\bibinfo {author} {\bibfnamefont {Rui-Xing}\
  \bibnamefont {Zhang}}\ and\ \bibinfo {author} {\bibfnamefont
  {S.}~\bibnamefont {Das~Sarma}},\ }\bibfield  {title} {\enquote {\bibinfo
  {title} {Intrinsic time-reversal-invariant topological superconductivity in
  thin films of iron-based superconductors},}\ }\href {\doibase
  10.1103/PhysRevLett.126.137001} {\bibfield  {journal} {\bibinfo  {journal}
  {Phys. Rev. Lett.}\ }\textbf {\bibinfo {volume} {126}},\ \bibinfo {pages}
  {137001} (\bibinfo {year} {2021})}\BibitemShut {NoStop}%
\bibitem [{\citenamefont {Feng}\ \emph
  {et~al.}(2022{\natexlab{b}})\citenamefont {Feng}, \citenamefont {Zhang},\
  and\ \citenamefont {Yan}}]{Feng2022TSC}%
  \BibitemOpen
  \bibfield  {author} {\bibinfo {author} {\bibfnamefont {Guan-Hao}\
  \bibnamefont {Feng}}, \bibinfo {author} {\bibfnamefont {Hong-Hao}\
  \bibnamefont {Zhang}}, \ and\ \bibinfo {author} {\bibfnamefont {Zhongbo}\
  \bibnamefont {Yan}},\ }\bibfield  {title} {\enquote {\bibinfo {title}
  {Time-reversal invariant topological gapped phases in bilayer dirac
  materials},}\ }\href {\doibase 10.1103/PhysRevB.106.064509} {\bibfield
  {journal} {\bibinfo  {journal} {Phys. Rev. B}\ }\textbf {\bibinfo {volume}
  {106}},\ \bibinfo {pages} {064509} (\bibinfo {year}
  {2022}{\natexlab{b}})}\BibitemShut {NoStop}%
\bibitem [{\citenamefont {Langbehn}\ \emph {et~al.}(2017)\citenamefont
  {Langbehn}, \citenamefont {Peng}, \citenamefont {Trifunovic}, \citenamefont
  {von Oppen},\ and\ \citenamefont {Brouwer}}]{Langbehn2017}%
  \BibitemOpen
  \bibfield  {author} {\bibinfo {author} {\bibfnamefont {Josias}\ \bibnamefont
  {Langbehn}}, \bibinfo {author} {\bibfnamefont {Yang}\ \bibnamefont {Peng}},
  \bibinfo {author} {\bibfnamefont {Luka}\ \bibnamefont {Trifunovic}}, \bibinfo
  {author} {\bibfnamefont {Felix}\ \bibnamefont {von Oppen}}, \ and\ \bibinfo
  {author} {\bibfnamefont {Piet~W.}\ \bibnamefont {Brouwer}},\ }\bibfield
  {title} {\enquote {\bibinfo {title} {Reflection-symmetric second-order
  topological insulators and superconductors},}\ }\href {\doibase
  10.1103/PhysRevLett.119.246401} {\bibfield  {journal} {\bibinfo  {journal}
  {Phys. Rev. Lett.}\ }\textbf {\bibinfo {volume} {119}},\ \bibinfo {pages}
  {246401} (\bibinfo {year} {2017})}\BibitemShut {NoStop}%
\bibitem [{\citenamefont {Geier}\ \emph {et~al.}(2018)\citenamefont {Geier},
  \citenamefont {Trifunovic}, \citenamefont {Hoskam},\ and\ \citenamefont
  {Brouwer}}]{Geier2018}%
  \BibitemOpen
  \bibfield  {author} {\bibinfo {author} {\bibfnamefont {Max}\ \bibnamefont
  {Geier}}, \bibinfo {author} {\bibfnamefont {Luka}\ \bibnamefont
  {Trifunovic}}, \bibinfo {author} {\bibfnamefont {Max}\ \bibnamefont
  {Hoskam}}, \ and\ \bibinfo {author} {\bibfnamefont {Piet~W.}\ \bibnamefont
  {Brouwer}},\ }\bibfield  {title} {\enquote {\bibinfo {title} {Second-order
  topological insulators and superconductors with an order-two crystalline
  symmetry},}\ }\href {\doibase 10.1103/PhysRevB.97.205135} {\bibfield
  {journal} {\bibinfo  {journal} {Phys. Rev. B}\ }\textbf {\bibinfo {volume}
  {97}},\ \bibinfo {pages} {205135} (\bibinfo {year} {2018})}\BibitemShut
  {NoStop}%
\bibitem [{\citenamefont {Khalaf}(2018)}]{Khalaf2018}%
  \BibitemOpen
  \bibfield  {author} {\bibinfo {author} {\bibfnamefont {Eslam}\ \bibnamefont
  {Khalaf}},\ }\bibfield  {title} {\enquote {\bibinfo {title} {Higher-order
  topological insulators and superconductors protected by inversion
  symmetry},}\ }\href {\doibase 10.1103/PhysRevB.97.205136} {\bibfield
  {journal} {\bibinfo  {journal} {Phys. Rev. B}\ }\textbf {\bibinfo {volume}
  {97}},\ \bibinfo {pages} {205136} (\bibinfo {year} {2018})}\BibitemShut
  {NoStop}%
\bibitem [{\citenamefont {Zhu}(2018)}]{Zhu2018hosc}%
  \BibitemOpen
  \bibfield  {author} {\bibinfo {author} {\bibfnamefont {Xiaoyu}\ \bibnamefont
  {Zhu}},\ }\bibfield  {title} {\enquote {\bibinfo {title} {{Tunable Majorana
  corner states in a two-dimensional second-order topological superconductor
  induced by magnetic fields}},}\ }\href {\doibase 10.1103/PhysRevB.97.205134}
  {\bibfield  {journal} {\bibinfo  {journal} {Phys. Rev. B}\ }\textbf {\bibinfo
  {volume} {97}},\ \bibinfo {pages} {205134} (\bibinfo {year}
  {2018})}\BibitemShut {NoStop}%
\bibitem [{\citenamefont {Yan}\ \emph {et~al.}(2018)\citenamefont {Yan},
  \citenamefont {Song},\ and\ \citenamefont {Wang}}]{Yan2018hosc}%
  \BibitemOpen
  \bibfield  {author} {\bibinfo {author} {\bibfnamefont {Zhongbo}\ \bibnamefont
  {Yan}}, \bibinfo {author} {\bibfnamefont {Fei}\ \bibnamefont {Song}}, \ and\
  \bibinfo {author} {\bibfnamefont {Zhong}\ \bibnamefont {Wang}},\ }\bibfield
  {title} {\enquote {\bibinfo {title} {{Majorana Corner Modes in a
  High-Temperature Platform}},}\ }\href {\doibase
  10.1103/PhysRevLett.121.096803} {\bibfield  {journal} {\bibinfo  {journal}
  {Phys. Rev. Lett.}\ }\textbf {\bibinfo {volume} {121}},\ \bibinfo {pages}
  {096803} (\bibinfo {year} {2018})}\BibitemShut {NoStop}%
\bibitem [{\citenamefont {Wang}\ \emph
  {et~al.}(2018{\natexlab{a}})\citenamefont {Wang}, \citenamefont {Lin},\ and\
  \citenamefont {Hughes}}]{Wang2018weak}%
  \BibitemOpen
  \bibfield  {author} {\bibinfo {author} {\bibfnamefont {Yuxuan}\ \bibnamefont
  {Wang}}, \bibinfo {author} {\bibfnamefont {Mao}\ \bibnamefont {Lin}}, \ and\
  \bibinfo {author} {\bibfnamefont {Taylor~L.}\ \bibnamefont {Hughes}},\
  }\bibfield  {title} {\enquote {\bibinfo {title} {Weak-pairing higher order
  topological superconductors},}\ }\href {\doibase 10.1103/PhysRevB.98.165144}
  {\bibfield  {journal} {\bibinfo  {journal} {Phys. Rev. B}\ }\textbf {\bibinfo
  {volume} {98}},\ \bibinfo {pages} {165144} (\bibinfo {year}
  {2018}{\natexlab{a}})}\BibitemShut {NoStop}%
\bibitem [{\citenamefont {Wang}\ \emph
  {et~al.}(2018{\natexlab{b}})\citenamefont {Wang}, \citenamefont {Liu},
  \citenamefont {Lu},\ and\ \citenamefont {Zhang}}]{Wang2018hosc}%
  \BibitemOpen
  \bibfield  {author} {\bibinfo {author} {\bibfnamefont {Qiyue}\ \bibnamefont
  {Wang}}, \bibinfo {author} {\bibfnamefont {Cheng-Cheng}\ \bibnamefont {Liu}},
  \bibinfo {author} {\bibfnamefont {Yuan-Ming}\ \bibnamefont {Lu}}, \ and\
  \bibinfo {author} {\bibfnamefont {Fan}\ \bibnamefont {Zhang}},\ }\bibfield
  {title} {\enquote {\bibinfo {title} {{High-Temperature Majorana Corner
  States}},}\ }\href {\doibase 10.1103/PhysRevLett.121.186801} {\bibfield
  {journal} {\bibinfo  {journal} {Phys. Rev. Lett.}\ }\textbf {\bibinfo
  {volume} {121}},\ \bibinfo {pages} {186801} (\bibinfo {year}
  {2018}{\natexlab{b}})}\BibitemShut {NoStop}%
\bibitem [{\citenamefont {Liu}\ \emph {et~al.}(2018)\citenamefont {Liu},
  \citenamefont {He},\ and\ \citenamefont {Nori}}]{Liu2018hosc}%
  \BibitemOpen
  \bibfield  {author} {\bibinfo {author} {\bibfnamefont {Tao}\ \bibnamefont
  {Liu}}, \bibinfo {author} {\bibfnamefont {James~Jun}\ \bibnamefont {He}}, \
  and\ \bibinfo {author} {\bibfnamefont {Franco}\ \bibnamefont {Nori}},\
  }\bibfield  {title} {\enquote {\bibinfo {title} {Majorana corner states in a
  two-dimensional magnetic topological insulator on a high-temperature
  superconductor},}\ }\href {\doibase 10.1103/PhysRevB.98.245413} {\bibfield
  {journal} {\bibinfo  {journal} {Phys. Rev. B}\ }\textbf {\bibinfo {volume}
  {98}},\ \bibinfo {pages} {245413} (\bibinfo {year} {2018})}\BibitemShut
  {NoStop}%
\bibitem [{\citenamefont {Wu}\ \emph {et~al.}(2019)\citenamefont {Wu},
  \citenamefont {Yan},\ and\ \citenamefont {Huang}}]{Wu2019hosc}%
  \BibitemOpen
  \bibfield  {author} {\bibinfo {author} {\bibfnamefont {Zhigang}\ \bibnamefont
  {Wu}}, \bibinfo {author} {\bibfnamefont {Zhongbo}\ \bibnamefont {Yan}}, \
  and\ \bibinfo {author} {\bibfnamefont {Wen}\ \bibnamefont {Huang}},\
  }\bibfield  {title} {\enquote {\bibinfo {title} {{Higher-order topological
  superconductivity: Possible realization in Fermi gases and
  Sr$_{2}$RuO$_{4}$}},}\ }\href {\doibase 10.1103/PhysRevB.99.020508}
  {\bibfield  {journal} {\bibinfo  {journal} {Phys. Rev. B}\ }\textbf {\bibinfo
  {volume} {99}},\ \bibinfo {pages} {020508} (\bibinfo {year}
  {2019})}\BibitemShut {NoStop}%
\bibitem [{\citenamefont {Yan}(2019)}]{Yan2019hosca}%
  \BibitemOpen
  \bibfield  {author} {\bibinfo {author} {\bibfnamefont {Zhongbo}\ \bibnamefont
  {Yan}},\ }\bibfield  {title} {\enquote {\bibinfo {title} {{Higher-Order
  Topological Odd-Parity Superconductors}},}\ }\href {\doibase
  10.1103/PhysRevLett.123.177001} {\bibfield  {journal} {\bibinfo  {journal}
  {Phys. Rev. Lett.}\ }\textbf {\bibinfo {volume} {123}},\ \bibinfo {pages}
  {177001} (\bibinfo {year} {2019})}\BibitemShut {NoStop}%
\bibitem [{\citenamefont {Volpez}\ \emph {et~al.}(2019)\citenamefont {Volpez},
  \citenamefont {Loss},\ and\ \citenamefont {Klinovaja}}]{Volpez2019SOTSC}%
  \BibitemOpen
  \bibfield  {author} {\bibinfo {author} {\bibfnamefont {Yanick}\ \bibnamefont
  {Volpez}}, \bibinfo {author} {\bibfnamefont {Daniel}\ \bibnamefont {Loss}}, \
  and\ \bibinfo {author} {\bibfnamefont {Jelena}\ \bibnamefont {Klinovaja}},\
  }\bibfield  {title} {\enquote {\bibinfo {title} {{Second-Order Topological
  Superconductivity in $\ensuremath{\pi}$-Junction {Rashba} Layers}},}\ }\href
  {\doibase 10.1103/PhysRevLett.122.126402} {\bibfield  {journal} {\bibinfo
  {journal} {Phys. Rev. Lett.}\ }\textbf {\bibinfo {volume} {122}},\ \bibinfo
  {pages} {126402} (\bibinfo {year} {2019})}\BibitemShut {NoStop}%
\bibitem [{\citenamefont {Zhang}\ \emph {et~al.}(2019)\citenamefont {Zhang},
  \citenamefont {Cole}, \citenamefont {Wu},\ and\ \citenamefont
  {Das~Sarma}}]{Zhang2019hoscb}%
  \BibitemOpen
  \bibfield  {author} {\bibinfo {author} {\bibfnamefont {Rui-Xing}\
  \bibnamefont {Zhang}}, \bibinfo {author} {\bibfnamefont {William~S.}\
  \bibnamefont {Cole}}, \bibinfo {author} {\bibfnamefont {Xianxin}\
  \bibnamefont {Wu}}, \ and\ \bibinfo {author} {\bibfnamefont {S.}~\bibnamefont
  {Das~Sarma}},\ }\bibfield  {title} {\enquote {\bibinfo {title} {{Higher-Order
  Topology and Nodal Topological Superconductivity in {Fe(Se,Te)}
  Heterostructures}},}\ }\href {\doibase 10.1103/PhysRevLett.123.167001}
  {\bibfield  {journal} {\bibinfo  {journal} {Phys. Rev. Lett.}\ }\textbf
  {\bibinfo {volume} {123}},\ \bibinfo {pages} {167001} (\bibinfo {year}
  {2019})}\BibitemShut {NoStop}%
\bibitem [{\citenamefont {Pan}\ \emph {et~al.}(2019)\citenamefont {Pan},
  \citenamefont {Yang}, \citenamefont {Chen}, \citenamefont {Xu}, \citenamefont
  {Liu},\ and\ \citenamefont {Liu}}]{Pan2019SOTSC}%
  \BibitemOpen
  \bibfield  {author} {\bibinfo {author} {\bibfnamefont {Xiao-Hong}\
  \bibnamefont {Pan}}, \bibinfo {author} {\bibfnamefont {Kai-Jie}\ \bibnamefont
  {Yang}}, \bibinfo {author} {\bibfnamefont {Li}~\bibnamefont {Chen}}, \bibinfo
  {author} {\bibfnamefont {Gang}\ \bibnamefont {Xu}}, \bibinfo {author}
  {\bibfnamefont {Chao-Xing}\ \bibnamefont {Liu}}, \ and\ \bibinfo {author}
  {\bibfnamefont {Xin}\ \bibnamefont {Liu}},\ }\bibfield  {title} {\enquote
  {\bibinfo {title} {{Lattice-Symmetry-Assisted Second-Order Topological
  Superconductors and {Majorana} Patterns}},}\ }\href {\doibase
  10.1103/PhysRevLett.123.156801} {\bibfield  {journal} {\bibinfo  {journal}
  {Phys. Rev. Lett.}\ }\textbf {\bibinfo {volume} {123}},\ \bibinfo {pages}
  {156801} (\bibinfo {year} {2019})}\BibitemShut {NoStop}%
\bibitem [{\citenamefont {Zhu}(2019)}]{Zhu2019mixed}%
  \BibitemOpen
  \bibfield  {author} {\bibinfo {author} {\bibfnamefont {Xiaoyu}\ \bibnamefont
  {Zhu}},\ }\bibfield  {title} {\enquote {\bibinfo {title} {{Second-Order
  Topological Superconductors with Mixed Pairing}},}\ }\href {\doibase
  10.1103/PhysRevLett.122.236401} {\bibfield  {journal} {\bibinfo  {journal}
  {Phys. Rev. Lett.}\ }\textbf {\bibinfo {volume} {122}},\ \bibinfo {pages}
  {236401} (\bibinfo {year} {2019})}\BibitemShut {NoStop}%
\bibitem [{\citenamefont {Hsu}\ \emph {et~al.}(2020)\citenamefont {Hsu},
  \citenamefont {Cole}, \citenamefont {Zhang},\ and\ \citenamefont
  {Sau}}]{Hsu2020hosc}%
  \BibitemOpen
  \bibfield  {author} {\bibinfo {author} {\bibfnamefont {Yi-Ting}\ \bibnamefont
  {Hsu}}, \bibinfo {author} {\bibfnamefont {William~S.}\ \bibnamefont {Cole}},
  \bibinfo {author} {\bibfnamefont {Rui-Xing}\ \bibnamefont {Zhang}}, \ and\
  \bibinfo {author} {\bibfnamefont {Jay~D.}\ \bibnamefont {Sau}},\ }\bibfield
  {title} {\enquote {\bibinfo {title} {{Inversion-Protected Higher-Order
  Topological Superconductivity in Monolayer {WTe$_{2}$}}},}\ }\href {\doibase
  10.1103/PhysRevLett.125.097001} {\bibfield  {journal} {\bibinfo  {journal}
  {Phys. Rev. Lett.}\ }\textbf {\bibinfo {volume} {125}},\ \bibinfo {pages}
  {097001} (\bibinfo {year} {2020})}\BibitemShut {NoStop}%
\bibitem [{\citenamefont {Wu}\ \emph {et~al.}(2020{\natexlab{a}})\citenamefont
  {Wu}, \citenamefont {Hou}, \citenamefont {Li}, \citenamefont {Luo},
  \citenamefont {Shi},\ and\ \citenamefont {Zhang}}]{Wu2020SOTSC}%
  \BibitemOpen
  \bibfield  {author} {\bibinfo {author} {\bibfnamefont {Ya-Jie}\ \bibnamefont
  {Wu}}, \bibinfo {author} {\bibfnamefont {Junpeng}\ \bibnamefont {Hou}},
  \bibinfo {author} {\bibfnamefont {Yun-Mei}\ \bibnamefont {Li}}, \bibinfo
  {author} {\bibfnamefont {Xi-Wang}\ \bibnamefont {Luo}}, \bibinfo {author}
  {\bibfnamefont {Xiaoyan}\ \bibnamefont {Shi}}, \ and\ \bibinfo {author}
  {\bibfnamefont {Chuanwei}\ \bibnamefont {Zhang}},\ }\bibfield  {title}
  {\enquote {\bibinfo {title} {{In-Plane Zeeman-Field-Induced Majorana Corner
  and Hinge Modes in an $s$-Wave Superconductor Heterostructure}},}\ }\href
  {\doibase 10.1103/PhysRevLett.124.227001} {\bibfield  {journal} {\bibinfo
  {journal} {Phys. Rev. Lett.}\ }\textbf {\bibinfo {volume} {124}},\ \bibinfo
  {pages} {227001} (\bibinfo {year} {2020}{\natexlab{a}})}\BibitemShut
  {NoStop}%
\bibitem [{\citenamefont {Kheirkhah}\ \emph {et~al.}(2020)\citenamefont
  {Kheirkhah}, \citenamefont {Yan}, \citenamefont {Nagai},\ and\ \citenamefont
  {Marsiglio}}]{Majid2020hoscb}%
  \BibitemOpen
  \bibfield  {author} {\bibinfo {author} {\bibfnamefont {Majid}\ \bibnamefont
  {Kheirkhah}}, \bibinfo {author} {\bibfnamefont {Zhongbo}\ \bibnamefont
  {Yan}}, \bibinfo {author} {\bibfnamefont {Yuki}\ \bibnamefont {Nagai}}, \
  and\ \bibinfo {author} {\bibfnamefont {Frank}\ \bibnamefont {Marsiglio}},\
  }\bibfield  {title} {\enquote {\bibinfo {title} {{First- and Second-Order
  Topological Superconductivity and Temperature-Driven Topological Phase
  Transitions in the Extended {Hubbard} Model with Spin-Orbit Coupling}},}\
  }\href {\doibase 10.1103/PhysRevLett.125.017001} {\bibfield  {journal}
  {\bibinfo  {journal} {Phys. Rev. Lett.}\ }\textbf {\bibinfo {volume} {125}},\
  \bibinfo {pages} {017001} (\bibinfo {year} {2020})}\BibitemShut {NoStop}%
\bibitem [{\citenamefont {Wu}\ \emph {et~al.}(2020{\natexlab{b}})\citenamefont
  {Wu}, \citenamefont {Benalcazar}, \citenamefont {Li}, \citenamefont
  {Thomale}, \citenamefont {Liu},\ and\ \citenamefont
  {Hu}}]{wu2020boundaryobstructedb}%
  \BibitemOpen
  \bibfield  {author} {\bibinfo {author} {\bibfnamefont {Xianxin}\ \bibnamefont
  {Wu}}, \bibinfo {author} {\bibfnamefont {Wladimir~A.}\ \bibnamefont
  {Benalcazar}}, \bibinfo {author} {\bibfnamefont {Yinxiang}\ \bibnamefont
  {Li}}, \bibinfo {author} {\bibfnamefont {Ronny}\ \bibnamefont {Thomale}},
  \bibinfo {author} {\bibfnamefont {Chao-Xing}\ \bibnamefont {Liu}}, \ and\
  \bibinfo {author} {\bibfnamefont {Jiangping}\ \bibnamefont {Hu}},\ }\bibfield
   {title} {\enquote {\bibinfo {title} {{Boundary-Obstructed Topological
  High-{T}$_{c}$ Superconductivity in Iron Pnictides}},}\ }\href {\doibase
  10.1103/PhysRevX.10.041014} {\bibfield  {journal} {\bibinfo  {journal} {Phys.
  Rev. X}\ }\textbf {\bibinfo {volume} {10}},\ \bibinfo {pages} {041014}
  (\bibinfo {year} {2020}{\natexlab{b}})}\BibitemShut {NoStop}%
\bibitem [{\citenamefont {Qin}\ \emph {et~al.}(2022)\citenamefont {Qin},
  \citenamefont {Fang}, \citenamefont {Zhang},\ and\ \citenamefont
  {Hu}}]{Qin2022hosc}%
  \BibitemOpen
  \bibfield  {author} {\bibinfo {author} {\bibfnamefont {Shengshan}\
  \bibnamefont {Qin}}, \bibinfo {author} {\bibfnamefont {Chen}\ \bibnamefont
  {Fang}}, \bibinfo {author} {\bibfnamefont {Fu-Chun}\ \bibnamefont {Zhang}}, \
  and\ \bibinfo {author} {\bibfnamefont {Jiangping}\ \bibnamefont {Hu}},\
  }\bibfield  {title} {\enquote {\bibinfo {title} {{Topological
  Superconductivity in an Extended $s$-Wave Superconductor and Its Implication
  to Iron-Based Superconductors}},}\ }\href {\doibase
  10.1103/PhysRevX.12.011030} {\bibfield  {journal} {\bibinfo  {journal} {Phys.
  Rev. X}\ }\textbf {\bibinfo {volume} {12}},\ \bibinfo {pages} {011030}
  (\bibinfo {year} {2022})}\BibitemShut {NoStop}%
\bibitem [{\citenamefont {Zhu}\ \emph {et~al.}(2022)\citenamefont {Zhu},
  \citenamefont {Li},\ and\ \citenamefont {Yan}}]{Zhu2022sublattice}%
  \BibitemOpen
  \bibfield  {author} {\bibinfo {author} {\bibfnamefont {Di}~\bibnamefont
  {Zhu}}, \bibinfo {author} {\bibfnamefont {Bo-Xuan}\ \bibnamefont {Li}}, \
  and\ \bibinfo {author} {\bibfnamefont {Zhongbo}\ \bibnamefont {Yan}},\
  }\bibfield  {title} {\enquote {\bibinfo {title} {{Sublattice-sensitive
  Majorana modes}},}\ }\href {\doibase 10.1103/PhysRevB.106.245418} {\bibfield
  {journal} {\bibinfo  {journal} {Phys. Rev. B}\ }\textbf {\bibinfo {volume}
  {106}},\ \bibinfo {pages} {245418} (\bibinfo {year} {2022})}\BibitemShut
  {NoStop}%
\bibitem [{\citenamefont {Li}\ \emph {et~al.}(2021)\citenamefont {Li},
  \citenamefont {Geier}, \citenamefont {Ingham},\ and\ \citenamefont
  {Scammell}}]{Li2022hosc}%
  \BibitemOpen
  \bibfield  {author} {\bibinfo {author} {\bibfnamefont {Tommy}\ \bibnamefont
  {Li}}, \bibinfo {author} {\bibfnamefont {Max}\ \bibnamefont {Geier}},
  \bibinfo {author} {\bibfnamefont {Julian}\ \bibnamefont {Ingham}}, \ and\
  \bibinfo {author} {\bibfnamefont {Harley~D}\ \bibnamefont {Scammell}},\
  }\bibfield  {title} {\enquote {\bibinfo {title} {Higher-order topological
  superconductivity from repulsive interactions in kagome and honeycomb
  systems},}\ }\href {\doibase 10.1088/2053-1583/ac4060} {\bibfield  {journal}
  {\bibinfo  {journal} {2D Materials}\ }\textbf {\bibinfo {volume} {9}},\
  \bibinfo {pages} {015031} (\bibinfo {year} {2021})}\BibitemShut {NoStop}%
\bibitem [{\citenamefont {Scammell}\ \emph {et~al.}(2022)\citenamefont
  {Scammell}, \citenamefont {Ingham}, \citenamefont {Geier},\ and\
  \citenamefont {Li}}]{Scammell2022hosc}%
  \BibitemOpen
  \bibfield  {author} {\bibinfo {author} {\bibfnamefont {Harley~D.}\
  \bibnamefont {Scammell}}, \bibinfo {author} {\bibfnamefont {Julian}\
  \bibnamefont {Ingham}}, \bibinfo {author} {\bibfnamefont {Max}\ \bibnamefont
  {Geier}}, \ and\ \bibinfo {author} {\bibfnamefont {Tommy}\ \bibnamefont
  {Li}},\ }\bibfield  {title} {\enquote {\bibinfo {title} {Intrinsic first- and
  higher-order topological superconductivity in a doped topological
  insulator},}\ }\href {\doibase 10.1103/PhysRevB.105.195149} {\bibfield
  {journal} {\bibinfo  {journal} {Phys. Rev. B}\ }\textbf {\bibinfo {volume}
  {105}},\ \bibinfo {pages} {195149} (\bibinfo {year} {2022})}\BibitemShut
  {NoStop}%
\bibitem [{\citenamefont {Tsuei}\ and\ \citenamefont
  {Kirtley}(2000)}]{Tsuei2000}%
  \BibitemOpen
  \bibfield  {author} {\bibinfo {author} {\bibfnamefont {C.~C.}\ \bibnamefont
  {Tsuei}}\ and\ \bibinfo {author} {\bibfnamefont {J.~R.}\ \bibnamefont
  {Kirtley}},\ }\bibfield  {title} {\enquote {\bibinfo {title} {{Pairing
  symmetry in cuprate superconductors}},}\ }\href {\doibase
  10.1103/RevModPhys.72.969} {\bibfield  {journal} {\bibinfo  {journal} {Rev.
  Mod. Phys.}\ }\textbf {\bibinfo {volume} {72}},\ \bibinfo {pages} {969--1016}
  (\bibinfo {year} {2000})}\BibitemShut {NoStop}%
\bibitem [{\citenamefont {Qi}\ \emph {et~al.}(2006)\citenamefont {Qi},
  \citenamefont {Wu},\ and\ \citenamefont {Zhang}}]{qi2006QWZ}%
  \BibitemOpen
  \bibfield  {author} {\bibinfo {author} {\bibfnamefont {Xiao-Liang}\
  \bibnamefont {Qi}}, \bibinfo {author} {\bibfnamefont {Yong-Shi}\ \bibnamefont
  {Wu}}, \ and\ \bibinfo {author} {\bibfnamefont {Shou-Cheng}\ \bibnamefont
  {Zhang}},\ }\bibfield  {title} {\enquote {\bibinfo {title} {{Topological
  quantization of the spin Hall effect in two-dimensional paramagnetic
  semiconductors}},}\ }\href {\doibase 10.1103/PhysRevB.74.085308} {\bibfield
  {journal} {\bibinfo  {journal} {Phys. Rev. B}\ }\textbf {\bibinfo {volume}
  {74}},\ \bibinfo {pages} {085308} (\bibinfo {year} {2006})}\BibitemShut
  {NoStop}%
\bibitem [{\citenamefont {Monthoux}\ \emph {et~al.}(2007)\citenamefont
  {Monthoux}, \citenamefont {Pines},\ and\ \citenamefont
  {Lonzarich}}]{Monthoux2007}%
  \BibitemOpen
  \bibfield  {author} {\bibinfo {author} {\bibfnamefont {P.}~\bibnamefont
  {Monthoux}}, \bibinfo {author} {\bibfnamefont {D.}~\bibnamefont {Pines}}, \
  and\ \bibinfo {author} {\bibfnamefont {G.~G.}\ \bibnamefont {Lonzarich}},\
  }\bibfield  {title} {\enquote {\bibinfo {title} {Superconductivity without
  phonons},}\ }\href {\doibase 10.1038/nature06480} {\bibfield  {journal}
  {\bibinfo  {journal} {Nature}\ }\textbf {\bibinfo {volume} {450}},\ \bibinfo
  {pages} {1177--1183} (\bibinfo {year} {2007})}\BibitemShut {NoStop}%
\bibitem [{\citenamefont {Gor'kov}\ and\ \citenamefont
  {Rashba}(2001)}]{Gorkov2001}%
  \BibitemOpen
  \bibfield  {author} {\bibinfo {author} {\bibfnamefont {Lev~P.}\ \bibnamefont
  {Gor'kov}}\ and\ \bibinfo {author} {\bibfnamefont {Emmanuel~I.}\ \bibnamefont
  {Rashba}},\ }\bibfield  {title} {\enquote {\bibinfo {title} {Superconducting
  2d system with lifted spin degeneracy: Mixed singlet-triplet state},}\ }\href
  {\doibase 10.1103/PhysRevLett.87.037004} {\bibfield  {journal} {\bibinfo
  {journal} {Phys. Rev. Lett.}\ }\textbf {\bibinfo {volume} {87}},\ \bibinfo
  {pages} {037004} (\bibinfo {year} {2001})}\BibitemShut {NoStop}%
\bibitem [{\citenamefont {Sun}\ \emph {et~al.}(2018)\citenamefont {Sun},
  \citenamefont {Wang}, \citenamefont {Xu}, \citenamefont {Yi}, \citenamefont
  {Zhang}, \citenamefont {Wu}, \citenamefont {Deng}, \citenamefont {Liu},
  \citenamefont {Chen},\ and\ \citenamefont {Pan}}]{Sun2018}%
  \BibitemOpen
  \bibfield  {author} {\bibinfo {author} {\bibfnamefont {Wei}\ \bibnamefont
  {Sun}}, \bibinfo {author} {\bibfnamefont {Bao-Zong}\ \bibnamefont {Wang}},
  \bibinfo {author} {\bibfnamefont {Xiao-Tian}\ \bibnamefont {Xu}}, \bibinfo
  {author} {\bibfnamefont {Chang-Rui}\ \bibnamefont {Yi}}, \bibinfo {author}
  {\bibfnamefont {Long}\ \bibnamefont {Zhang}}, \bibinfo {author}
  {\bibfnamefont {Zhan}\ \bibnamefont {Wu}}, \bibinfo {author} {\bibfnamefont
  {Youjin}\ \bibnamefont {Deng}}, \bibinfo {author} {\bibfnamefont {Xiong-Jun}\
  \bibnamefont {Liu}}, \bibinfo {author} {\bibfnamefont {Shuai}\ \bibnamefont
  {Chen}}, \ and\ \bibinfo {author} {\bibfnamefont {Jian-Wei}\ \bibnamefont
  {Pan}},\ }\bibfield  {title} {\enquote {\bibinfo {title} {{Highly
  Controllable and Robust 2D Spin-Orbit Coupling for Quantum Gases}},}\ }\href
  {\doibase 10.1103/PhysRevLett.121.150401} {\bibfield  {journal} {\bibinfo
  {journal} {Phys. Rev. Lett.}\ }\textbf {\bibinfo {volume} {121}},\ \bibinfo
  {pages} {150401} (\bibinfo {year} {2018})}\BibitemShut {NoStop}%
\bibitem [{\citenamefont {Chen}\ \emph {et~al.}(2023)\citenamefont {Chen},
  \citenamefont {Huang},\ and\ \citenamefont {Wu}}]{Chench2023}%
  \BibitemOpen
  \bibfield  {author} {\bibinfo {author} {\bibfnamefont {Canhao}\ \bibnamefont
  {Chen}}, \bibinfo {author} {\bibfnamefont {Guan-Hua}\ \bibnamefont {Huang}},
  \ and\ \bibinfo {author} {\bibfnamefont {Zhigang}\ \bibnamefont {Wu}},\
  }\bibfield  {title} {\enquote {\bibinfo {title} {{Intrinsic anomalous Hall
  effect across the magnetic phase transition of a spin-orbit-coupled
  Bose-Einstein condensate}},}\ }\href {\doibase
  10.1103/PhysRevResearch.5.023070} {\bibfield  {journal} {\bibinfo  {journal}
  {Phys. Rev. Res.}\ }\textbf {\bibinfo {volume} {5}},\ \bibinfo {pages}
  {023070} (\bibinfo {year} {2023})}\BibitemShut {NoStop}%
\bibitem [{\citenamefont {Sato}(2010)}]{sato2010odd}%
  \BibitemOpen
  \bibfield  {author} {\bibinfo {author} {\bibfnamefont {Masatoshi}\
  \bibnamefont {Sato}},\ }\bibfield  {title} {\enquote {\bibinfo {title}
  {Topological odd-parity superconductors},}\ }\href {\doibase
  10.1103/PhysRevB.81.220504} {\bibfield  {journal} {\bibinfo  {journal} {Phys.
  Rev. B}\ }\textbf {\bibinfo {volume} {81}},\ \bibinfo {pages} {220504}
  (\bibinfo {year} {2010})}\BibitemShut {NoStop}%
\bibitem [{\citenamefont {Teo}\ \emph {et~al.}(2008)\citenamefont {Teo},
  \citenamefont {Fu},\ and\ \citenamefont {Kane}}]{teo2008}%
  \BibitemOpen
  \bibfield  {author} {\bibinfo {author} {\bibfnamefont {Jeffrey C.~Y.}\
  \bibnamefont {Teo}}, \bibinfo {author} {\bibfnamefont {Liang}\ \bibnamefont
  {Fu}}, \ and\ \bibinfo {author} {\bibfnamefont {C.~L.}\ \bibnamefont
  {Kane}},\ }\bibfield  {title} {\enquote {\bibinfo {title} {{Surface states
  and topological invariants in three-dimensional topological insulators:
  Application to ${\text{Bi}}_{1\ensuremath{-}x}{\text{Sb}}_{x}$}},}\ }\href
  {\doibase 10.1103/PhysRevB.78.045426} {\bibfield  {journal} {\bibinfo
  {journal} {Phys. Rev. B}\ }\textbf {\bibinfo {volume} {78}},\ \bibinfo
  {pages} {045426} (\bibinfo {year} {2008})}\BibitemShut {NoStop}%
\bibitem [{\citenamefont {Zhang}\ \emph
  {et~al.}(2013{\natexlab{b}})\citenamefont {Zhang}, \citenamefont {Kane},\
  and\ \citenamefont {Mele}}]{Zhang2013mirror}%
  \BibitemOpen
  \bibfield  {author} {\bibinfo {author} {\bibfnamefont {Fan}\ \bibnamefont
  {Zhang}}, \bibinfo {author} {\bibfnamefont {C.~L.}\ \bibnamefont {Kane}}, \
  and\ \bibinfo {author} {\bibfnamefont {E.~J.}\ \bibnamefont {Mele}},\
  }\bibfield  {title} {\enquote {\bibinfo {title} {{Topological Mirror
  Superconductivity}},}\ }\href {\doibase 10.1103/PhysRevLett.111.056403}
  {\bibfield  {journal} {\bibinfo  {journal} {Phys. Rev. Lett.}\ }\textbf
  {\bibinfo {volume} {111}},\ \bibinfo {pages} {056403} (\bibinfo {year}
  {2013}{\natexlab{b}})}\BibitemShut {NoStop}%
\bibitem [{\citenamefont {Ryu}\ \emph {et~al.}(2010)\citenamefont {Ryu},
  \citenamefont {Schnyder}, \citenamefont {Furusaki},\ and\ \citenamefont
  {Ludwig}}]{Ryu2010}%
  \BibitemOpen
  \bibfield  {author} {\bibinfo {author} {\bibfnamefont {Shinsei}\ \bibnamefont
  {Ryu}}, \bibinfo {author} {\bibfnamefont {Andreas~P}\ \bibnamefont
  {Schnyder}}, \bibinfo {author} {\bibfnamefont {Akira}\ \bibnamefont
  {Furusaki}}, \ and\ \bibinfo {author} {\bibfnamefont {Andreas W~W}\
  \bibnamefont {Ludwig}},\ }\bibfield  {title} {\enquote {\bibinfo {title}
  {Topological insulators and superconductors: tenfold way and dimensional
  hierarchy},}\ }\href {\doibase 10.1088/1367-2630/12/6/065010} {\bibfield
  {journal} {\bibinfo  {journal} {New Journal of Physics}\ }\textbf {\bibinfo
  {volume} {12}},\ \bibinfo {pages} {065010} (\bibinfo {year}
  {2010})}\BibitemShut {NoStop}%
\bibitem [{\citenamefont {Zhu}\ \emph {et~al.}(2023)\citenamefont {Zhu},
  \citenamefont {Kheirkhah},\ and\ \citenamefont {Yan}}]{Zhu2023sublattice}%
  \BibitemOpen
  \bibfield  {author} {\bibinfo {author} {\bibfnamefont {Di}~\bibnamefont
  {Zhu}}, \bibinfo {author} {\bibfnamefont {Majid}\ \bibnamefont {Kheirkhah}},
  \ and\ \bibinfo {author} {\bibfnamefont {Zhongbo}\ \bibnamefont {Yan}},\
  }\bibfield  {title} {\enquote {\bibinfo {title} {Sublattice-enriched
  tunability of bound states in second-order topological insulators and
  superconductors},}\ }\href {\doibase 10.1103/PhysRevB.107.085407} {\bibfield
  {journal} {\bibinfo  {journal} {Phys. Rev. B}\ }\textbf {\bibinfo {volume}
  {107}},\ \bibinfo {pages} {085407} (\bibinfo {year} {2023})}\BibitemShut
  {NoStop}%
\bibitem [{\citenamefont {Jackiw}\ and\ \citenamefont
  {Rebbi}(1976)}]{jackiw1976b}%
  \BibitemOpen
  \bibfield  {author} {\bibinfo {author} {\bibfnamefont {R.}~\bibnamefont
  {Jackiw}}\ and\ \bibinfo {author} {\bibfnamefont {C.}~\bibnamefont {Rebbi}},\
  }\bibfield  {title} {\enquote {\bibinfo {title} {Solitons with fermion number
  $1/2$},}\ }\href {\doibase 10.1103/PhysRevD.13.3398} {\bibfield  {journal}
  {\bibinfo  {journal} {Phys. Rev. D}\ }\textbf {\bibinfo {volume} {13}},\
  \bibinfo {pages} {3398--3409} (\bibinfo {year} {1976})}\BibitemShut {NoStop}%
\bibitem [{\citenamefont {Schindler}\ \emph {et~al.}(2018)\citenamefont
  {Schindler}, \citenamefont {Cook}, \citenamefont {Vergniory}, \citenamefont
  {Wang}, \citenamefont {Parkin}, \citenamefont {Bernevig},\ and\ \citenamefont
  {Neupert}}]{Schindler2018}%
  \BibitemOpen
  \bibfield  {author} {\bibinfo {author} {\bibfnamefont {Frank}\ \bibnamefont
  {Schindler}}, \bibinfo {author} {\bibfnamefont {Ashley~M.}\ \bibnamefont
  {Cook}}, \bibinfo {author} {\bibfnamefont {Maia~G.}\ \bibnamefont
  {Vergniory}}, \bibinfo {author} {\bibfnamefont {Zhijun}\ \bibnamefont
  {Wang}}, \bibinfo {author} {\bibfnamefont {Stuart S.~P.}\ \bibnamefont
  {Parkin}}, \bibinfo {author} {\bibfnamefont {B.~Andrei}\ \bibnamefont
  {Bernevig}}, \ and\ \bibinfo {author} {\bibfnamefont {Titus}\ \bibnamefont
  {Neupert}},\ }\bibfield  {title} {\enquote {\bibinfo {title} {Higher-order
  topological insulators},}\ }\href {\doibase 10.1126/sciadv.aat0346}
  {\bibfield  {journal} {\bibinfo  {journal} {Science Advances}\ }\textbf
  {\bibinfo {volume} {4}},\ \bibinfo {pages} {eaat0346} (\bibinfo {year}
  {2018})}\BibitemShut {NoStop}%
\bibitem [{\citenamefont {Plekhanov}\ \emph {et~al.}(2021)\citenamefont
  {Plekhanov}, \citenamefont {M\"uller}, \citenamefont {Volpez}, \citenamefont
  {Kennes}, \citenamefont {Schoeller}, \citenamefont {Loss},\ and\
  \citenamefont {Klinovaja}}]{Plekhanov2021}%
  \BibitemOpen
  \bibfield  {author} {\bibinfo {author} {\bibfnamefont {Kirill}\ \bibnamefont
  {Plekhanov}}, \bibinfo {author} {\bibfnamefont {Niclas}\ \bibnamefont
  {M\"uller}}, \bibinfo {author} {\bibfnamefont {Yanick}\ \bibnamefont
  {Volpez}}, \bibinfo {author} {\bibfnamefont {Dante~M.}\ \bibnamefont
  {Kennes}}, \bibinfo {author} {\bibfnamefont {Herbert}\ \bibnamefont
  {Schoeller}}, \bibinfo {author} {\bibfnamefont {Daniel}\ \bibnamefont
  {Loss}}, \ and\ \bibinfo {author} {\bibfnamefont {Jelena}\ \bibnamefont
  {Klinovaja}},\ }\bibfield  {title} {\enquote {\bibinfo {title} {Quadrupole
  spin polarization as signature of second-order topological
  superconductors},}\ }\href {\doibase 10.1103/PhysRevB.103.L041401} {\bibfield
   {journal} {\bibinfo  {journal} {Phys. Rev. B}\ }\textbf {\bibinfo {volume}
  {103}},\ \bibinfo {pages} {L041401} (\bibinfo {year} {2021})}\BibitemShut
  {NoStop}%
\bibitem [{\citenamefont {He}\ \emph {et~al.}(2014)\citenamefont {He},
  \citenamefont {Ng}, \citenamefont {Lee},\ and\ \citenamefont
  {Law}}]{He2014spin}%
  \BibitemOpen
  \bibfield  {author} {\bibinfo {author} {\bibfnamefont {James~J.}\
  \bibnamefont {He}}, \bibinfo {author} {\bibfnamefont {T.~K.}\ \bibnamefont
  {Ng}}, \bibinfo {author} {\bibfnamefont {Patrick~A.}\ \bibnamefont {Lee}}, \
  and\ \bibinfo {author} {\bibfnamefont {K.~T.}\ \bibnamefont {Law}},\
  }\bibfield  {title} {\enquote {\bibinfo {title} {Selective equal-spin andreev
  reflections induced by majorana fermions},}\ }\href {\doibase
  10.1103/PhysRevLett.112.037001} {\bibfield  {journal} {\bibinfo  {journal}
  {Phys. Rev. Lett.}\ }\textbf {\bibinfo {volume} {112}},\ \bibinfo {pages}
  {037001} (\bibinfo {year} {2014})}\BibitemShut {NoStop}%
\bibitem [{\citenamefont {Sun}\ \emph {et~al.}(2016)\citenamefont {Sun},
  \citenamefont {Zhang}, \citenamefont {Hu}, \citenamefont {Li}, \citenamefont
  {Wang}, \citenamefont {Ma}, \citenamefont {Xu}, \citenamefont {Gao},
  \citenamefont {Guan}, \citenamefont {Li}, \citenamefont {Liu}, \citenamefont
  {Qian}, \citenamefont {Zhou}, \citenamefont {Fu}, \citenamefont {Li},
  \citenamefont {Zhang},\ and\ \citenamefont {Jia}}]{Sun2016Majorana}%
  \BibitemOpen
  \bibfield  {author} {\bibinfo {author} {\bibfnamefont {Hao-Hua}\ \bibnamefont
  {Sun}}, \bibinfo {author} {\bibfnamefont {Kai-Wen}\ \bibnamefont {Zhang}},
  \bibinfo {author} {\bibfnamefont {Lun-Hui}\ \bibnamefont {Hu}}, \bibinfo
  {author} {\bibfnamefont {Chuang}\ \bibnamefont {Li}}, \bibinfo {author}
  {\bibfnamefont {Guan-Yong}\ \bibnamefont {Wang}}, \bibinfo {author}
  {\bibfnamefont {Hai-Yang}\ \bibnamefont {Ma}}, \bibinfo {author}
  {\bibfnamefont {Zhu-An}\ \bibnamefont {Xu}}, \bibinfo {author} {\bibfnamefont
  {Chun-Lei}\ \bibnamefont {Gao}}, \bibinfo {author} {\bibfnamefont {Dan-Dan}\
  \bibnamefont {Guan}}, \bibinfo {author} {\bibfnamefont {Yao-Yi}\ \bibnamefont
  {Li}}, \bibinfo {author} {\bibfnamefont {Canhua}\ \bibnamefont {Liu}},
  \bibinfo {author} {\bibfnamefont {Dong}\ \bibnamefont {Qian}}, \bibinfo
  {author} {\bibfnamefont {Yi}~\bibnamefont {Zhou}}, \bibinfo {author}
  {\bibfnamefont {Liang}\ \bibnamefont {Fu}}, \bibinfo {author} {\bibfnamefont
  {Shao-Chun}\ \bibnamefont {Li}}, \bibinfo {author} {\bibfnamefont {Fu-Chun}\
  \bibnamefont {Zhang}}, \ and\ \bibinfo {author} {\bibfnamefont {Jin-Feng}\
  \bibnamefont {Jia}},\ }\bibfield  {title} {\enquote {\bibinfo {title}
  {{Majorana Zero Mode Detected with Spin Selective Andreev Reflection in the
  Vortex of a Topological Superconductor}},}\ }\href {\doibase
  10.1103/PhysRevLett.116.257003} {\bibfield  {journal} {\bibinfo  {journal}
  {Phys. Rev. Lett.}\ }\textbf {\bibinfo {volume} {116}},\ \bibinfo {pages}
  {257003} (\bibinfo {year} {2016})}\BibitemShut {NoStop}%
\bibitem [{\citenamefont {Jeon}\ \emph {et~al.}(2017)\citenamefont {Jeon},
  \citenamefont {Xie}, \citenamefont {Li}, \citenamefont {Wang}, \citenamefont
  {Bernevig},\ and\ \citenamefont {Yazdani}}]{Jeon2017}%
  \BibitemOpen
  \bibfield  {author} {\bibinfo {author} {\bibfnamefont {Sangjun}\ \bibnamefont
  {Jeon}}, \bibinfo {author} {\bibfnamefont {Yonglong}\ \bibnamefont {Xie}},
  \bibinfo {author} {\bibfnamefont {Jian}\ \bibnamefont {Li}}, \bibinfo
  {author} {\bibfnamefont {Zhijun}\ \bibnamefont {Wang}}, \bibinfo {author}
  {\bibfnamefont {B.~Andrei}\ \bibnamefont {Bernevig}}, \ and\ \bibinfo
  {author} {\bibfnamefont {Ali}\ \bibnamefont {Yazdani}},\ }\bibfield  {title}
  {\enquote {\bibinfo {title} {{Distinguishing a Majorana zero mode using
  spin-resolved measurements}},}\ }\href {\doibase 10.1126/science.aan3670}
  {\bibfield  {journal} {\bibinfo  {journal} {Science}\ }\textbf {\bibinfo
  {volume} {358}},\ \bibinfo {pages} {772--776} (\bibinfo {year}
  {2017})}\BibitemShut {NoStop}%
\end{thebibliography}%

\end{document}